\documentclass[twocolumn,showpacs,superscriptaddress]{revtex4}

\usepackage{graphicx}
\usepackage{amsmath}
\usepackage{amsfonts}
\usepackage{amssymb}
\usepackage{epsf}
\usepackage{hyperref}
\usepackage{color}

\newcommand{\eps}[0]{\varepsilon}

\newcommand{\ic}{\mathrm{i}}

\newcommand{\ud}{\mathrm{d}}

\usepackage{ulem}
\usepackage{color}

\begin{document}

\title{Routes towards the experimental observation of the large fluctuations due to chaos assisted tunneling effects with cold atoms}

\author{R. Dubertrand}
\affiliation{Institut  de  Physique  Nucl\'eaire,  Atomique  et  de  Spectroscopie,
Universit\'e  de  Li\`ege,  B\^at.  B15,  B  -  4000  Li\`ege,  Belgium}
\affiliation{Laboratoire de Physique Th\'eorique, IRSAMC, Universit\'e de Toulouse, CNRS, UPS, France}
\author{J. Billy}
\affiliation{Laboratoire Collisions, Agr\'egats, R\'eactivit\'e, IRSAMC, Universit\'e de Toulouse, CNRS, UPS, France}
\author{D. Gu\'ery-Odelin}
\affiliation{Laboratoire Collisions, Agr\'egats, R\'eactivit\'e, IRSAMC, Universit\'e de Toulouse, CNRS, UPS, France}
\author{B. Georgeot} 
\affiliation{Laboratoire de Physique Th\'eorique, IRSAMC, Universit\'e de Toulouse, CNRS, UPS, France}
\author{G. Lemari\'e}
\affiliation{Laboratoire de Physique Th\'eorique, IRSAMC, Universit\'e de Toulouse, CNRS, UPS, France}
\date{\today}

\begin{abstract}
In the presence of a complex classical dynamics associated with a mixed phase space, a quantum wave function can tunnel between two stable islands through the chaotic sea, an effect that has no classical counterpart. This phenomenon, referred to as chaos assisted tunneling, is characterized by large fluctuations of the tunneling rate when a parameter is varied. 
To date the full extent of this effect as well as the associated statistical distribution have never been observed in a quantum system. Here we analyze the possibility of characterizing these effects accurately in a cold atom experiment.
Using realistic values of the parameters of an experimental setup, we examine through analytical estimates and extensive numerical simulations a specific system that can be implemented with cold atoms, the atomic modulated pendulum. We assess the efficiency of three possible routes to observe in detail chaos assisted tunneling properties. Our main conclusion is that due to the fragility of the symmetry between positive and negative momenta as a function of quasimomentum, it is very challenging to use tunneling between classical islands centered on fixed points with opposite momentum. We show that it is more promising to use islands symmetric in position space, and characterize the regime where it could be done. The proposed experiment could be realized with current state-of-the-art technology.
\end{abstract}
\maketitle

\section{Introduction}

Tunneling is 
a cornerstone of quantum mechanics associated with the wave nature of particles.
Standard tunneling has been known from the early years of the quantum theory, and is usually studied through the crossing of an energy barrier in one dimension. In this case, it is possible to solve analytically many simple systems, and the tunneling amplitudes are simple functions decreasing exponentially with the height of the barrier and the inverse of the Planck constant $\hbar$ (see e.g. \cite{landau}). 

The tunnel effect occurs in many other contexts with specific properties, and this even in the absence of repulsive barriers. For instance, the Landau-Zener transition through the gap of two adjacent energy bands of a periodic lattice is nothing but a tunnel effect in the reciprocal space \cite{LZ}.  These effective barriers can be transposed in real space (spatial gaps) by shaping the envelope of the lattice \cite{Santos,Carusotto,EPLSG,PRAGAPS}. Similarly, it is well known that  two zones of a classical phase space separated by an invariant {manifold} of a dynamical system cannot be connected classically. However, in quantum mechanics it is possible to cross the invariant {manifold} through a process called dynamical tunneling \cite{heller, semicl}. 

In an integrable system where constants of motion exist and the classical trajectories are confined due to invariant structures of lower dimensionality, the simple picture of one dimensional potential barrier tunneling can be readily extended, with simple exponential dependence of the tunneling amplitudes on parameters such as $\hbar$. However, in general for multi-dimensional systems new phenomena arise. Indeed, one-dimensional time independent system are always integrable, while generically in higher dimensions, systems exhibit various levels of chaos. The most generic case corresponds to {\it mixed systems}, where chaotic and regular zones coexist in phase space \cite{physrep}. In this situation, one can consider tunneling between two symmetric regular islands separated by a chaotic sea. In this case, tunneling becomes a much more subtle process called chaos assisted tunneling \cite{tomul}. Intuitively, it can be understood by the fact that the chaotic sea being ergodic, instead of tunneling directly from one island to the other, it is preferable to tunnel to the chaotic sea and from there to the other island. However, this process, being mediated by quantum states in the chaotic sea, depends on the energies of these states, which vary with the parameters of the system. Instead of simple monotonous exponential laws, the tunneling leads to strongly fluctuating quantities, which can vary by orders of magnitude over small {changes} of parameters.
  
In the case of tunneling between symmetric islands, tunneling manifests itself through an energy splitting between symmetric and antisymmetric states. The energy splitting distribution in this context 
was determined in \cite{leyvraz} and a more refined semiclassical version was developed in \cite{narim}. A version of chaos assisted tunneling using the theory of increase of a perturbation through chaos was explained in
\cite{aberg}. Last the presence of chaos can be also seen by considering the large fluctuations of the tunneling rate from one stable island to the surrounding chaotic sea \cite{peter}. The effects of chaos assisted tunneling have been very challenging to confirm in experiments. The main experimental signatures have been obtained using microwave cavities  \cite{mushroom,microwave1,microwave} with a recent significant progress to access very small energy splittings \cite{barbara}.

Applications of chaos assisted tunneling to cold atom systems were discussed in 
 \cite{delande1}, and applications of chaotic tunneling to cold atom systems were studied in \cite{delande2}. In  \cite{artuso} the effects of nonlinearities due
to mean-field interactions (Gross-Pitaevski) on dynamical tunneling were studied.
In the mean time there were two pioneering experiments on chaos assisted tunneling with cold atoms \cite{RaizenScience,PhillipsNature}. A theoretical analysis of the results of the latter experiment has been performed in \cite{PhillipsPRA}. The main conclusion is that, although some tunneling fluctuations have indeed been observed, the very strong fluctuations of the tunneling period associated to chaos assisted tunneling in a cold atom experiment have remained elusive. {Finally, there has been another recent proposal for the observation of dynamical, and possibly chaos assisted, tunneling in a cold atom experiment \cite{lenz}, which investigates a different regime from that of the present proposal.}

In this paper we want to assess theoretically  different routes to observe chaos assisted tunneling with cold atoms, using as main system the atomic modulated pendulum such as in \cite{RaizenScience,PhillipsNature}.  We will discuss the direct measure of the tunneling oscillations from islands symmetric in momenta as in  \cite{RaizenScience,PhillipsNature}, and also an alternative proposed in \cite{delande1} to use a Landau-Zener scheme to improve the efficiency of this {approach}. Our results using analytical estimates and extensive numerical simulations show that both these routes are experimentally challenging to get a clear signature of the theory of chaos assisted tunneling. We propose a third route using tunneling between classical islands symmetric in the coordinate axis, and show that this scheme overcomes most of the drawbacks of the preceding ones.
We devote a special effort to make quantitative predictions, which can be directly used in the experiment performed by experimental groups.

The paper is organized as follows. In Sect.~\ref{model} we present the model of the atomic modulated pendulum used throughout this work. We discuss its classical and quantum dynamics, present the theoretical basis of chaos assisted tunneling in this context, and specify the experimental parameters. 
The simplest method to directly observe chaos assisted tunneling between islands symmetric in momentum is exposed in Sect.~\ref{tunnelp_oscill}. We show that this method is difficult to implement accurately due to the instability of the symmetry $p \rightarrow -p$ with respect to quasimomentum changes. A more efficient method proposed in \cite{delande1} using a Landau-Zener scheme is described in Sect.~\ref{tunnelp_LZ}. Analytical estimates and extensive numerical simulations allow us to predict that this method will also be very difficult to implement in order to get accurate results. Then an alternative approach to observe standard tunneling using islands symmetric in space is detailed in Sect.~\ref{tunnelx}. An experimental protocol is given in Sect.~\ref{protocol_exp} in order to observe chaos assisted tunneling in a cold atom experiment in this situation. Last a conclusion is drawn in Sect.~\ref{conclusion}.

\section{Chaos assisted tunneling in a cold atom experiment}
\label{model}

\subsection{The atomic modulated pendulum}

The atomic modulated pendulum describes atom{s} in a time modulated standing wave, {which} can be tuned to regimes where chaotic dynamics and tunneling coexist. The cold atom experiments use atoms of mass $M$, identified as two-level atoms. 
They are embedded in a quasi $1-$dimensional guide, say along the $X$ axis.
Two counter propagating laser beams create an optical lattice of spatial period $d$. This length scale introduces natural units of velocity and energy:
\begin{equation}
  \label{scales}
v_L=\frac{h}{M d},\;\;{\rm and }\;\;\quad E_L=M\frac{v_L^2}{2}\ ,
\end{equation}
where 
$h$ is the Planck constant. The energy $E_L$ is the lattice characteristic energy scale \footnote{$E_L=E_R/4$ where $E_R$ refers to the recoil energy \cite{LZ}.}.

The optical lattice is modulated in time in a sinusoidal way with an amplitude $\varepsilon$ and an angular frequency $\omega$, so that the Hamiltonian reads \cite{delande1}:{
\begin{equation}
    H(P,X,\mathcal T)= \frac{P^2}{2M}-\frac{U_0}{2} \left(1+\varepsilon\cos\omega \mathcal T \right) \; \cos\left( \frac{2\pi X}{d} \right),
\end{equation}}
where $U_0$ is the depth of the periodic potential for $\varepsilon=0$. We introduce the dimensionless variables:
{
\begin{equation}
  \label{rescale}
  p=\dfrac{2\pi}{M\omega d} P,\quad x=\frac{2\pi X}{d},\quad t=\omega \mathcal T\ ,
\end{equation}}
and the dimensionless parameter{
\begin{equation}
  \label{defgamma}
  \gamma=\left(\dfrac{E_L}{\hbar\omega}\right)^2 \frac{U_0}{E_L}
=\left(\dfrac{E_L}{\hbar\omega}\right)^2 s\ .
\end{equation}}
The parameter $s=U_0/E_L$ can be seen as the dimensionless depth of the lattice wells. With these variables the effective Planck constant is
\begin{equation}
  \label{defhbar_eff}
  \hbar_{\rm eff}\equiv -\ic [ x,p ]=\frac{2E_L}{\hbar\omega}\ ,
\end{equation}
a quantity that can be tuned with the modulation frequency.
Using the new variables, the dimensionless Hamiltonian reads:{
\begin{equation}
 H(p,x,t)=\frac{p^2}{2}-\gamma (1+\varepsilon \cos t)\cos x\ .
\label{Hclas}
\end{equation} }
which will be used throughout the paper.

\subsection{Classical dynamics}

\begin{figure*}
  \centering
  \includegraphics[width=0.5\textwidth]{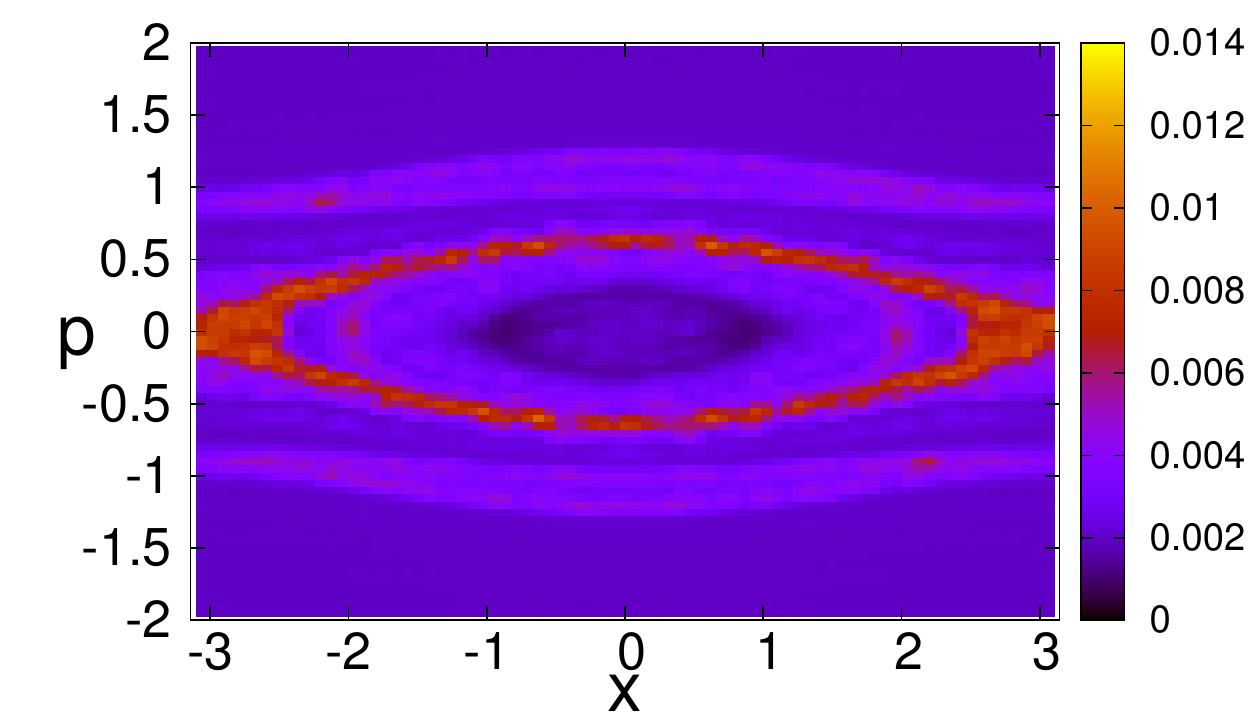}
  \includegraphics[width=0.375\textwidth]{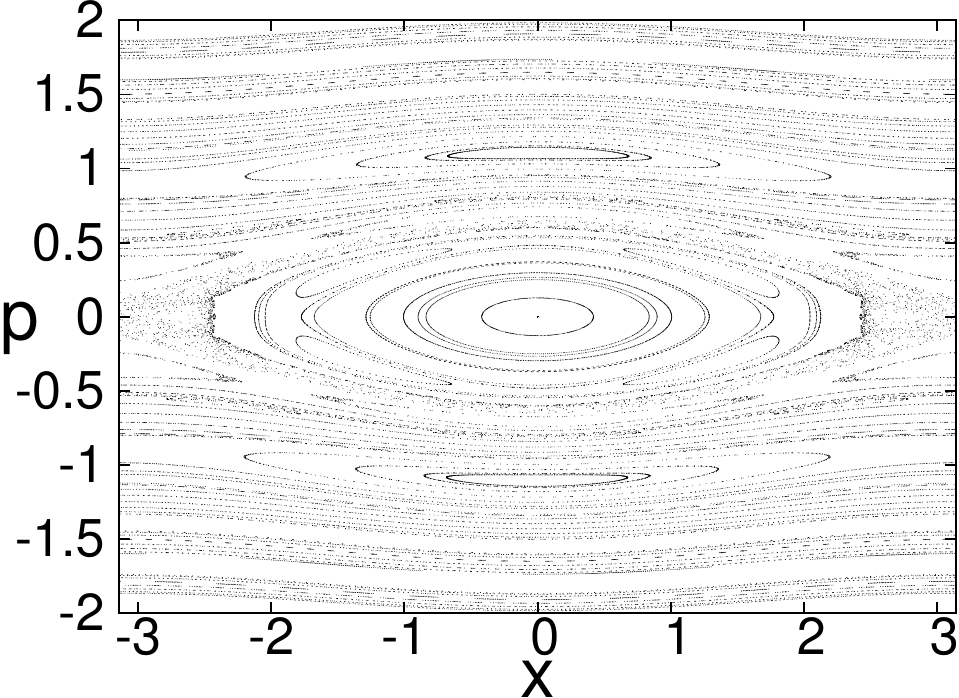}\\
  \includegraphics[width=0.5\textwidth]{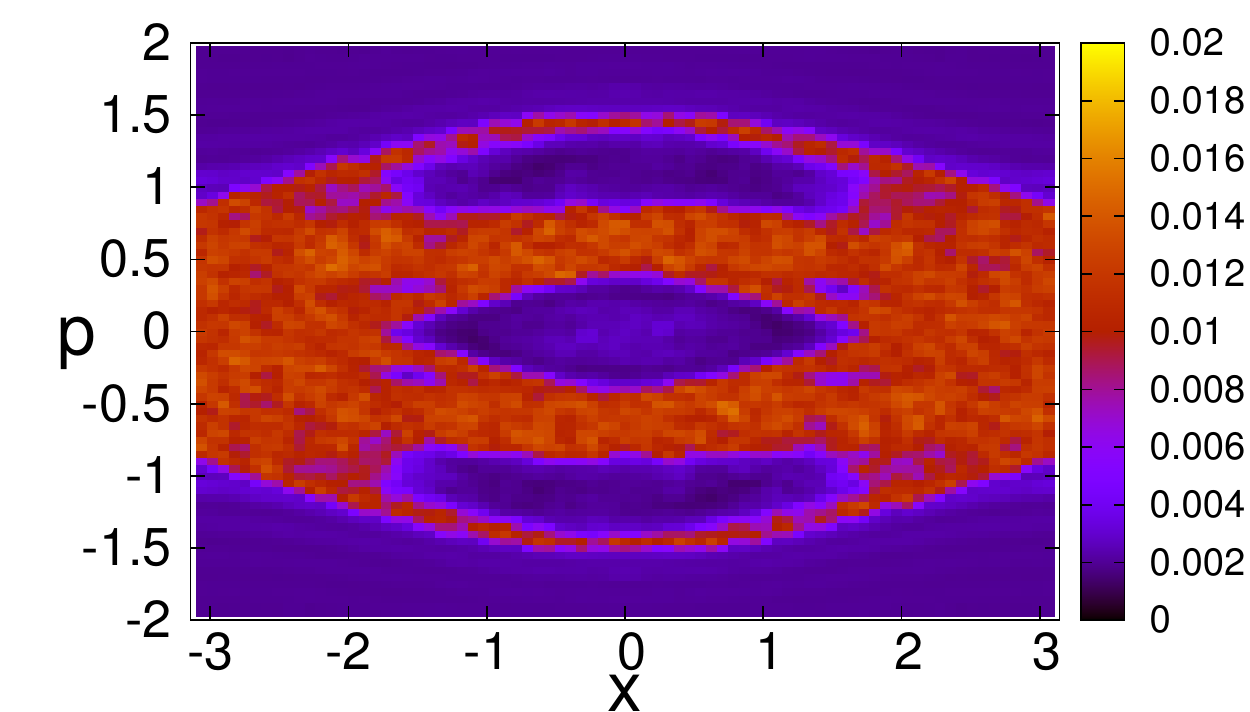}
  \includegraphics[width=0.375\textwidth]{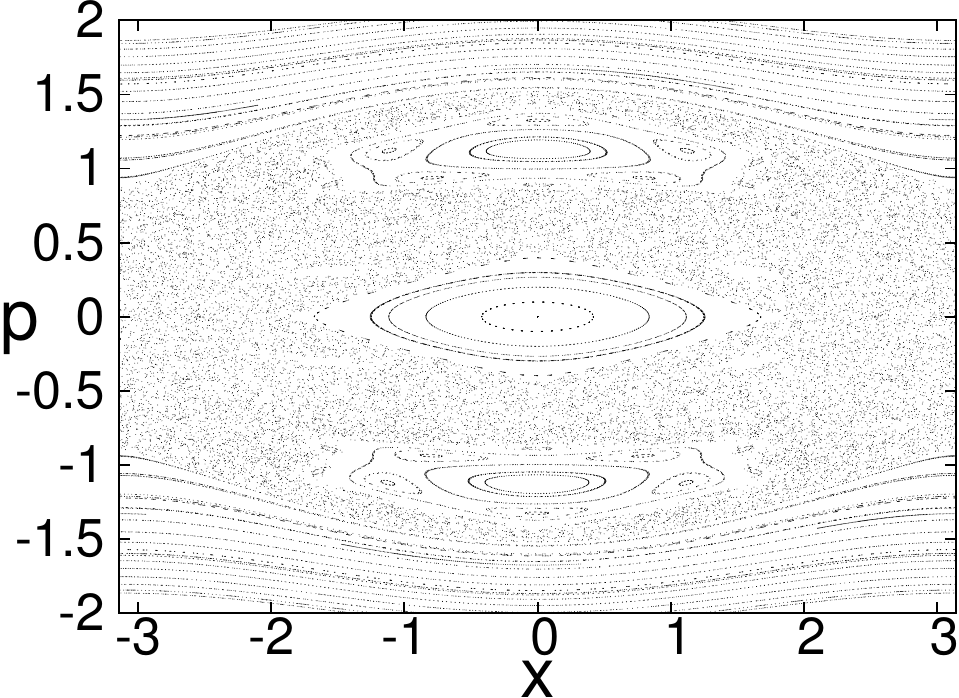}
\caption{(Color online) Comparison of the Lyapunov chart (left) and the Poincar\'e SOS (right). Top $\varepsilon=0.1$ and $\gamma=0.1$. Bottom: $\varepsilon=0.9$ and $\gamma=0.1$. 
For every value of $(\gamma,\varepsilon)$ the phase space is subdivided into boxes. The local Lyapunov exponent, which measures the divergence of nearby trajectories, is evaluated in each box. One obtains a Lyapunov chart as shown in the left column. The color (gray) scale measures the magnitude of the local Lyapunov exponent. 
The more usual Poincar\'e surface of section (SOS) is shown in the right column. 
The SOS describes iterates of trajectories stroboscopically at each period of the Hamiltonian (\ref{Hclas}). Regular zones correspond to lines in this representation, while chaotic zones are visible as clouds of points. The figure shows that the Lyapunov chart correctly identifies both chaotic and regular regions.
}
\label{SOS_vs_Lyap}
\end{figure*}

Classically, the Hamiltonian (\ref{Hclas}) depends on two parameters $\gamma$ and $\varepsilon$ that can be modified through the properties of the lattice and its modulation. 
The classical dynamics follows the Hamilton equations:
\begin{eqnarray}
    \dfrac{\ud x}{\ud t} & = & \dfrac{\partial H}{\partial p}=p, \nonumber \\
  \dfrac{\ud p}{\ud t} & = & -\dfrac{\partial H}{\partial x}=-\gamma(1+\eps\cos t)\sin x.
\label{clasdyn}
\end{eqnarray}

For $\varepsilon=0$ or $\gamma=0$ the system is integrable (energy is a constant of motion). When $\varepsilon$ and $\gamma$ increase from zero, the system is no longer integrable and the phase space displays chaotic and regular zones. Three {main} regular islands appear around resonances {at $(x,p)=(0,0)$, $(0,p_*)$ and $(0,-p_*)$, with $p_*\approx 1$}, one being static and the two other ones corresponding to effective potentials moving at constant velocities to the right or to the left, see Fig.~\ref{SOS_vs_Lyap}. It can be seen analytically by rewriting the second part of (\ref{Hclas}) as a sum of three cosines, namely $\cos x$, $\cos (x-t)$, $\cos (x+t)$. For certain choices of $(\gamma,\varepsilon)$ it is thus possible to have two stable islands, which are symmetric under $p\mapsto -p$.

In order to study chaos assisted tunneling, it is important to be in the case where these stable islands are separated by a chaotic region in phase space. {The relevant parameter regions in the $(\gamma,\varepsilon)$ plane can be precisely determined as follows. 
{An} important measure of "chaoticity" is the Lyapunov exponent, which corresponds to the rate of divergence of nearby trajectories. 
For every value of $(\gamma,\varepsilon)$ we {thus} subdivide our phase space into boxes, {where t}he local Lyapunov exponent is evaluated. 
A comparison between such an obtained Lyapunov chart and the more usual Poincar\'e surface of section (SOS) is shown in Fig.~\ref{SOS_vs_Lyap}, which illustrates the accuracy of our method. 
Fig.~\ref{SOS_vs_Lyap} contains both a regime of very weak chaos (very small chaotic zones) and another regime where the two symmetric islands are separated by a large chaotic sea. In both cases the Lyapunov chart correctly identifies the chaotic zones.

This enables us to use the Lyapunov chart to 
evaluate the area of the chaotic region in the phase space. First we looked at several Lyapunov charts and compared it to the usual Poincar\'e SOS. This allows to define a threshold value ($\lambda_c=0.012$) to claim that one box contains mainly chaotic trajectories. Then we count the number of boxes, inside which the local Lyapunov exponent exceeds that threshold.
This is what we call here the degree of chaos, or "chaoticity", of our system. 
Figure~\ref{degree_chaos} shows how it varies in the parameter space.
As one can expect, the system is more and more chaotic when either $\gamma$ or $\varepsilon$ increases. This representation of the system enables one to identify easily the different regimes in the parameter space, and to spot the parameter zones where the dynamics will make chaos assisted tunneling effects especially prominent.

\begin{figure}
  \centering
  \includegraphics[width=\linewidth]{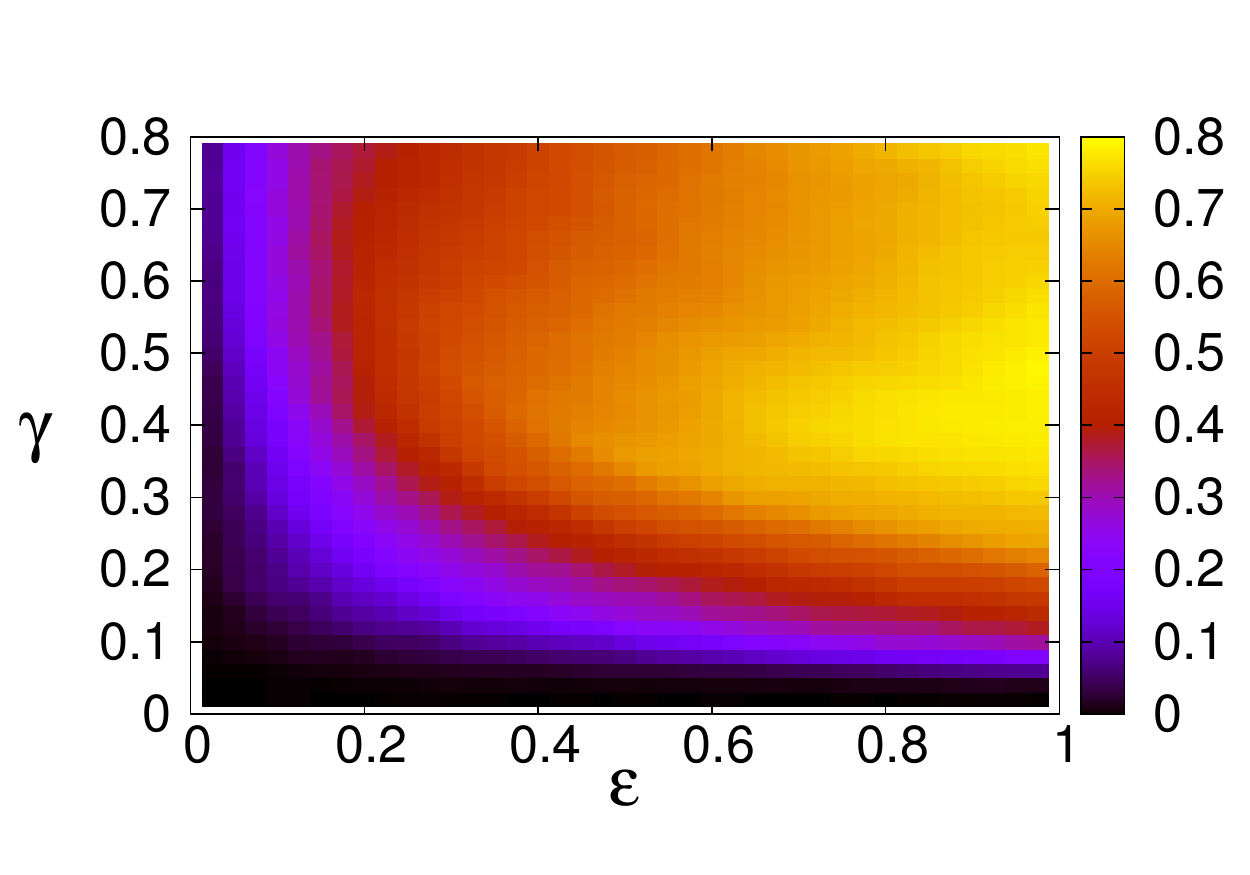}
  \caption{(Color online) "Measure of the chaoticity" in the classical modulated pendulum. The color (gray scale) gives a heuristic estimate of the proportion of chaotic regions in the phase space cell (see text).}
  \label{degree_chaos}
\end{figure}

\begin{figure}
  \centering
  \includegraphics[width=\linewidth]{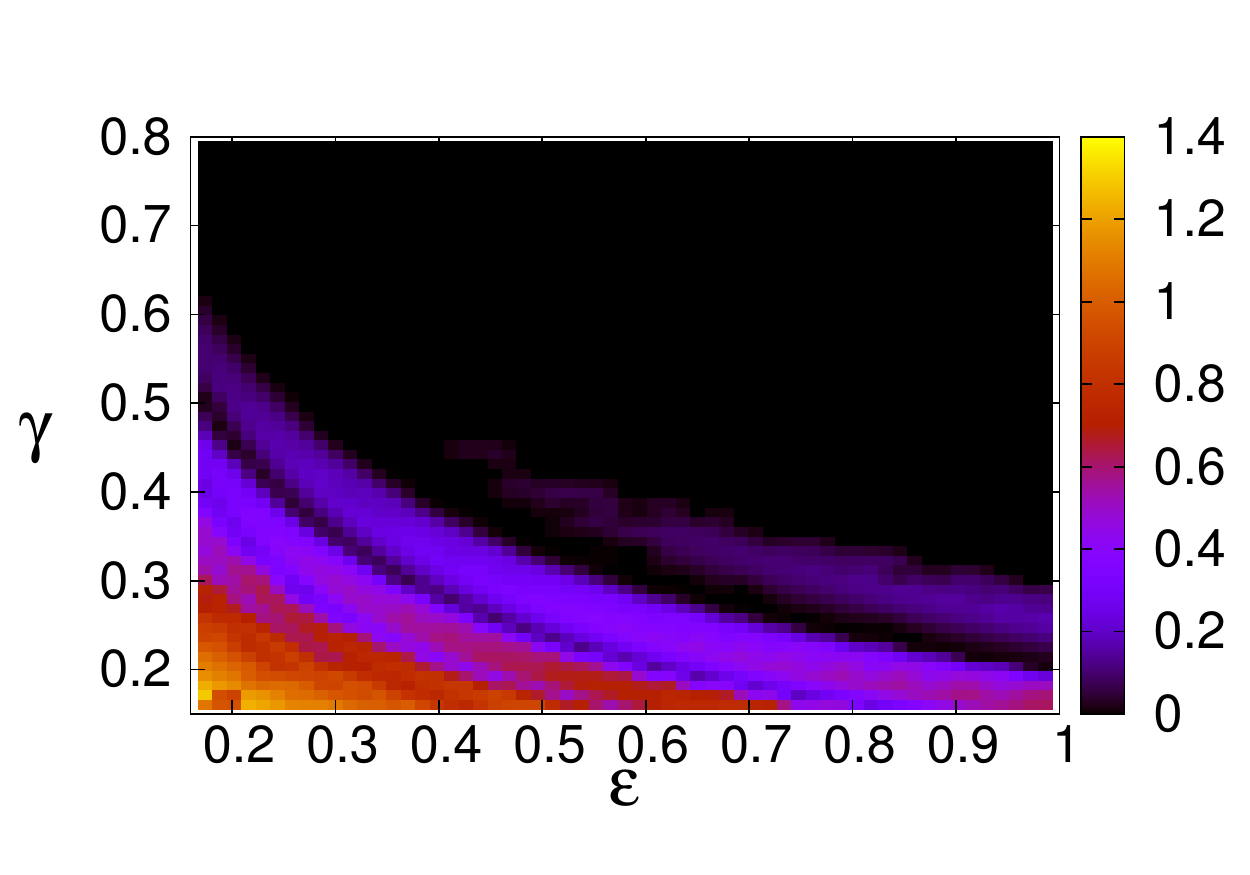}
  \caption{(Color online) Size of the islands along the $p$ axis. The color gives a numerical estimate of the area of the stable islands centered at the points with phase coordinates $(0,\pm p_*)$. First a sampling along the momentum axis is performed to identify the fixed point in the upper half phase space with high precision. Then a conformal mapping is used to change coordinates into polar coordinates $(\theta,I)$ centered at this fixed point. Launching trajectories far from the origin in this frame allows to draw the boundary of the island. The area is eventually given by the integral of $I(\theta)$ along the boundary. The color (gray) scale stands for the value of this area.}
  \label{size_islandsp}
\end{figure}

Additionally, in order to study tunneling between islands, we want to determine the range in the parameter plane {$(\gamma, \varepsilon)$} such that there are two large symmetric stable islands, centered around $p_*\simeq\pm1$. Figure~\ref{size_islandsp} shows how the area of these islands varies in the parameter plane. In this paper we will detail the case of several choices of classical parameters especially relevant to ensure both a large central chaotic region and a pair of large stable islands along the $p$ axis{: $(\gamma, \varepsilon)=(0.25, 0.4)$ and $(0.29,0.29)$ mainly.} An example of the Poincar\'e SOS corresponding to {$(\gamma, \varepsilon)=(0.25, 0.4)$} is plotted in Fig.~\ref{g0.25_eps0.4}.

\begin{figure}
  \centering
  \includegraphics[width=\linewidth]{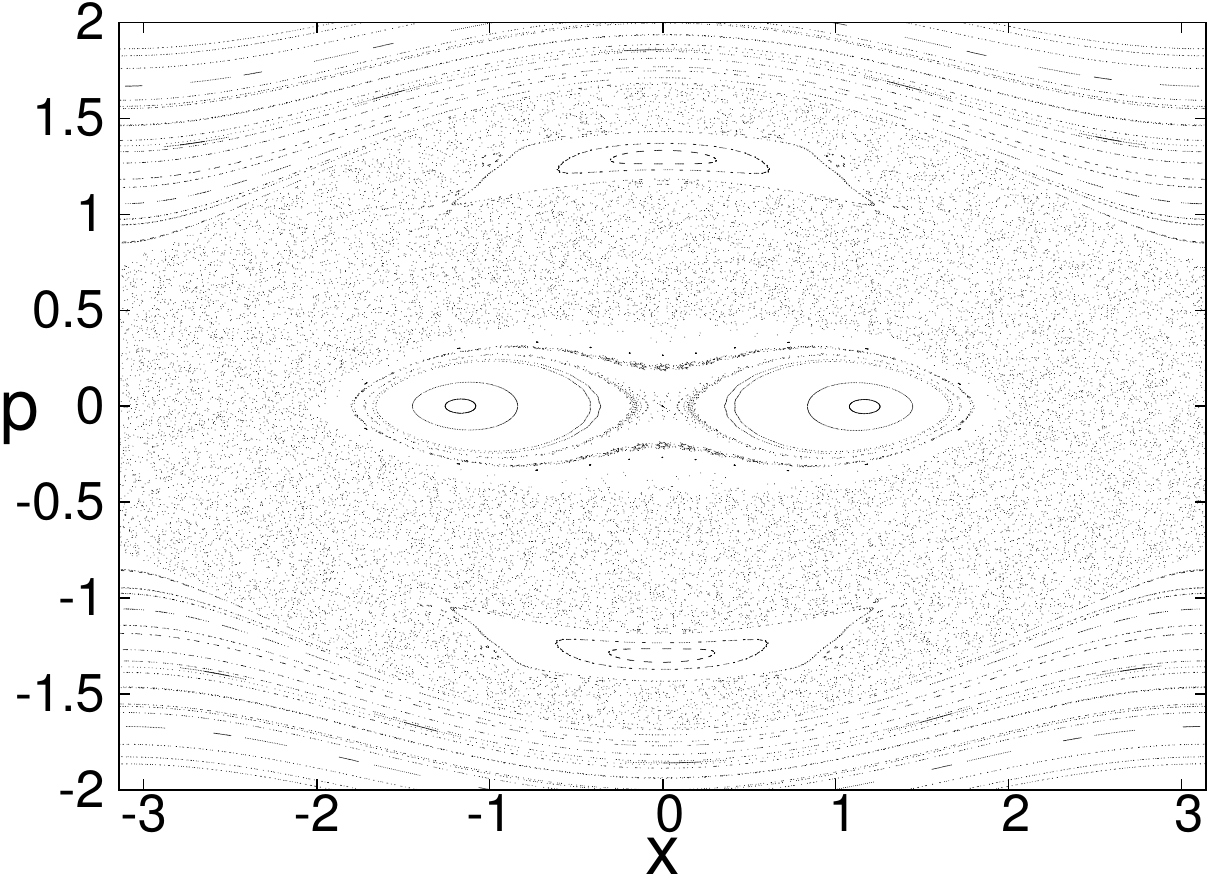}
  \caption{Poincar\'e SOS for the Hamiltonian (\ref{Hclas}) for $\gamma=0.25$ and $\varepsilon=0.4$. }
  \label{g0.25_eps0.4}
\end{figure}

\subsection{Quantum dynamics}

The Hamiltonian (\ref{Hclas}) is periodic in both time and space. Due to the first symmetry, it is more convenient to analyze the time dynamics following Floquet theory. One considers the quantum propagator $\hat{U}$ over one period associated to the Hamiltonian (\ref{Hclas}). {This evolution operator $\hat{U}$ is unitary and admits Floquet eigenstates $\vert \psi_n \rangle$, and quasienergies $E_n$ defined by
\begin{equation}
  \label{eig_eqU}
  \hat{U}\left|\psi_n\right>= e^{-\ic 2\pi E_n/\hbar_{\rm eff}} \left|\psi_n\right>\ .
\end{equation}}
The $2\pi$ factor comes from the fact that the time period of (\ref{Hclas}) is $2\pi$. 
The second symmetry, the periodicity in space, implies that the quantum propagator $\hat{U}$ commutes with a discrete family of translation (in space) operators. These operators are all multiples of the elementary translation operator $\hat{T}=e^{-\ic 2\pi \hat{p}/\hbar_{\rm eff}}$. Again here, the $2\pi$ factor comes from the fact that the spatial period of (\ref{Hclas}) is $2\pi$.
Therefore one can find an eigenbasis{, the Bloch wave basis,} that makes both $\hat{U}$ and $\hat{T}$ diagonal. 
Each Bloch wave $\left|\psi_{n,\beta}\right>$ is labeled by a real number, {the quasimomentum $\beta$,} and is the solution of the eigenvalue problem:
\begin{equation}
  \hat{U}_\beta\left|\psi_{n,\beta}\right>=e^{-\ic 2\pi {E_n}(\beta)/\hbar_{\rm eff}} \left|\psi_{n,\beta}\right>,
\label{eigvectUbeta}
\end{equation}
where $\hat{U}_\beta$ is the quantum propagator $\hat{U}$ 
with the substitution $\hat{p}\mapsto\hat{p}+\hbar_{\rm eff}\beta$. 
{Numerically, $\hat{U}_\beta$ can be obtained efficiently using the split-step Fourier method \cite{splitstep}.} When $\beta$ spans the first Brillouin zone, which is $[-1/2;1/2[$ here, the computation of {$E_n(\beta)$} for every $n$ leads to the band diagram of the problem.

The Hamiltonian (\ref{Hclas}) is invariant under two reflection symmetries. The first one is $x\mapsto -x$ and survives in the Floquet-Bloch picture. The second, which is crucial for our subsequent analysis, is the symmetry under $p\mapsto -p$. One can see immediately {that the substitution $\hat{p}\mapsto\hat{p}+\hbar_{\rm eff}\beta$ in the Hamiltonian \eqref{Hclas} breaks this symmetry whenever $\beta\ne 0$}. Therefore {the momentum symmetry} appears to be very sensitive to the quasimomentum. This high sensitivity will play a key role in the simulation of cold atom experiment output and will be discussed in more details below.

\subsection{Chaos assisted tunneling}

\begin{figure}
\includegraphics[width=1.\linewidth]{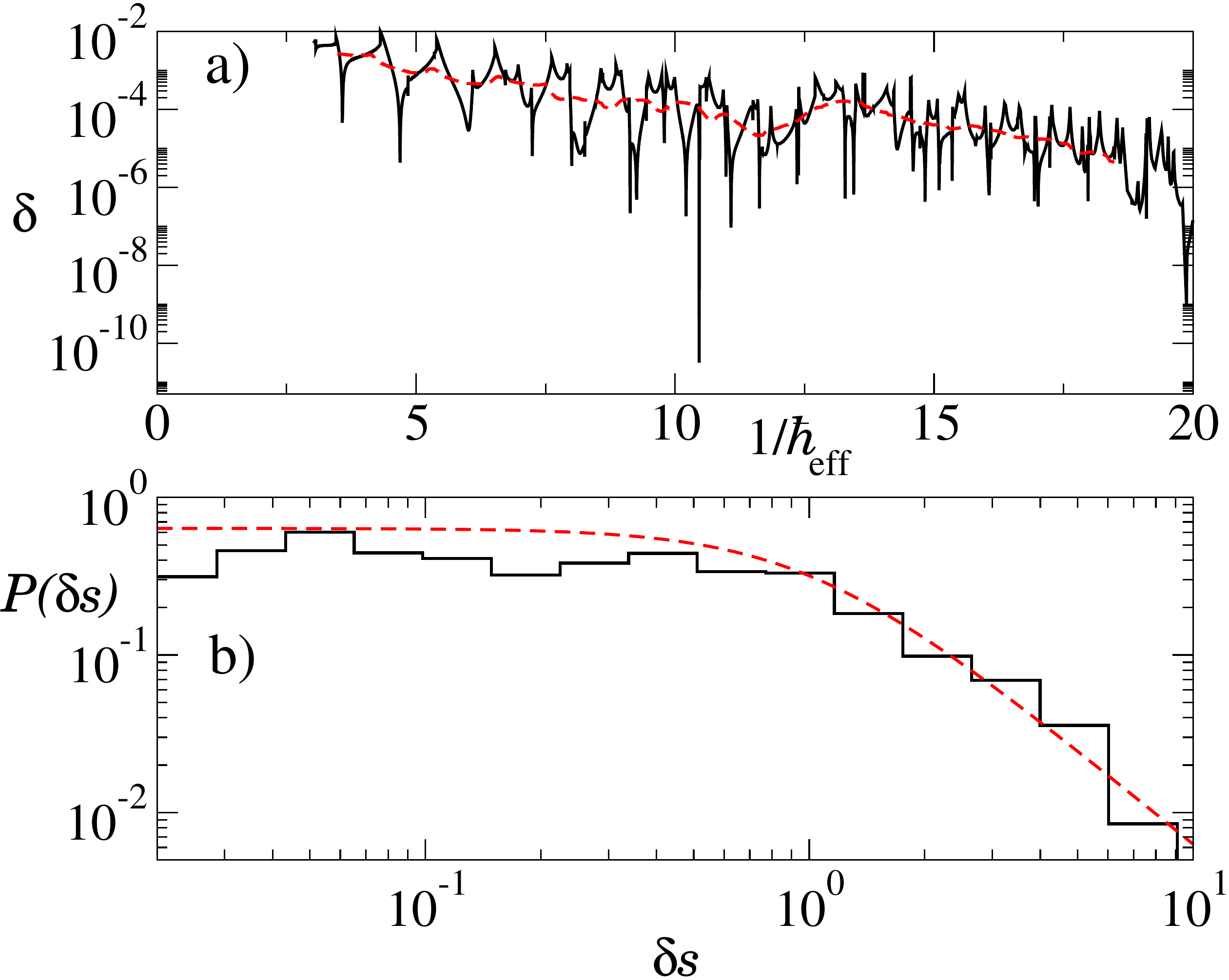}
  \caption{(Color online) a) Black line: Energy splitting (\ref{def_splitting}) as a function of $1/\hbar_{\rm eff}$ for the Hamiltonian (\ref{Hclas}) with $\varepsilon=0.4$ and $\gamma=0.25$ associated to the symmetric islands {along} the momentum axis. Dashed red line: typical value. b) Black line: Histogram of the fluctuation of the splitting renormalized by its typical value, $\delta s=\delta/\delta_{\rm typ}$. The dashed red line stands for the prediction (\ref{cauchy}).}
\label{split_eps0.4_g0.25}
\end{figure}
{Quantum tunneling couples classically disconnected regions of phase space separated by invariant curves. In the case of symmetric stable islands,} quantum eigenstates in the islands come in pairs of symmetric and antisymmetric states. 
One eigenstate, say $|\psi_+\rangle$,  corresponds to a symmetric superposition of wave packets on each island whereas the other $|\psi_-\rangle$ corresponds to an antisymmetric superposition. {Because} the islands are coupled by {quantum tunneling}, the energy levels associated to these two eigenstates, denoted by {$E_+$ and $E_-$} respectively, do not exactly coincide. They differ by a small quantity, which we will call from now on the (energy) splitting:{
\begin{equation}
  \label{def_splitting}
  \delta=|E_+-E_-|\ .
\end{equation}}
In an integrable system where no chaotic zone is present, the results of standard one-dimensional tunneling can be extended; tunneling splittings decrease exponentially as a function of the ratio between the action integral 
along the tunneling barrier 
and {$\hbar_{\rm eff}$}. In this case, tunneling splittings vary smoothly and monotonically with the parameters of the system. 

In the case where a chaotic zone separates the two islands, standard tunneling becomes chaos assisted tunneling. Instead of a simple tunneling directly from one island to the other, the preferred mechanism is the coupling to an ergodic state inside the chaotic sea, which makes the tunneling splittings dependent on the precise phase space and energy 
position of these chaotic states. Thus the main hallmark of chaos assisted tunneling is the very strong fluctuations, by orders of magnitude, of the splitting over small parameter ranges \cite{tomul,leyvraz}. These reproducible variations cannot be described any more by smooth monotonous functions, but represent a quantum interference signature of chaotic dynamics between the stable islands. In this sense, chaos assisted tunneling is reminiscent of the universal conductance fluctuations observed in mesoscopic systems \cite{ucf1,ucf2}.


More precisely one can adopt a statistical approach to account for these large variations and define the probability distribution of the fluctuation $\delta s=\delta/\delta_{\rm typ}$ of the splitting from its typical value. In \cite{leyvraz} it was shown that under very general assumptions this distribution corresponds to the Cauchy distribution described by: 
\begin{equation}
  \label{cauchy}
  P(\delta s)=\frac{2}{\pi}\frac{1}{1+\delta s^2}\ .
\end{equation}
Although predicted since several decades, this distribution for the splitting fluctuations has never been observed experimentally in a quantum system.

In Fig.~\ref{split_eps0.4_g0.25} we show the variations of the splitting when the semiclassical parameter $\hbar_{\rm eff}$ is varied. {We consider the parameter values $(\gamma,\varepsilon)=(0.25,0.4)$, with the corresponding Poincar\'e SOS displayed in Fig.~\ref{g0.25_eps0.4}. The symmetric stable islands are centered around the fixed points with coordinates $(0,\pm p_*)$, and $p_*\simeq 1.296$. The energy splitting is computed for $\beta=0$ where the momentum symmetry $p\mapsto -p$ is preserved.} The fluctuations are in good agreement with the prediction (\ref{cauchy}).

We note that an important aspect of chaos assisted tunneling is that, in order to be observed and characterized, the value of $\hbar_{\rm eff}$ should be small enough. Indeed, chaos assisted tunneling is mainly a semiclassical effect, and requires quantum states to be unambiguously associated with the chaotic or integrable parts of phase space. In order to reach this regime,  $\hbar_{\rm eff}$ should be significantly smaller than the phase space area $\mathcal A$ of the islands {($\mathcal A \approx 0.530$ in the case of Fig.~\ref{split_eps0.4_g0.25})} and of the chaotic sea. 

\subsection{Range of parameters accessible to an experiment}

In the preceding subsections we studied the atomic modulated pendulum and showed that it can display effects of chaos assisted tunneling. In a cold atom setting, this system is built from the interaction of atoms with stationary laser waves
\cite{RaizenScience,PhillipsNature,delande1}.
In this last subsection we detail the typical {accessible} values of the corresponding {experimental} parameters.
As an example we consider a cold atom setup with $^{87}$Rb atoms ($M=1.45 \times 10^{-25}$ kg) and an optical lattice of spacing $d=532$ nm realized by two counter propagating laser beams at 1064 nm.

The depth of the lattice $U_0$, which is proportional to $\gamma$ following (\ref{defgamma}), depends on the atom and on the characteristics of each Gaussian laser beams used to create the optical lattice (the power $P$, the wavelength and the waist $w_0$). For the parameters given above, the dimensionless depth of the lattice $s$ is:
\begin{equation}
\label{power}
s=\frac{U_{0}}{E_L}=1.03 \times 10^6 \frac{P [{\rm W}]}{(w_0[\mu{\rm m}])^2 }.
\end{equation}
Values of $s$ in the range between 10 and 50 can be achieved with state-of-the-art monomode fibered lasers.

The effective Planck constant of the problem defined in (\ref{defhbar_eff}) is tuned via the angular frequency of modulation:
\begin{equation}
  \label{hbar_eff_2}
   \hbar_{\rm eff}=\frac{4\pi^2\hbar}{M\omega d^2}\ .
\end{equation}
From these estimates one can specify the range of parameters for our model (see Table~\ref{param_exp}).
\begin{table}[!h]
  \begin{tabular}[c]{|c|c|}
    \hline
    $\gamma$
& $0...1.1$\\\hline
$\varepsilon$ & $0...1$\\\hline
$\hbar_{\rm eff}$  & 0.1..0.3\\\hline
  \end{tabular}
\caption{Range for the parameters of the model (\ref{Hclas}) in a cold atom experiment with $\omega$  ranging from $\sim 2\pi \times 3\ 10^5$ to $\sim 2\pi \times 10^6$ Hz.}
\label{param_exp}
\end{table}
We note that the achievable values of $\hbar_{\rm eff}$ cannot be too small in this setting. Indeed, decreasing $\hbar_{\rm eff}$ leads to increasing modulation frequencies. But, in order to keep a constant value of $\gamma$, the power of the laser beams creating the optical lattice should be increased accordingly, see (\ref{defgamma}).

A key quantity for the analysis of our result is the estimate of the width of the quasimomentum distribution of the atomic cloud. Indeed, the symmetry $p\mapsto -p$, which was used in \cite{RaizenScience,PhillipsNature,delande1}, is very sensitive to variations in quasimomentum (see a more detailed discussion below).
An achievable width for the velocity distribution of the atoms using delta kick cooling technique is
\cite{deltakick,JosseCBS}:
\begin{equation}
  \Delta v=170~\mu {\rm m.s}^{-1}\ .
\label{Dv_init}
\end{equation}
This allows us to estimate the width of the quasimomentum distribution of the atomic cloud:
\begin{equation}
  \label{betamin}
  \Delta\beta =\dfrac{ Md\Delta v}{h}\approx 0.02\ .
\end{equation}
Note that the value in the experiments of \cite{RaizenScience} was $\Delta\beta \approx 0.05$.

\section{First route to chaos assisted tunneling: measure of oscillation period of the average momentum}
\label{tunnelp_oscill}

It has been explained in the previous Section that chaos assisted tunneling 
leads to very large fluctuations of the energy splitting when a parameter of the system is varied.
In this Section we want to describe the most direct way to observe this effect \cite{PhillipsNature,RaizenScience}. {I}f one is able to prepare an initial wave packet sitting on one of 
the symmetric stable islands, quantum tunneling will allow it to go back and forth between those islands. The period of the oscillations, denoted by $T_{\rm tunn}$, is simply related to the energy splitting $\delta$ (\ref{def_splitting}) via
\begin{equation}
  T_{\rm tunn}=\dfrac{\hbar_{\rm eff}}{\delta}\ .
\label{deltatoT}
\end{equation}
Chaos assisted tunneling should therefore be observable by considering tunneling oscillations and the associated large variations of the tunneling period $T_{\rm tunn}$, when a parameter is varied.

\subsection{Fragility of the $p\mapsto -p$ symmetry. Analogy of the double well potential.}
\label{fragilsym}

We start here with reviewing briefly the tunneling effect in a double well potential. In the case of a standard double well potential, there is an energy barrier of finite height, which separates both wells. 
This situation is partly similar to the one of the atomic modulated pendulum. Besides it gives a more intuitive picture and may help to understand the crucial role of the symmetry.

Let us investigate the lowest energy states of the double well. 
For a {\it symmetric} double well, there is a special doublet of eigenstates, corresponding to a symmetric and antisymmetric superposition in both wells.  If the initial wave packet sits initially in one well only, the state is expanded along both these eigenstates. Then the time dynamics corresponds to a superposition of these eigenstates with oscillating coefficients. This is what gives rise to the seminal tunneling oscillations of period $T_{\rm tunn}$: the whole wave packet sits alternatively in one or the other well.

The situation is slightly different when the double well potential is not symmetric. In the basis of states localized in the right/left wells, the Hamiltonian can be written as:
\begin{equation}
  \label{H_double_well_asym}
  \hat{H}=\left(
\begin{array}{cc}
  E_0+A & \delta/2 \\
 \delta/2 & E_0
\end{array}\right)\ ,
\end{equation}
where $A$ stand for the asymmetry between the wells. The asymmetry leads to two important consequences. 
First the tunneling period is changed to:
\begin{equation}
  \label{T_Delta}
  T_A=\frac{\hbar_{\rm eff}}{\sqrt{\delta^2+A^2}}\ .
\end{equation}
Second and more importantly, the asymmetry induces a decrease of the proportion of atoms which can tunnel:
\begin{equation}
  \label{pop_tunnel_asym}
  \frac{\delta^2}{\delta^2+A^2} \le 1\ .
\end{equation}

One very important point of the atomic modulated pendulum is that there is a parameter, which tunes the asymmetry between both islands around $\beta=0$. This can be seen in the band diagram, see e.g. Fig.~\ref{band_diag1} left below, which represents in red square points the levels which have the highest overlap with the stable island around the fixed point $(0,p_*)$ in phase space. The energy of such level is related to the quasimomentum through $
\langle p \rangle=\dfrac{1}{\hbar_{\rm eff}} \dfrac{\partial E}{\partial \beta}\ 
$ \cite{delande1}, where the average momentum is here $\langle p \rangle\approx p_*$.
Therefore, at finite quasimomentum, the energies of the states localized in the upper or lower regular islands differ in analogy with (\ref{H_double_well_asym}) by an amount:
\begin{equation}\label{eq:asymmetryA}
A\approx2 p_*\hbar_{\rm eff}\beta\ .
\end{equation}
Whenever the quasimomentum is nonzero, it breaks the symmetry between the islands by the asymmetry parameter $A$. Not only the tunneling period is changed following (\ref{T_Delta}), but the number of tunneling atoms is also decreased as indicated by (\ref{pop_tunnel_asym}).

\subsection{Tunneling oscillations in momentum direction. High sensitivity towards quasimomentum.}

\begin{figure}
  \centering
  \includegraphics[width=\linewidth]{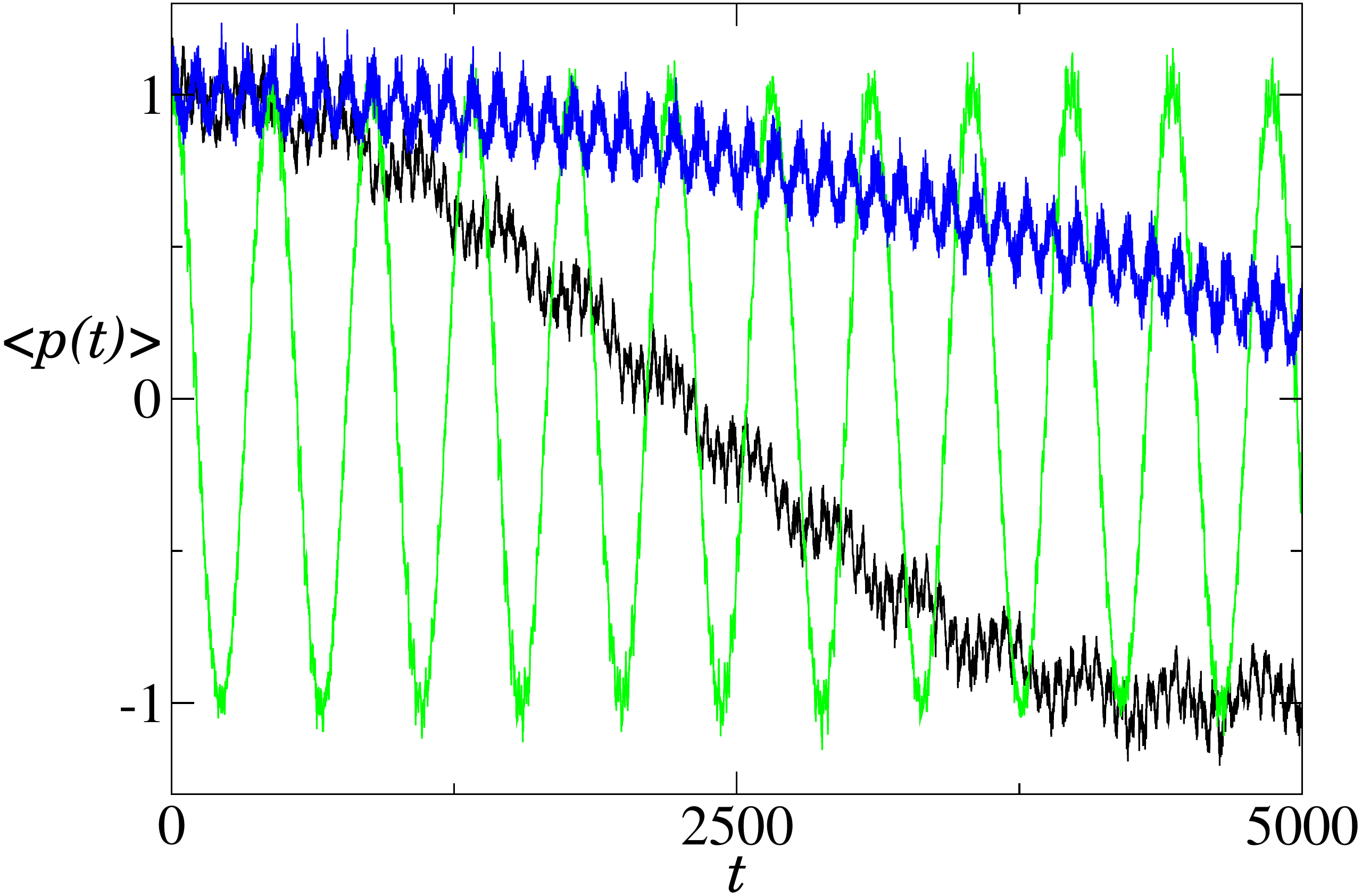}
  \caption{(Color online) Large fluctuations of tunneling oscillations for $\beta=0$ under the quantum dynamics dictated by (\ref{Hclas}) for $\gamma=0.25$ and $\varepsilon=0.4$. The average momentum is plotted for $\beta=0$ as a function of time expressed in number of periods.
The initial state is a coherent state centered  at the phase space point $(x,p)=(0,1.296)$ with width $\Delta p=1$. The different curves stand for different values of $\hbar_{\rm eff}$ leading to different tunneling periods $T_{\rm tunn}$. Black curve (bottom at $t=5000$): $\hbar_{\rm eff}=0.1248$, $T_{\rm tunn}\sim 9000$. 
Green curve (middle at $t=5000$): $\hbar_{\rm eff}=0.1619$, $T_{\rm tunn}\sim 440$. Blue curve (top at $t=5000$): $\hbar_{\rm eff}=0.2131$, $T_{\rm tunn}\sim 12000$.}
  \label{osc_tunnel}
\end{figure}

\begin{figure}
  \centering
  \includegraphics[width=.49\textwidth]{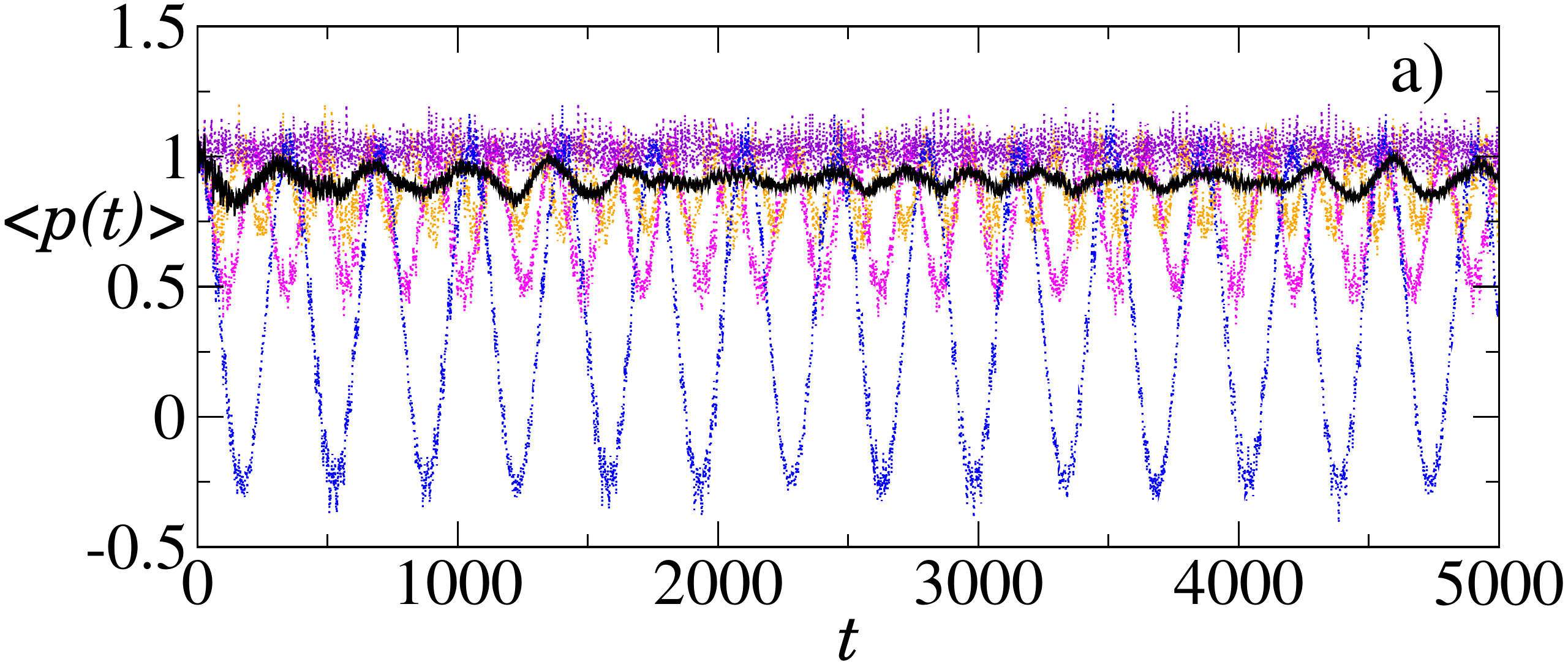}
  \includegraphics[width=.49\textwidth]{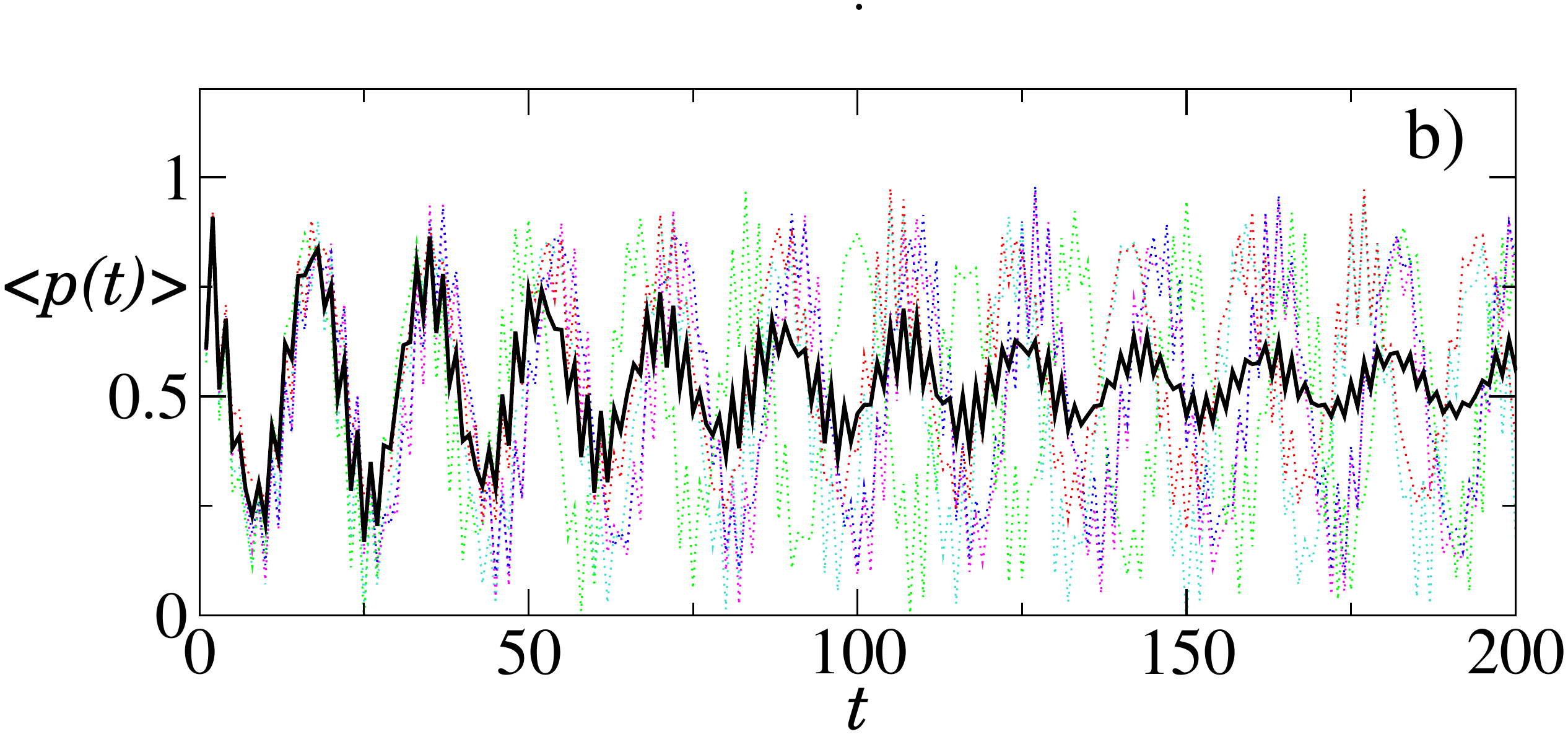}
  \caption{(Color online) High sensitivity of tunneling oscillations vs momentum under the quantum dynamics dictated by (\ref{Hclas}) for $\gamma=0.25$ and $\varepsilon=0.4$. The average momentum is plotted as a function of time expressed in number of periods.
Here the initial state is a coherent state with an initial momentum $p=1.296$ and $\Delta p=1$. All the dotted curves correspond to {different values of} $\beta\neq 0$. The full black curve is a an average over the quasimomenta with a uniform distribution $\rho(\beta)$ for $\beta$ between $-0.02$ and $0.02$. a) $\hbar_{\rm eff}=0.1619$: {the asymmetry induces a sharp exponential reduction of the proportion of atoms, that participate to tunneling}. b) $\hbar_{\rm eff}=1$: {the tunneling splitting is large and thus tunneling oscillations are still observable}. }
  \label{osc_tunnel3}
\end{figure}

We are interested here in the tunneling oscillations between two stable islands, whose centers are along the momentum axis, with phase space coordinates $(x,p)=(0,\pm p_*)$. The main goal is to observe large fluctuations of the splitting as predicted by the theory. We stress that this prediction has been formulated in the semiclassical regime which requires $\hbar_{\rm eff}\ll {\cal A}$, where ${\cal A}$ is the area of the classical stable island in phase space (on the order of $1$ for the atomic modulated pendulum). In particular this is different from the pioneering experiments \cite{PhillipsNature,RaizenScience}. As they both observed tunneling oscillations, we want to review here this possible route in order to detect signature of chaos assisted tunneling. In particular, we will account for the fragility of the symmetry $p\mapsto -p$ in the semiclassical regime as described in the previous subsection.

There are two important issues worth being detailed. The first is to have an initial state, which is well supported inside the stable island. This means that 
the widths of the initial wave packet $\Delta x$ and $\Delta p$ along the position and momentum directions respectively also have to be precisely under control.
The second issue is to allow for a parameter to vary in order to see fluctuations by orders of magnitude of the tunneling splitting (\ref{def_splitting}).
From the definition (\ref{defhbar_eff}) a natural parameter to be varied in a cold atom experiment is 
the effective Planck constant $\hbar_{\rm eff}$ through the modulation frequency.

First we consider the dynamics for an initial wave packet with a zero quasimomentum:
$$\left|\psi(t=0)\right>= \sum_n c_{n,\beta=0} \left|\psi_{n,\beta=0}\right>\ ,$$
where the eigenstates $\left|\psi_{n,\beta}\right>$ are defined by (\ref{eigvectUbeta}) and $c_{n,\beta}$ are the overlap coefficients. The constraint on the quasimomentum ensures that the symmetry $p\mapsto -p$ is preserved as detailed in the previous paragraph.
In order to avoid further complications associated to the initial state preparation, the initial state is taken as a coherent state with prescribed widths $\Delta x$ and $\Delta p$. 
In our numerical simulations, the initial state sits on the upper island around the fixed point $(x,p)=(0,p_*)$.
Such an initial state is very well approximated by a superposition of the two eigenstates $|\psi_\pm\rangle$ of the propagator, which are supported by both islands as mentioned previously. The quantum dynamics under (\ref{Hclas}) will consist in an oscillating superposition of these states, leading to the standard tunneling effect. The choice of these symmetric islands means that, instead of having tunneling oscillations of the mean position as in a double well, tunneling will result here to oscillations of the mean momentum, $\langle p(t) \rangle$.

In Fig.~\ref{osc_tunnel} this effect is displayed by plotting the mean momentum $\langle p(t)\rangle$ for $\beta=0$ as a function of time $t$. 
This observable oscillates between the extremal value $p_*$ and $-p_*$. The presence of chaos can be clearly shown by the non monotonic and large variations of the tunneling period when $\hbar_{\rm eff}$ is slightly varied. 
The large fluctuations due to chaos assisted tunneling can therefore be seen in the tunneling oscillations with such $\beta=0$ initial state.

Next we want to investigate the effect of a distribution of quasimomenta. The goal is to evaluate the contributions of the wave packet with non zero quasimomentum, which break the symmetry $p\mapsto -p$, and their effect on the tunneling oscillations. This is important in an experimental perspective as 
an atomic cloud in a cold atom experiment is rather described by a superposition of different quasimomenta. This superposition can be modeled by a quasimomentum distribution $\rho(\beta)$ such that:
$$\left|\psi(t=0)\right>= \int\rho(\beta) \sum_n c_{n,\beta} \left|\psi_{n,\beta}\right> \ud\beta \ .$$
Fig.~\ref{osc_tunnel3} shows that whenever $\beta\neq 0$, the tunneling oscillations are much less visible. 
This can be easily understood from the consequences of the asymmetry explained in the previous paragraph. The typical value of the splitting varies as a function of $\hbar_{\rm eff}$ like 
\begin{equation}
  \label{splitting_Leg}
  \delta\sim \exp(-S/\hbar_{\rm eff})\ ,
\end{equation}
with $S$ the classical action related to the tunneling barrier. The asymmetry parameter is linear in $\hbar_{\rm eff}$.
When the asymmetry is small, e.g. for $\hbar_{\rm eff}\sim 1$, one may see some tunneling oscillations (see Fig.~\ref{osc_tunnel3} bottom). This is indeed the regime, where such oscillations were observed in the previous experiments \cite{PhillipsNature,RaizenScience}. But going closer to the semiclassical regime and decreasing $\hbar_{\rm eff}$ will have much more dramatic consequences. The amount of the tunneling population using (\ref{pop_tunnel_asym}) will vary like:
\begin{equation}
  \frac{\delta^2}{\delta^2+A^2}\sim \frac{\exp(-2S/\hbar_{\rm eff})}{\hbar_{\rm eff}^2}\ ,
\end{equation}

In other words the asymmetry induced by a non zero quasimomentum in the semiclassical regime almost completely annihilates the tunneling oscillations as the amount of the wave packet that can tunnel becomes exponentially small. This is illustrated in Fig.~\ref{osc_tunnel3} top.
As a conclusion tunneling oscillations, despite their intuitive picture, do not stand for an efficient method to see large fluctuations of the splitting associated to chaos assisted tunneling in the semiclassical regime. 

\section{Second route to chaos assisted tunneling: Landau-Zener scheme}
\label{tunnelp_LZ}

As the previous method to detect chaos assisted tunneling does not work in the semiclassical regime to see large fluctuations of the splitting, we will now focus on another strategy. It consists in forcing the atoms to tunnel from one stable island to its symmetric partner following a Landau Zener (LZ) transition \cite{delande1}.
We will recall first the main idea of this method. 
Then we will present how to implement it in order to analyze numerical data relevant for experiments. Last we will discuss its efficiency in order to detect the large fluctuations of the splitting associated to chaos assisted tunneling.

\subsection{Main idea of the Landau Zener scheme}

\begin{figure}
  \includegraphics[width=\linewidth]{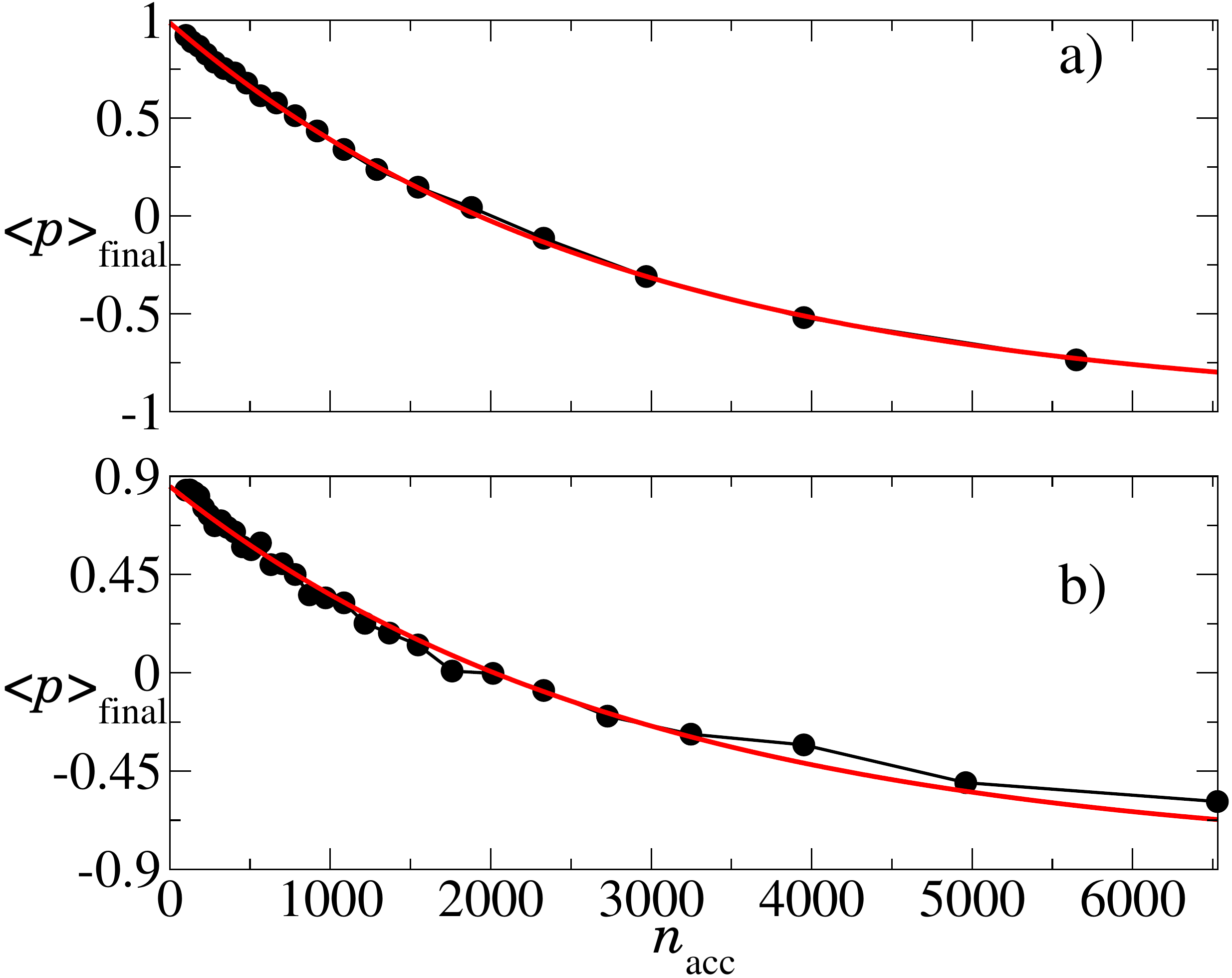}
  \caption{(Color online) Final value of the mean momentum $\langle p \rangle_{\rm final}$ as a function of the traveling time along the quasimomentum axis in the band diagram. The traveling time $n_{\rm acc}$ is expressed in number of time periods of modulation. Black circles: numerical data following our LZ scheme (see text). Red full line: Fit following Eq.~(\ref{fit_eq}). Parameters: $\varepsilon=0.4$ and $\gamma=0.24$. a) $\hbar_{\rm eff}=0.2$. b) $\hbar_{\rm eff}=0.053$.}
  \label{pfin_vsnacc}
\end{figure}

This scheme builds up on the previous observation and aims to force all the atoms to tunnel at $\beta=0$, a requirement that turns out to be crucial to ensure the tunneling between the classical stable islands. For this purpose, the quasimomentum is here made to vary linearly in time. This can be seen as if the atom is traveling along the quasimomentum axis in the band diagram. Approaching $\beta=0$, the atom will experience a LZ transition between each of the two bands corresponding 
to a wave packet localized in one dynamical well.
It is achieved experimentally by imposing a difference, linear in time, between the frequencies of the two counter-propagating lasers building the optical lattice \cite{Salomon}.

It is convenient to describe the dynamics of an atom with its band diagram. 
In the simplest case, and as illustrated in Fig.~\ref{band_diag1}, we will assume that the atom can only access to two bands lying on the stable islands around $(0,\pm p_*)$. The wave function is described in the basis of these eigenstates by the following vector:
\begin{equation}
  \label{psiLZdef}
  \left|\psi(t)\right>=\left(
  \begin{array}{c}
    \psi_1(t) \\\psi_2(t)
     \end{array}\right)\ ,
\end{equation}
where $\psi_1(t)$ is the quantum amplitude on the first (say lower stable island) energy band and $\psi_2(t)$ is the quantum amplitude on the second band (upper stable island).
In this basis the Schr\"odinger equation for the atom is:{
\begin{equation}
  \ic\hbar_{\rm eff} \frac{\ud \left|\psi(t)\right> }{\ud t}  =
  \left(
    \begin{array}{cc}
      E_0-p_* \hbar_{\rm eff}\beta_t& \delta/2\\
      \delta/2&E_0+p_* \hbar_{\rm eff}\beta_t
    \end{array}\right)
  \left|\psi(t)\right>\ ,\label{schro_LZ}
\end{equation}
where the quasimomentum $\beta_t\equiv \beta_0 - F t/\hbar_{\rm eff}$} drifts because of the constant force $F$ in the accelerated frame.  The crucial parameter is the energy splitting $\delta$ as defined in (\ref{def_splitting}). Here it is also the distance between the two energy bands at $\beta=0$.
The Schr\"odinger equation (\ref{schro_LZ}) can be more conveniently written under the form
\begin{equation}
  \label{LZ1}
  \ic \mathfrak{h} \frac{\ud}{\ud \mathfrak{t}} \left|\psi(\mathfrak{t})\right>=\left(
    \begin{array}{cc}
     \mathfrak{t} & 1\\
     1 & - \mathfrak{t}
    \end{array}\right)
  \left|\psi(\mathfrak{t})\right>\ .
\end{equation}
after shifting the origin of the phase of $\left|\psi(t)\right>$ and with the following rescaling
\begin{equation}
  \label{rescal001}
\mathfrak{t} =\dfrac{2 p_* F}{\delta} t,\quad \mathfrak{h}=\dfrac{4\hbar_{\rm eff} p_* F}{\delta^2}\ .
\end{equation}
It is assumed that, initially each atom has an overlap with only one energy band, say the second one: 
\begin{equation}
  \label{psiLZ_init}
  \langle \psi(\mathfrak{t}) | \psi(\mathfrak{t})\rangle \stackrel{\mathfrak{t}\to -\infty}{\longrightarrow} \left(
  \begin{array}{c}
    0 \\ 1
     \end{array}\right)\ ,
\end{equation}
This problem was analytically solved by Zener \cite{Zener}. The famous Landau-Zener formula is recovered when taking the asymptotic behavior of the solution:
\begin{equation}
  \label{psiLZ}
  \langle \psi(\mathfrak{t}) | \psi(\mathfrak{t})\rangle \stackrel{\mathfrak{t}\to +\infty}{\longrightarrow} \left(
  \begin{array}{c}
    1-e^{-\pi/\mathfrak{h}} \\ e^{-\pi/\mathfrak{h}}
     \end{array}\right)\ .
\end{equation}
The coupling between the energy bands $\delta$, as shown in (\ref{schro_LZ}), can be extracted when varying the speed of the gap crossing. As we assume a linear variation of the quasimomentum, this speed is simply given by:
\begin{equation}
  \label{v_beta}
  \dfrac{\ud \beta}{\ud t}=\dfrac{F}{\hbar_{\rm eff}}=\dfrac{\Delta \beta}{\Delta t}\ .
\end{equation}
where $\Delta \beta$ is the variation of the mean quasimomentum of the atom during the time $\Delta t$. We always assume that the quasimomentum of the atom cloud goes from $\beta_0$ to $-\beta_0$ so that $\Delta \beta=-2\beta_0$. Further note that in our units each time period has a length $2\pi$. The elapsed time $\Delta t$ is then $\Delta t=2\pi n_{\rm acc}$, where $n_{\rm acc}$ denotes the number of periods during which the atom experiences a constant force in the accelerated frame. Finally one has:
\begin{equation}
  \label{v_beta2}
  \dfrac{\ud \beta}{\ud t}=-\dfrac{\beta_0}{\pi n_{\rm acc}}\ .
\end{equation}
Using (\ref{rescal001}), (\ref{v_beta}) and (\ref{v_beta2}) the parameter $\mathfrak{h}$ entering (\ref{psiLZ}) can be rewritten:
\begin{equation}
  \label{h_vs_N}
  \dfrac{1}{\mathfrak{h}}=\dfrac{\pi \delta^2 n_{\rm acc}}{4\beta_0 p_*\hbar_{\rm eff}^2}.
\end{equation}
It means that the amount of atoms that jump to the other band, given by (\ref{psiLZ}), depends on their traveling speed along the $\beta$ axis (or here more conveniently on the traveling time $2\pi n_{\rm acc}$). 

Consider now the momentum distribution of the atomic cloud within the two band approximation:
\begin{equation}
  \label{def_avp}
  \langle p(t) \rangle= -p_* |\psi_1(t)|^2+p_*|\psi_2(t)|^2\ .
\end{equation}
Initially every atom is in the upper stable island band, i.e. $\langle p\rangle_{\rm init}=p_*$ as assumed in (\ref{psiLZ_init}). 
Then the final average momentum 
after an evolution during $n_{\rm acc}$ periods (assumed to be large enough) will have a precise functional dependence on $n_{\rm acc}$ following (\ref{psiLZ}) and (\ref{h_vs_N}):
\begin{equation}
  \label{avp_LZ}
  \langle p \rangle_{\rm final} = -p_* (1-2e^{-\alpha n_{\rm acc}}),\quad 
\alpha=\dfrac{\pi^2 \delta^2}{4\beta_0 p_*\hbar_{\rm eff}^2}\ .
\end{equation}
A fit of $\langle p \rangle_{\rm final}$ as a function of $n_{\rm acc}$ by the formula (\ref{avp_LZ}) leads to an estimate of the splitting $\delta$.

Two examples of such a fit are shown in Fig.~\ref{pfin_vsnacc}.
The numerical data are fitted by the following two parameter function, see (\ref{avp_LZ}):
\begin{equation}
  \label{fit_eq}
  y=-A (1-2e^{-Bx})\ . 
\end{equation}
The numerical splitting is finally computed following:
\begin{equation}
  \label{splittingnumLZ}
  \delta=\sqrt{\dfrac{4\beta_0 A B \hbar_{\rm eff}^2}{\pi^2}}\ .
\end{equation}
As a conclusion 
the use of an accelerated frame can be simulated to extract the energy splitting. Experimentally this requires to scan a fixed region of the band diagram with a varying speed. The fit by a Landau Zener formula (\ref{fit_eq}) leads to an estimate of the energy splitting through (\ref{splittingnumLZ}).

\subsection{Extraction of the splitting using a Landau Zener approach. Three examples}

\begin{figure}
  \includegraphics[width=\linewidth]{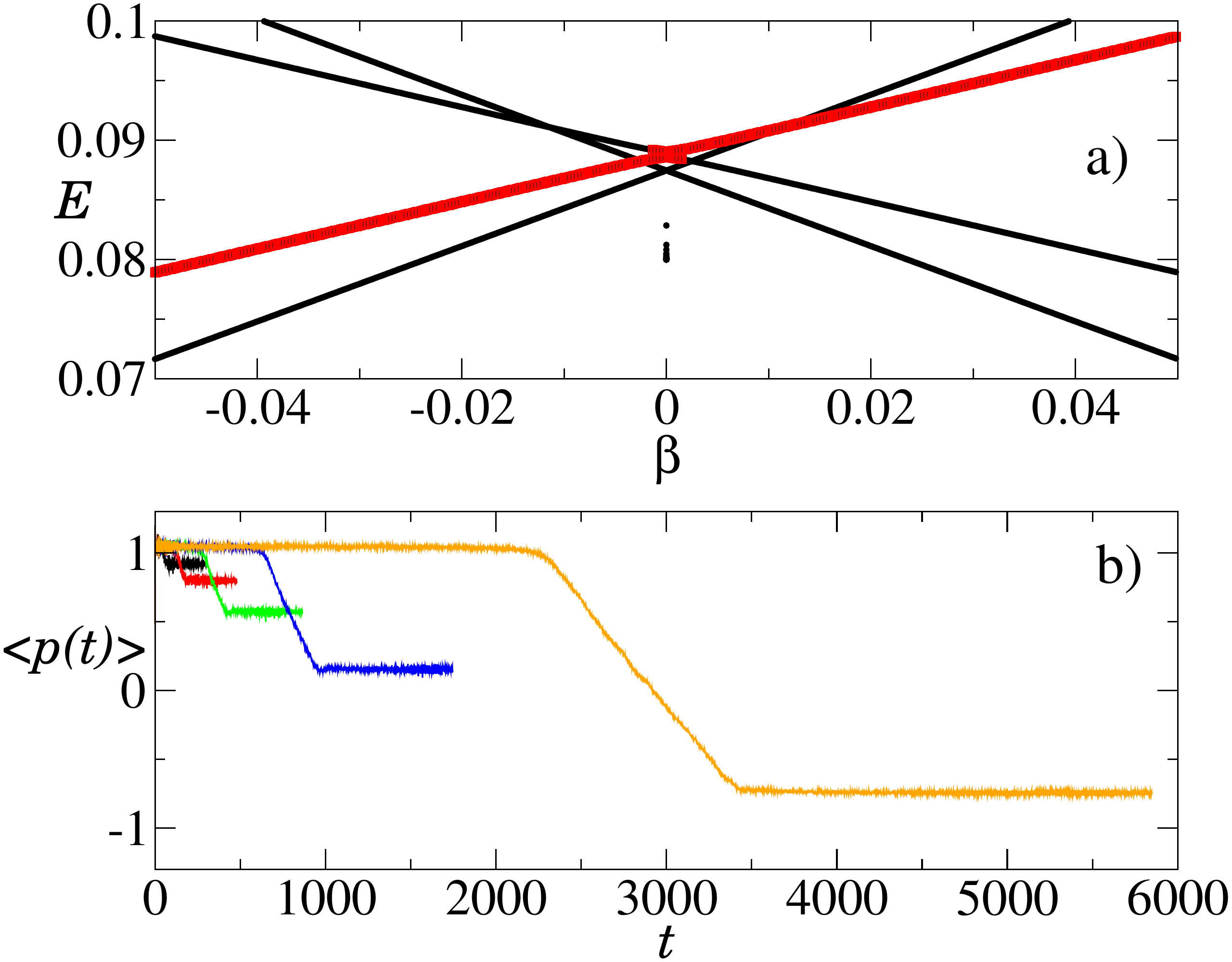}
  \caption{ (Color online) a) Band diagram for $\gamma=0.24$, $\varepsilon=0.4$ and $\hbar_{\rm eff}=0.2$. The red square branches stand for the levels, whose associated eigenstates in Husimi representation are mainly supported by the island around the point $(0,p_*)$ in phase space. The splitting $\delta$ as defined in (\ref{def_splitting}) is also the distance between the two red branches at $\beta=0$. b) Average momentum in the accelerated frame for {a distribution of atoms with $\beta_\text{ini} \in [0.04,0.06]$ traveling linearly along the quasimomentum direction from $\beta_0=0.05$ to $-\beta_0=-0.05$ as a function of the number {of} periods}. 
The different curves stand for different traveling time{s} $2\pi n_{\rm acc}$. From left to right at the final time: black: $n_{\rm acc}=100$. Red: $n_{\rm acc}=280$. Green: $n_{\rm acc}=665$. Blue: $n_{\rm acc}=1547$. Orange: $n_{\rm acc}=5649$.}
  \label{band_diag1}
\end{figure}

\begin{figure}
\includegraphics[width=\linewidth]{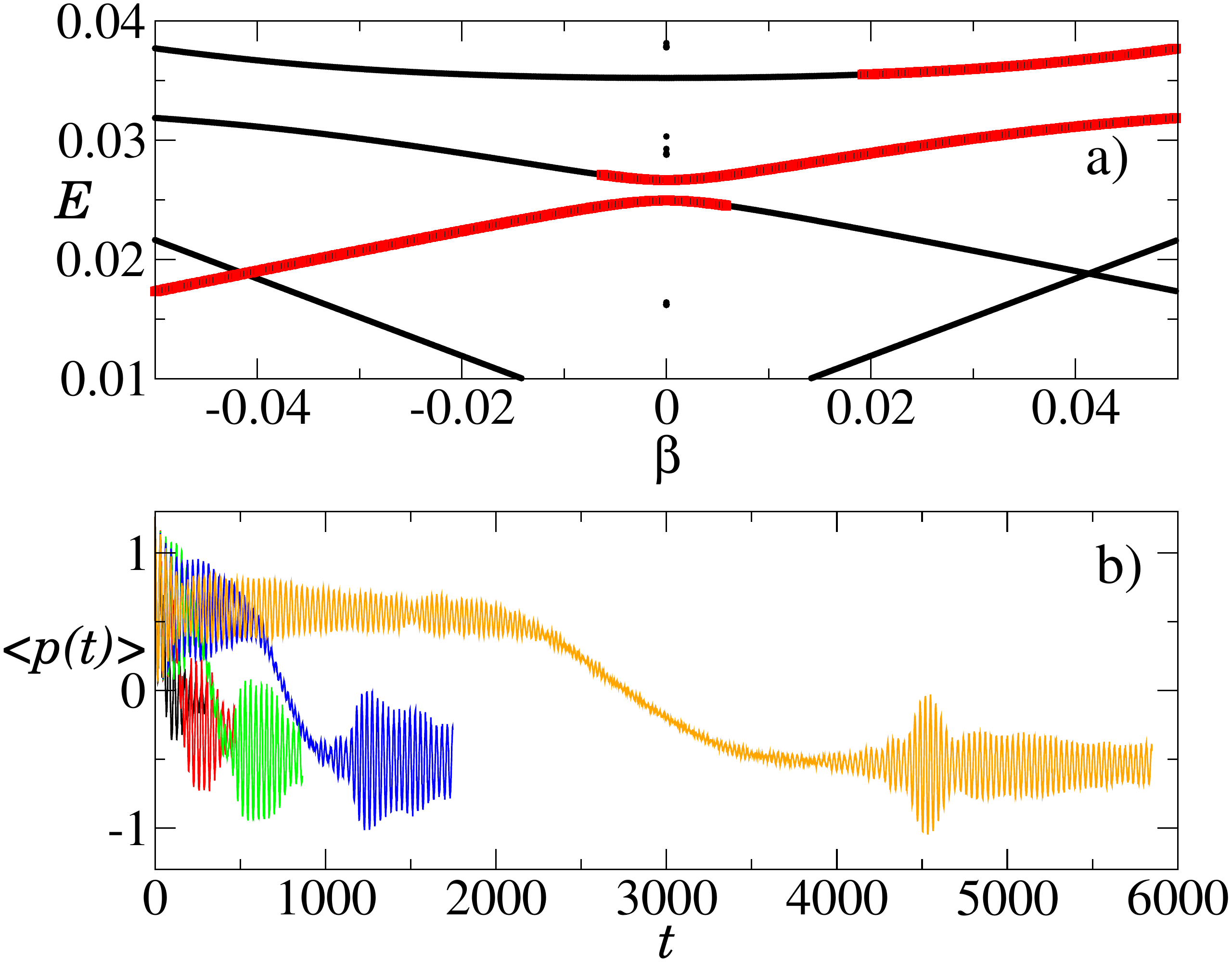}
  \caption{
(Color online) a) Band diagram for $\gamma=0.24$, $\varepsilon=0.4$ and $\hbar_{\rm eff}=0.18$. The red square branches stand for the levels, whose associated eigenstates in Husimi representation are mainly supported by the island around the point $(0,p_*)$ in phase space. The splitting $\delta$ as defined in (\ref{def_splitting}) is also the distance between the two red branches at $\beta=0$. b) Average momentum in the accelerated frame for {a distribution of atoms with $\beta_\text{ini} \in [0.04,0.06]$ traveling linearly along the quasimomentum direction from $\beta_0=0.05$ to $-\beta_0=-0.05$ as a function of the number {of} periods}. 
The different curves stand for different traveling time{s} $2\pi n_{\rm acc}$. From left to right at the final time: black: $n_{\rm acc}=100$. Red: $n_{\rm acc}=280$. Green: $n_{\rm acc}=665$. Blue: $n_{\rm acc}=1547$. Orange: $n_{\rm acc}=5649$.
  }
  \label{band_diag2}
\end{figure}

\begin{figure}
  \includegraphics[width=\linewidth]{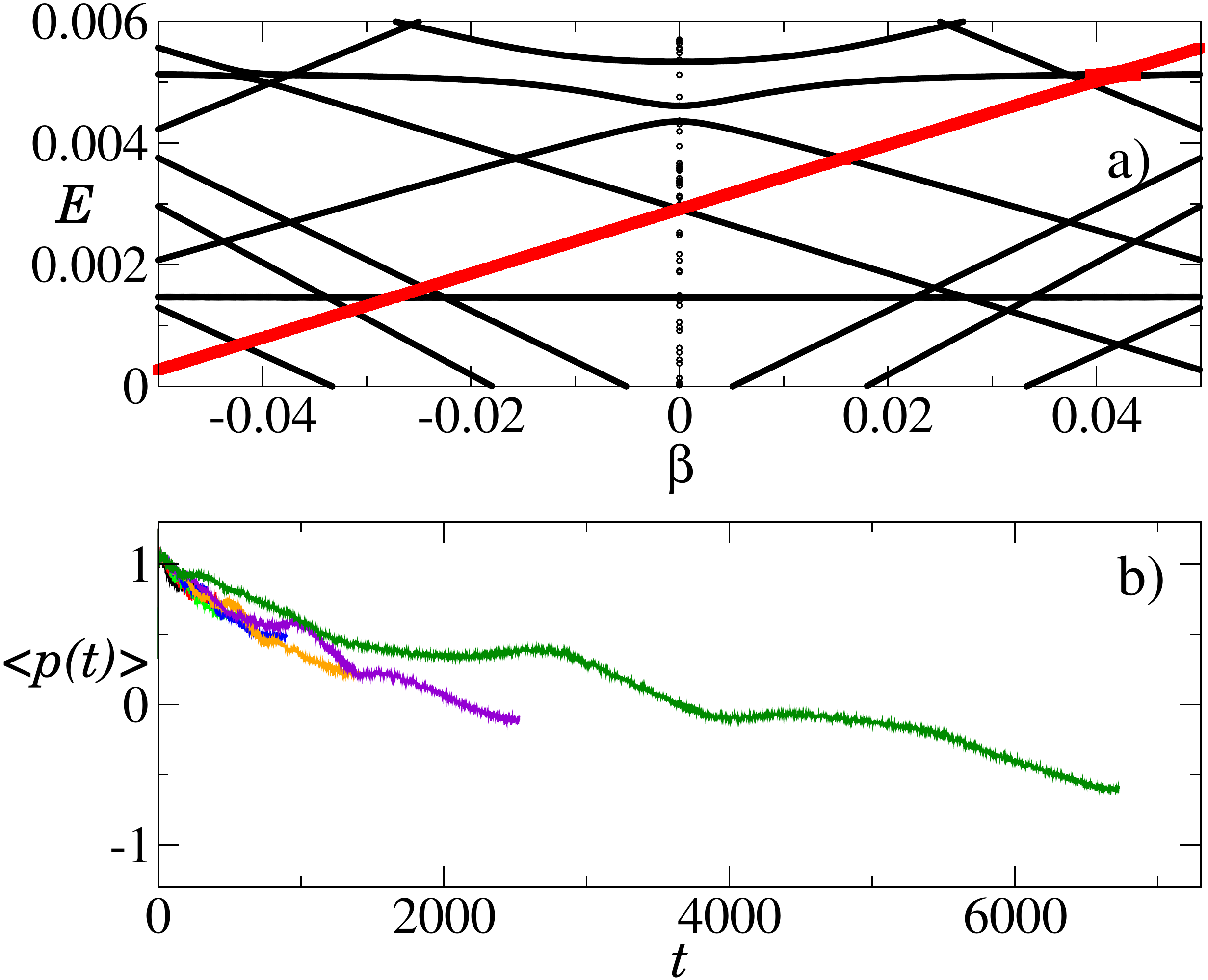}
  \caption{(Color online) a) Band diagram for $\gamma=0.24$, $\varepsilon=0.4$ and $\hbar_{\rm eff}=0.053$. The red square branches stand for the levels, whose associated eigenstates in Husimi representation are mainly supported by the island around the point $(0,p_*)$ in phase space. The splitting $\delta$ as defined in (\ref{def_splitting}) is also the distance between the two red branches at $\beta=0$. b) Average momentum in the accelerated frame for {a distribution of atoms with $\beta_\text{ini} \in [0.04,0.06]$ traveling linearly along the quasimomentum direction from $\beta_0=0.05$ to $-\beta_0=-0.05$ as a function of the number {of} periods}. Different curves stand different for traveling time{s} $2\pi n_{\rm acc}$. From left to right for the final time: black: $n_{\rm acc}=100$. Red: $n_{\rm acc}=211$. Green: $n_{\rm acc}=404$. Blue: $n_{\rm acc}=701$. Orange: $n_{\rm acc}=1216$. Violet: $n_{\rm acc}=2330$. Dark green: $n_{\rm acc}=6528$.}
  \label{band_diag3}
\end{figure}

We will now examine more systematically the LZ scheme described in the previous Subsection for $\gamma=0.24$ and $\varepsilon=0.4$. 
For $\hbar_{\rm eff}=0.2$ the local band diagram is displayed in Fig.~\ref{band_diag1}a. One can see that the two band approximation is very well justified in this case.
The atoms of the cloud have positive initial quasimomenta uniformly distributed in $[0.04,0.06]$. Due to the constant force in the accelerated frame, the quasimomenta decrease linearly with time and the atoms explore the band diagram (from right to left in Fig.~\ref{band_diag1}a).

The time evolution of the average momentum of the wave packet is displayed in Fig.~\ref{band_diag1}b. When it goes through the avoided crossing at $\beta=0$,
part of the wave-packet performs a Landau-Zener transition resulting in a rapid change of $\langle p(t) \rangle$ from $p_*$ to $\langle p \rangle_\text{final}$. The measurement of the final mean momentum $\langle p \rangle_\text{final}$ of the atoms leads to an estimate of the splitting from LZ formula (\ref{avp_LZ}).
For this case we found an excellent agreement between the exact splitting (extracted from the band diagram) and the one extracted from the LZ scheme:
\begin{equation}
  \label{compare_d}
  \delta^{{\rm(exact)}}=5.20\ 10^{-4},\qquad\delta^{{\rm (LZ)}}=5.36\ 10^{-4}\ .
\end{equation}

The same method was applied for exactly the same value of the parameters, except that $\hbar_{\rm eff}=0.18$. The band diagram is now more complex (see Fig.~\ref{band_diag2}a). Already here one can see that the two band approximation is less accurate as the atoms are initially not overlapping with only one band but with two bands. The presence of the second band leads to fast oscillations in the curve $\langle p(t) \rangle$. 
Nevertheless it is remarkable that the LZ scheme gives a very good estimate of the splitting:
\begin{equation}
  \label{compare_d2}
  \delta^{{\rm(exact)}}=1.71\ 10^{-3},\qquad\delta^{{\rm (LZ)}}=1.57\ 10^{-3}\ .
\end{equation}

Last we tried a smaller value of $\hbar_{\rm eff}$. The reason is that we want ultimately to explore the semiclassical regime, and analyze the fluctuations of the splitting (see Fig.~\ref{split_eps0.4_g0.25}). 
Our last example in Fig.~\ref{band_diag3} shows the case where the splitting is significantly reduced. It can be seen in the band diagram that there are several branch avoided crossings in the quasimomentum range that is explored. In particular, there is a branch avoided crossing around $\beta\simeq 0.04$, which is much larger than the one at $\beta=0$.
This leads to the presence of several plateaus for the curve $\langle p(t)\rangle$. 
Moreover, although the value of $n_{\rm acc}$ has been chosen larger than in Figs~\ref{band_diag1} and \ref{band_diag2}  in order to resolve more precisely the tiny splitting at $\beta=0$, the  splitting extracted from LZ formula remains far from the exact one:
\begin{equation}
  \label{compare_d3}
  \delta^{{\rm(exact)}}=2.79\ 10^{-6},\qquad\delta^{{\rm (LZ)}}=1.29\ 10^{-4}\ .
\end{equation}
This last example shows that the LZ method for the extraction of the splitting, especially small splitting, can be inaccurate. The precise reason for it will be discussed in details below.

\subsection{Statistics of Landau Zener splittings}

\begin{figure}
    \begin{center}
     \includegraphics[width=\linewidth]{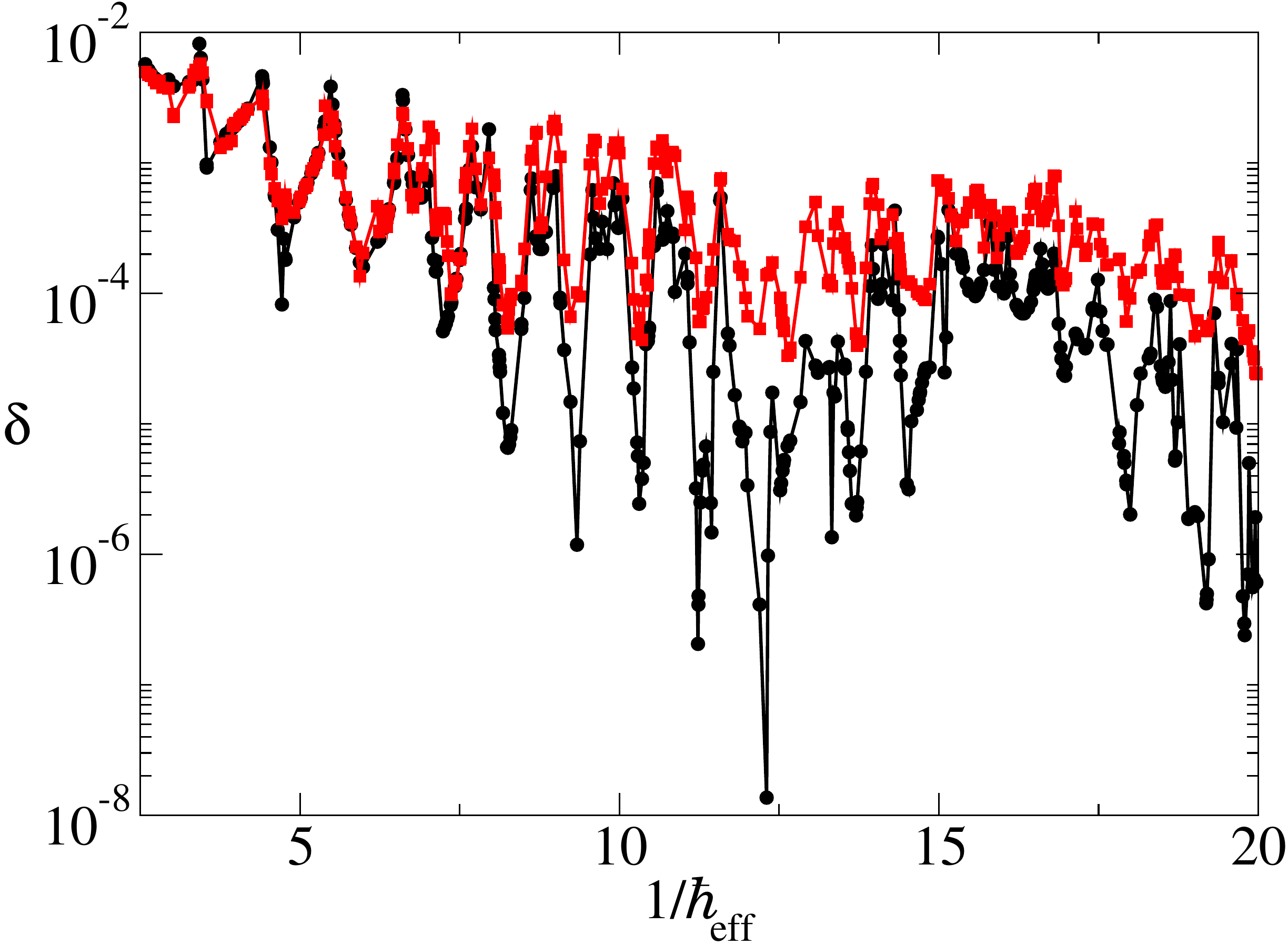}
    \end{center}
  \caption{(Color online) Energy splitting for the {atomic} modulated pendulum (\ref{Hclas}) with $\gamma=0.24$ and $\varepsilon=0.4$ as a function of $1/\hbar_{\rm eff}$. The splittings extracted from {the} LZ method (red squares) are compared with the exact ones (black circles). The $\hbar_{\rm eff}^{-1}$ axis is sampled with $441$ values. {Although the fluctuations of the LZ splitting appear similar to that of the exact splitting at $\beta=0$, the amplitude of the fluctuations and the typical LZ splitting are inaccurate.}}
  \label{splittings_LZ_Law}
\end{figure}

The same method 
was systematically used on a larger range of $\hbar_{\rm eff}$ in order to extract the fluctuations of  the splitting and compare it with the prediction (\ref{cauchy}). The results are displayed in Fig.~\ref{splittings_LZ_Law}. Although the fluctuations of the LZ splitting appear strongly correlated with that of the exact splitting at $\beta=0$, the amplitude of the fluctuations and the typical LZ splitting differ from the exact ones especially in the regime of small $\hbar_{\rm eff}$. We analyze in this subsection the reasons for these differences.

The main limitation of the LZ method for the splitting extraction is the following: when traveling in the band diagram, it is very likely that part of the wave packet will perform a Landau-Zener transition at $\beta\neq 0$ to a state in the chaotic sea instead of reaching the symmetric island at $\beta=0$. This limitation originates in the combination between the physics of chaos assisted tunneling and two limiting factors to be considered in the perspective of an experimental observation: 
the influence of the distribution in quasimomentum $\beta$ and the restriction on $\hbar_{\rm eff}$.

The first limiting factor comes from the fact that the range of $\beta$ to be explored in the band diagram is bounded from below by the momentum width of the atomic cloud, see (\ref{betamin}). In order to have a large number of atoms crossing the bands at $\beta=0$ one needs then $\beta_0 > \Delta\beta$. To comply with this requirement (see Table~\ref{param_exp}), the value $\beta_0=0.05$ was chosen in Figs.~\ref{band_diag1}, \ref{band_diag2} and \ref{band_diag3}. 
With this value it is possible (and very common from our numerics) that the wave packet explores more than two bands during its traveling in the band diagram. When $\beta$ spans the whole Brillouin zone, the band associated to the eigenstate supported by one island will form avoided crossings with all the eigenstates supported by the chaotic sea. The number of such states $N_{\rm ch}$ is given in the semiclassical regime (small $\hbar_{\rm eff}$) by the ratio of the area of the chaotic sea ${\cal A}_{\rm ch}$ and the effective Planck constant $\hbar_{\rm eff}$: $N_{\rm ch}\sim {\cal A}_{\rm ch}/\hbar_{\rm eff}$. Assuming that these avoided crossings are uniformly distributed on the $\beta$ axis, the typical number of avoided crossings within a smaller range of width $2\beta_0$ is:
\begin{equation}
  \label{Nch}
  N_{\rm ch}(\beta_0)\sim 2\beta_0{\cal A}_{\rm ch}/\hbar_{\rm eff}\ .
\end{equation}
Therefore a lower bound for the available range of $\beta_0$ generically leads to multiple avoided crossings in the band diagram as it was seen e.g. in Fig.~\ref{band_diag3}.

The second limiting factor we identified is the available range in $\hbar_{\rm eff}$. 
One important consequence of having moderate values of $\hbar_{\rm eff}$ is that multiple branch avoided crossings are more likely to occur (see Figs.\ref{band_diag2}a and \ref{band_diag3}a). Indeed, following (\ref{Nch}), the number of avoided crossings grows when $\hbar_{\rm eff}$ is decreased.
One could simulate a more complex band diagram and consider generalized Landau Zener formula for a larger number of bands. It is e.g. possible to write analytical prediction for the crossing probability in a three band approximation \cite{LZ_3bands}. The drawback of such an approach is that it introduces additional parameters (second splitting, distance between the two splittings), and we could not see how to extract them from observables in a cold atom experiment.

{In our opinion, the inaccuracy of the LZ approach can be summarized in a physical way by saying that,
when several band avoided crossings are seen by the wave packet traveling in the band diagram, the
splitting extracted by the LZ scheme roughly equals to an average of all the encountered splittings.}

\subsection{Criteria to assess the accuracy of Landau Zener scheme and extract the splitting at $\beta=0$}

\begin{figure}
  \vspace{1cm}
    \begin{center}
      \includegraphics[width=\linewidth]{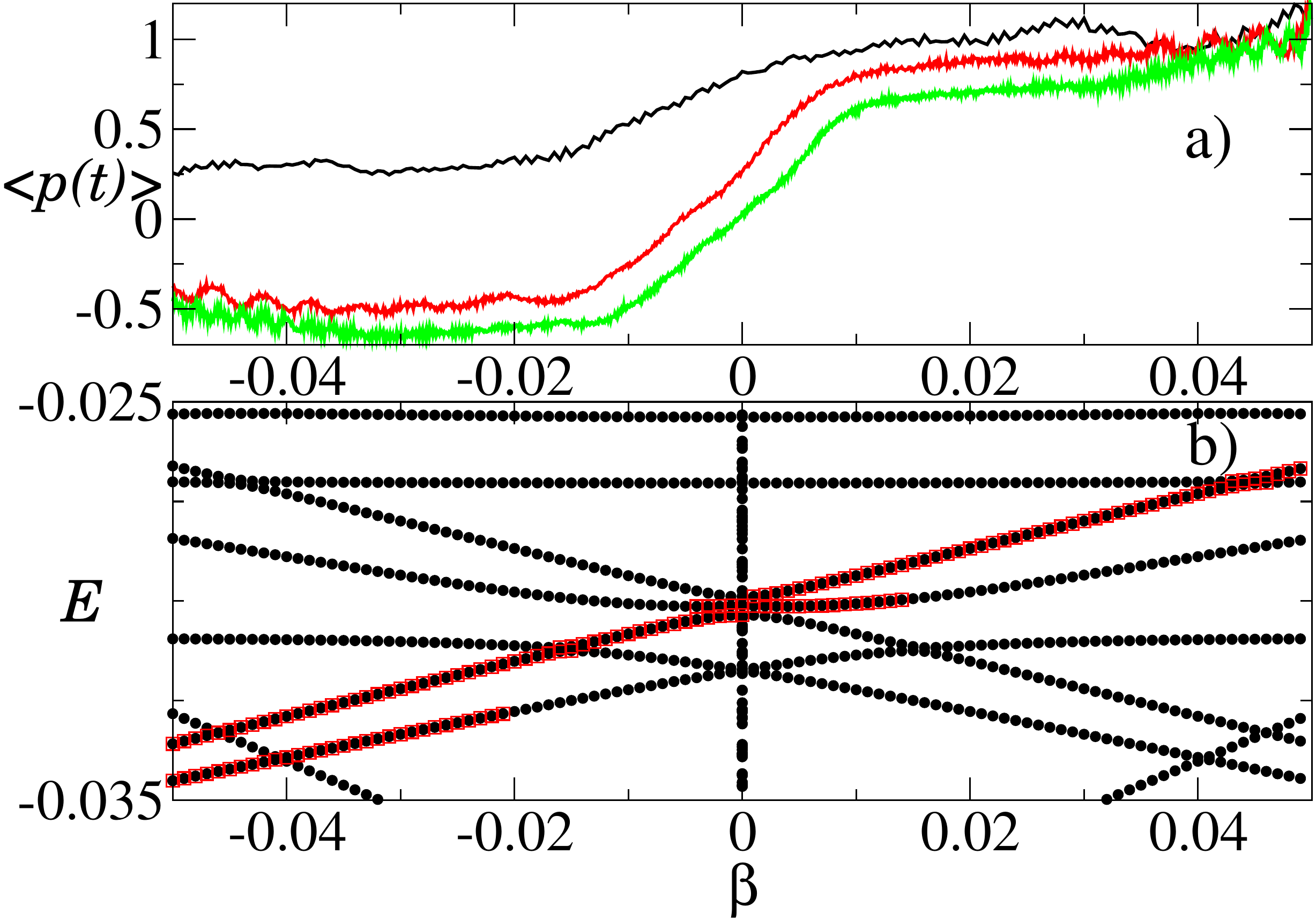}
    \end{center}
  \caption{(Color online) a) $\langle p(t)\rangle$ for $\varepsilon=0.4$, $\gamma=0.24$ and 
{$\hbar_{\rm eff}=0.0707$}. The horizontal axis has been converted into quasimomentum axis {using} the linear relation $\beta(t)=\beta_0-\beta_0 t/\pi n_{\rm acc}$.
Each curve stands for {different values} of $n_{\rm acc}$. Black (top on the left): $n_{\rm acc}=200$. Red (middle on the left): $n_{\rm acc}=947$. Green: (bottom on the left): $n_{\rm acc}=2147$. b) Band diagram. The red square branches stand for the levels, whose associated eigenstates in Husimi representation are mainly supported by the island around the point $(0,p_*)$ in phase space. One can see that $\langle p(t)\rangle$ changes appreciably when $\beta=0$ as expected for the tunneling between both symmetric islands.
Note that one has in this case $\delta^{{\rm(exact)}}=2.36\ 10^{-4}$ and $\delta^{{\rm (LZ)}}=3.38\ 10^{-4}$.  }
\label{hbar0.0707011}
\end{figure}

\begin{figure}
  \vspace{1cm}
    \begin{center}
      \includegraphics[width=\linewidth]{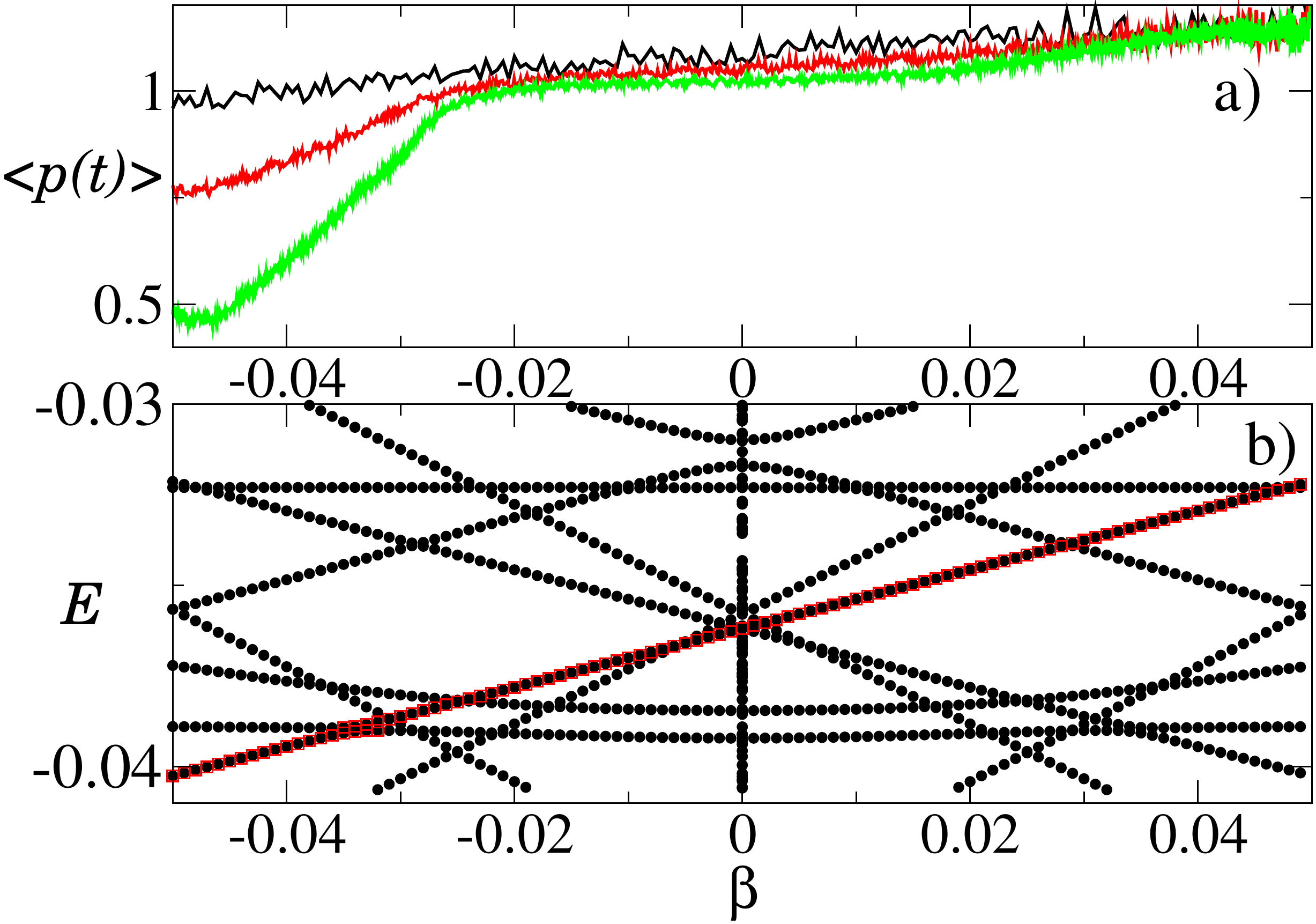}
    \end{center}
  \caption{
(Color online) a) $\langle p(t)\rangle$ for $\varepsilon=0.4$, $\gamma=0.24$ and $\hbar_{\rm eff}=0.0813$. The horizontal axis has been converted into quasimomentum axis {using} the linear relation $\beta(t)=\beta_0-\beta_0 t/\pi n_{\rm acc}$.
Each curve stands for {different values} of of $n_{\rm acc}$. Black (top on the left): $n_{\rm acc}=200$. Red (middle on the left): $n_{\rm acc}=881$. Green: (bottom on the left): $n_{\rm acc}=2430$. b) Band diagram. The red square branches stand for the levels, whose associated eigenstates in Husimi representation are mainly supported by the island around the point $(0,p_*)$ in phase space. One can see that $\langle p(t)\rangle$ changes appreciably when $\beta\neq 0$. This is in striking contrast to Fig.~\ref{hbar0.0707011}. It also means that the atom mainly tunnels to the chaotic sea.
One has here $\delta^{{\rm(exact)}}=1.36\ 10^{-8}$ and $\delta^{{\rm (LZ)}}=1.38\ 10^{-4}$.
}
\label{hbar0.0812794}
\end{figure}

\begin{figure}
    \includegraphics[width=\linewidth]{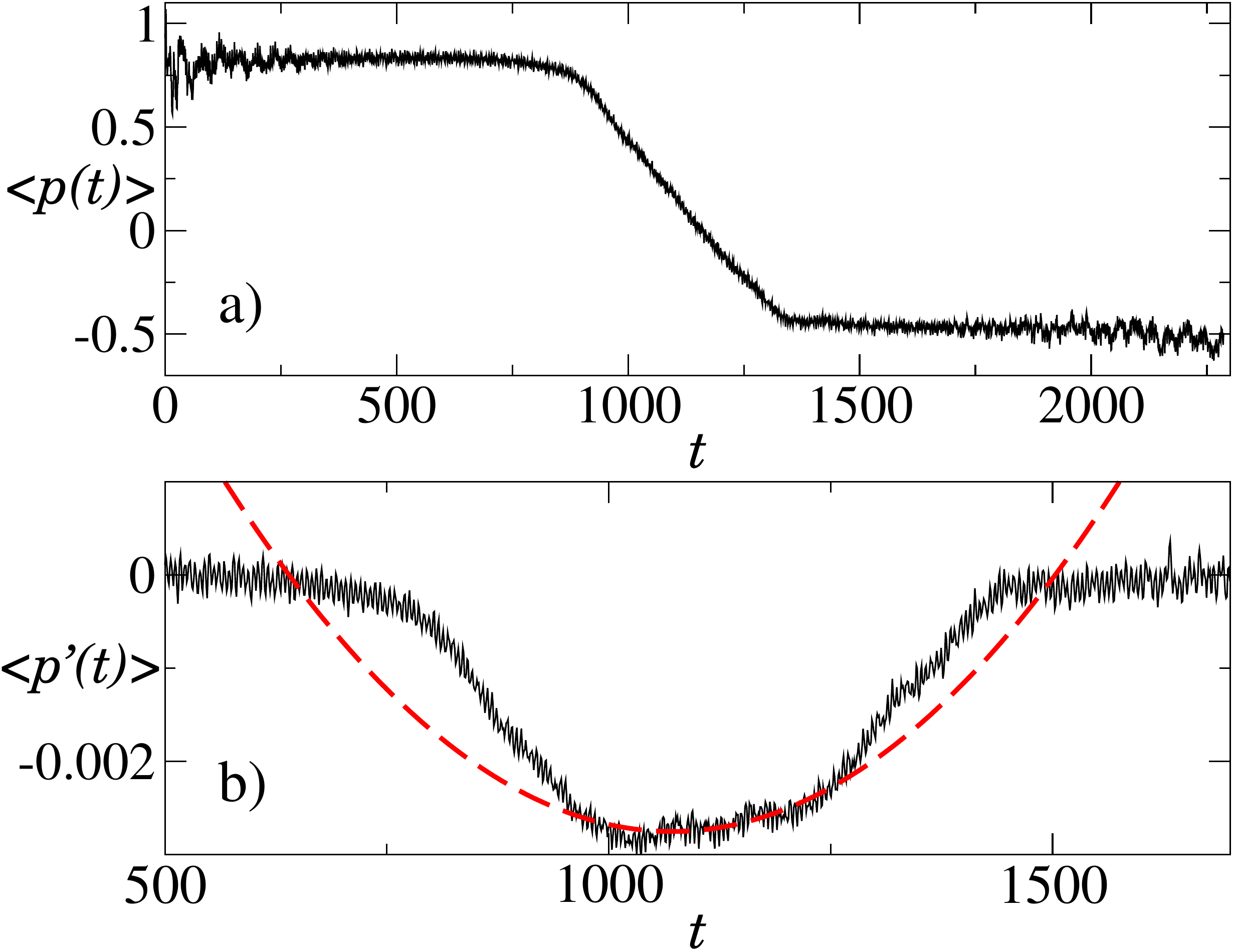}
  \caption{(Color online) a) $\langle p(t)\rangle$ for $\varepsilon=0.4$, $\gamma=0.24$ and $\hbar_{\rm eff}=0.1277$. b) Black line: time derivative of $\langle p(t)\rangle$ around $t=t_{\rm cross}=1143$. The red dashed line stands for a quadratic fit. The curvature of $\langle p(t)\rangle$ at $t=t_{\rm cross}$ is positive as expected.
Note that one has in this case $\delta^{{\rm(exact)}}=4.41\ 10^{-4}$ and $\delta^{{\rm (LZ)}}=4.80\ 10^{-4}$.}
\label{hbar0.127725}
\end{figure}

\begin{figure}
  \includegraphics[width=\linewidth]{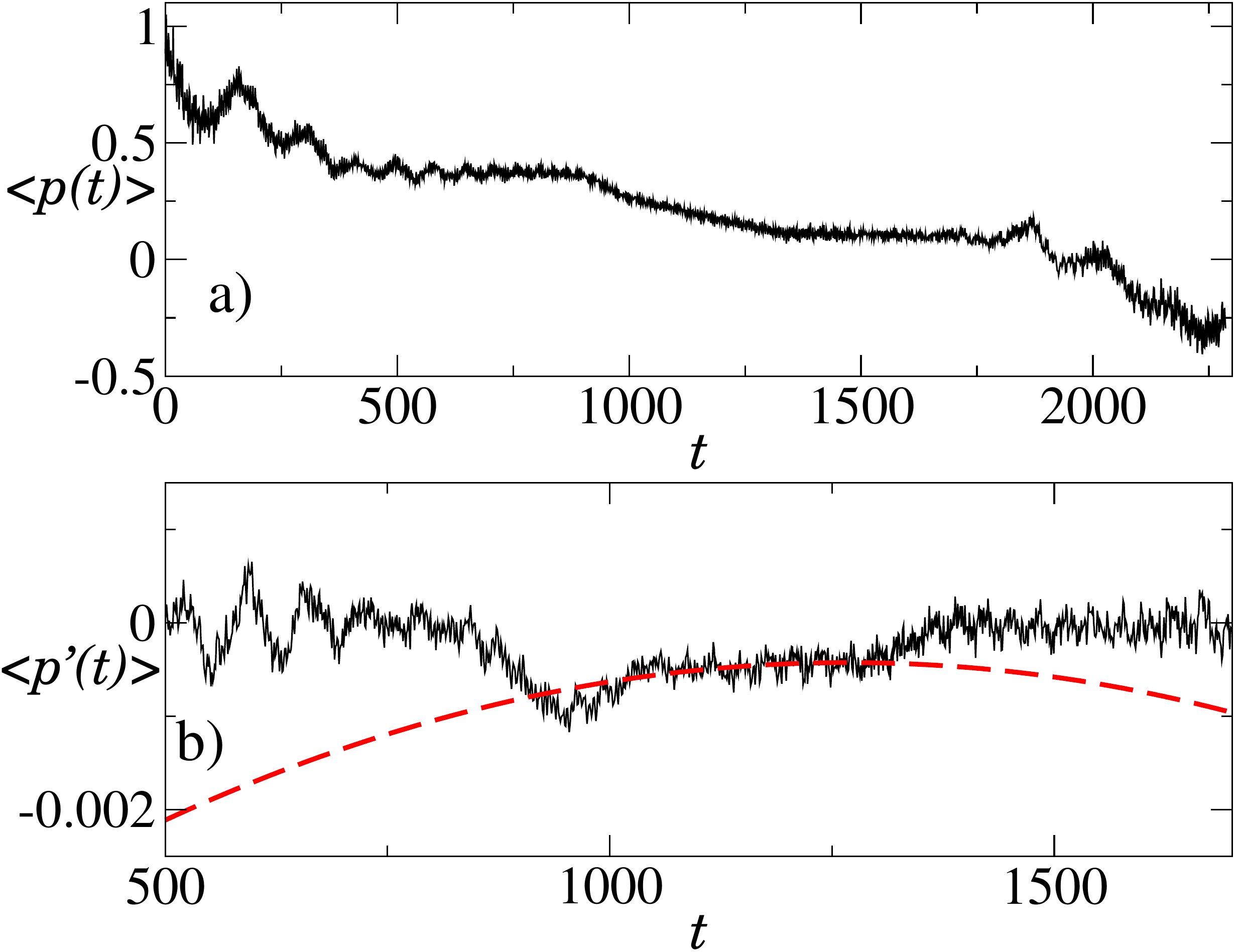}
  \caption{(Color online) a) $\langle p(t)\rangle$ for $\varepsilon=0.4$, $\gamma=0.24$ and $\hbar_{\rm eff}=0.1611$. b) Black  line: time derivative of $\langle p(t)\rangle$ around $t=t_{\rm cross}=1143$. The red dashed line stands for a quadratic fit. Now the curvature of $\langle p(t)\rangle$ around $t=t_{\rm cross}$ is negative.
Note that one has in this case $\delta^{{\rm(exact)}}=2.49\ 10^{-4}$ and $\delta^{{\rm (LZ)}}=3.81\ 10^{-4}$. The wrong sign of the curvature indicates a {significant} discrepancy between these two values.}
\label{hbar0.161057}
\end{figure}

Having described the two main limiting factors of the LZ method, we searched for a way to systematically control the accuracy of the LZ scheme. We designed several criteria in order to assess whether the LZ method is efficient for a given value of the set of parameters. All the criteria listed below share the same goal: to determine whether the atom, when traveling in the band diagram, does mainly tunnel at $\beta=0$.
The criteria are roughly based on a finer and finer analysis of the structure of $\langle p(t)\rangle$, as it is an accessible observable in a cold atom experiment.
We should stress here the importance of these criteria in order to analyze the results of an experimental run. While in the current numerical study, one can access to the exact splitting using the band diagram, this is no longer true in the experiment. Our strategy is then to build a procedure, which can tell whether the splitting extracted via the LZ scheme is close to the exact one without knowing the latter. 
In the rest of the paragraph, several criteria are presented, which were used to analyze our data and to select the LZ splittings which are surely in good agreement with the splittings at $\beta=0$. 

The first criterion consists in looking at the final value of $\langle p(t)\rangle$ for the largest $n_{\rm acc}$. 
When increasing $n_{\rm acc}$ we expect the final value $\langle p\rangle_{\rm final}$ to be small, theoretically close to $-p_*$ (see e.g. Fig.~\ref{band_diag1}b). 
This criterion has shown to be very useful to detect the situations where the splitting at $\beta=0$ is very small. 
Unfortunately this criterion is also very excluding. The numerically extracted value for $A$ in the fitting process, see (\ref{fit_eq}), is typically less than $1$ (the typical expected value) and setting a too large minimal acceptable value for $A$ excludes a large majority of LZ splittings.

The second criterion relies on a more precise analysis of the curve $\langle p(t)\rangle$ for the larger $n_{\rm acc}$ (lowest traveling speed along the $\beta$ axis).
From Eq.~(\ref{v_beta2}) and for a fixed $n_{\rm acc}$, one can determine exactly the time $t_{\rm cross}$, at which the wave packet must reach $\beta=0$. 
In other words, at $t=t_{\rm cross}$, the wave packet is assumed to tunnel from one stable island to its symmetric partner in the phase space.
In our analysis it was checked that 
$\langle p(t)\rangle$ changes appreciably around $t_{\rm cross}$. Two examples are shown in Fig.~\ref{hbar0.0707011} and Fig.~\ref{hbar0.0812794}. It can be clearly seen that this criterion can be useful to discard situations when the splitting extracted from the LZ scheme is far from the exact splitting at $\beta=0$.

A third criterion to check that the tunneling occurs between the symmetric stable islands, was performed by considering the local curvature of the curve $\langle p(t)\rangle$ around $t=t_{\rm cross}$.
Running a local quadratic fit of $\langle p'(t)\rangle$ around $t=t_{\rm cross}$ indeed gives access to its local curvature. 
In the expected scenario (see Fig.~\ref{band_diag1}b), the curve $\langle p(t)\rangle$ should be locally convex around $t=t_{\rm cross}$.
Checking the curvature has appeared to be efficient to sort out the situations where several crossings happen within the prescribed range of $\beta$, especially close to $\beta=0$. This criterion has proved to be another way to detect the values of $\hbar_{\rm eff}$ for which the band diagram is very complex. 
An example of such analysis of the convexity of $\langle p(t)\rangle$ is given in Figs.~\ref{hbar0.127725} and \ref{hbar0.161057}.

Eventually we used a last criterion in order to rule out too small splitting at $\beta=0$. This appeared to be necessary as these splittings would require a significantly longer traveling time. As already mentioned above, the number of modulations, which can be applied to the atomic cloud, is always limited in a real experiment. Therefore it is crucial to be able to exclude {\it a priori} the splittings, which cannot be detected within the given range of traveling time.
The criterion here is to look at the magnitude of the variation of the curve $\langle p(t)\rangle$. A small variation is surely related to a very tiny splitting. Physically this means that only a very small part of the wave packet may have tunneled during the experimental run.
This criterion has proved to be efficient, especially to rule out the values of $\hbar_{\rm eff}$ forming a dip as in Fig.~\ref{splittings_LZ_Law} around $\hbar_{\rm eff}^{-1}\simeq 4.8$. 

We have implemented the analysis described in the previous paragraphs in a systematic way. 
We tuned the above described criteria in order to extract the energy splittings from the LZ scheme such that the relative error is less than $20\% $ in comparison to the exact value at $\beta=0$. It reduced the number of extracted splittings from $441$ to $23$.
This dramatic fall of the number of LZ splittings puts too strong limitations in order to build a statistical description of the splitting fluctuations in the presence of chaos. 
It was checked that taking other values of the classical parameters $(\gamma,\varepsilon)$ leads to similar 'success' rate of the LZ method. {As a conclusion we estimate that this method has a very limited reliability in our context.}

\section{Third route to chaos assisted tunneling: spatial tunneling oscillations in a dynamical double well}
\label{tunnelx}

In the previous Sections, we showed that tunneling in the momentum space 
requires a localization of the initial wave packet in
both space and momentum with an extremely narrow initial momentum distribution, see (\ref{Dv_init}), 
to observe the strong fluctuations due to chaos assisted tunneling.

In this Section a different regime is explored, where the regular
islands are spatially symmetric (see Fig.~\ref{portraitph}). Such regime offers the following advantages. First of all, while the momentum
symmetry used in \cite{PhillipsNature, RaizenScience} is fragile, as detailed in Sect.~\ref{fragilsym}, the spatial
symmetry $x\mapsto -x$ is a robust property of the atomic modulated pendulum \eqref{Hclas}. Moreover, considering the tunneling in
space allows to enrich the problem to the case of a lattice of regular islands, and even to double or triple
well lattices (super-lattices). 

An example of double well super-lattice has been recently realized in \cite{Bloch}. However, with this technique it is difficult to change both the distance between the wells and the population imbalance. More importantly in the context considered here, the dynamics in this double well lattice is completely regular.
We propose another method, which is to dynamically generate a bi-periodic array with a controllable mixed classical dynamics. The study of tunneling in such configuration however requires to be able to extract sub-lattice spacing features
of the spatial distribution of the BEC, something impossible to achieve by direct observational means.
In the next section, we will describe a phase space rotation technique, which transfers spatial signatures to
momentum space. This technique has recently been validated experimentally \cite{GueryOdelin16}.

In this Section, we first analyze the classical dynamics giving rise to a dynamical double well super-lattice. Then the tunneling between the spatially symmetric wells is studied, especially in the regime where they are separated by a chaotic sea.

\subsection{Classical dynamics} 

\begin{figure*}
\centering
 \includegraphics[width=.4\textwidth]{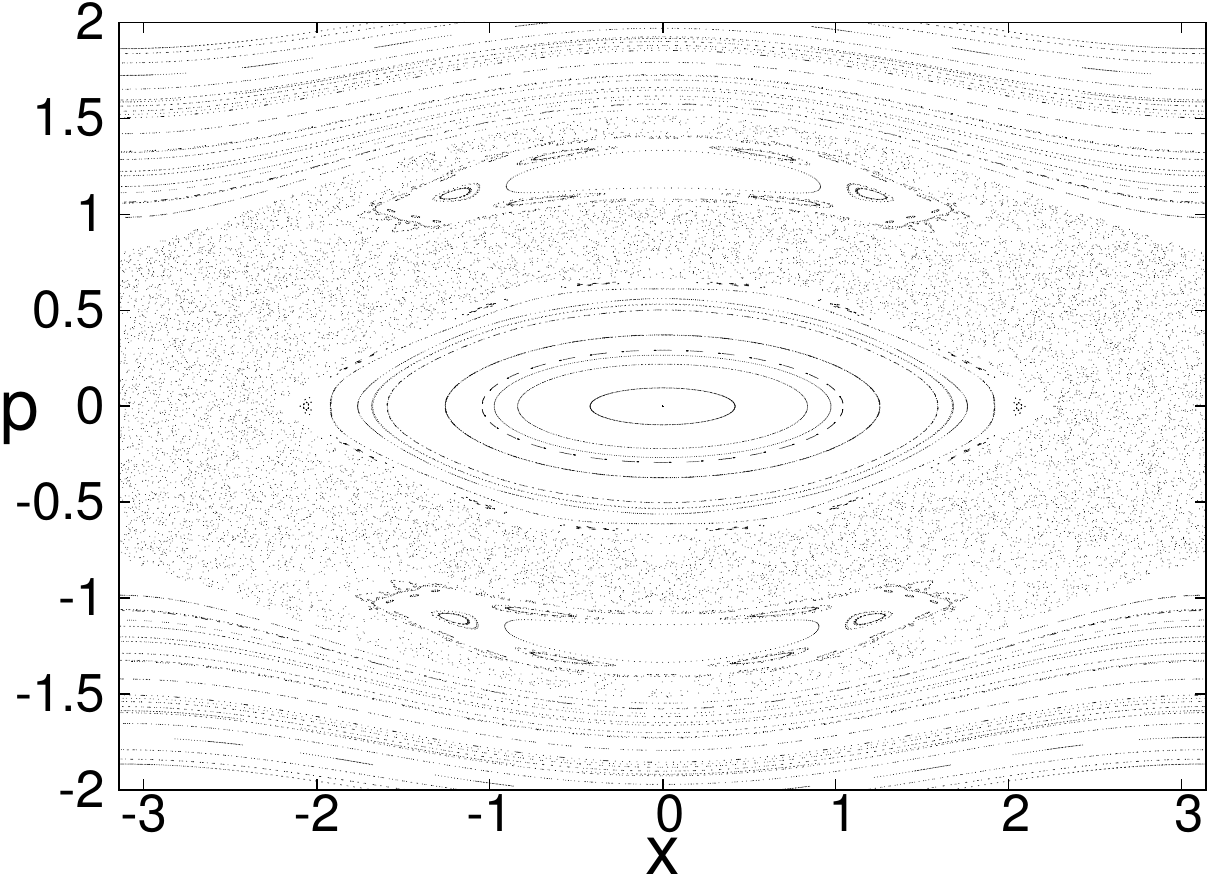}
 \includegraphics[width=.4\textwidth]{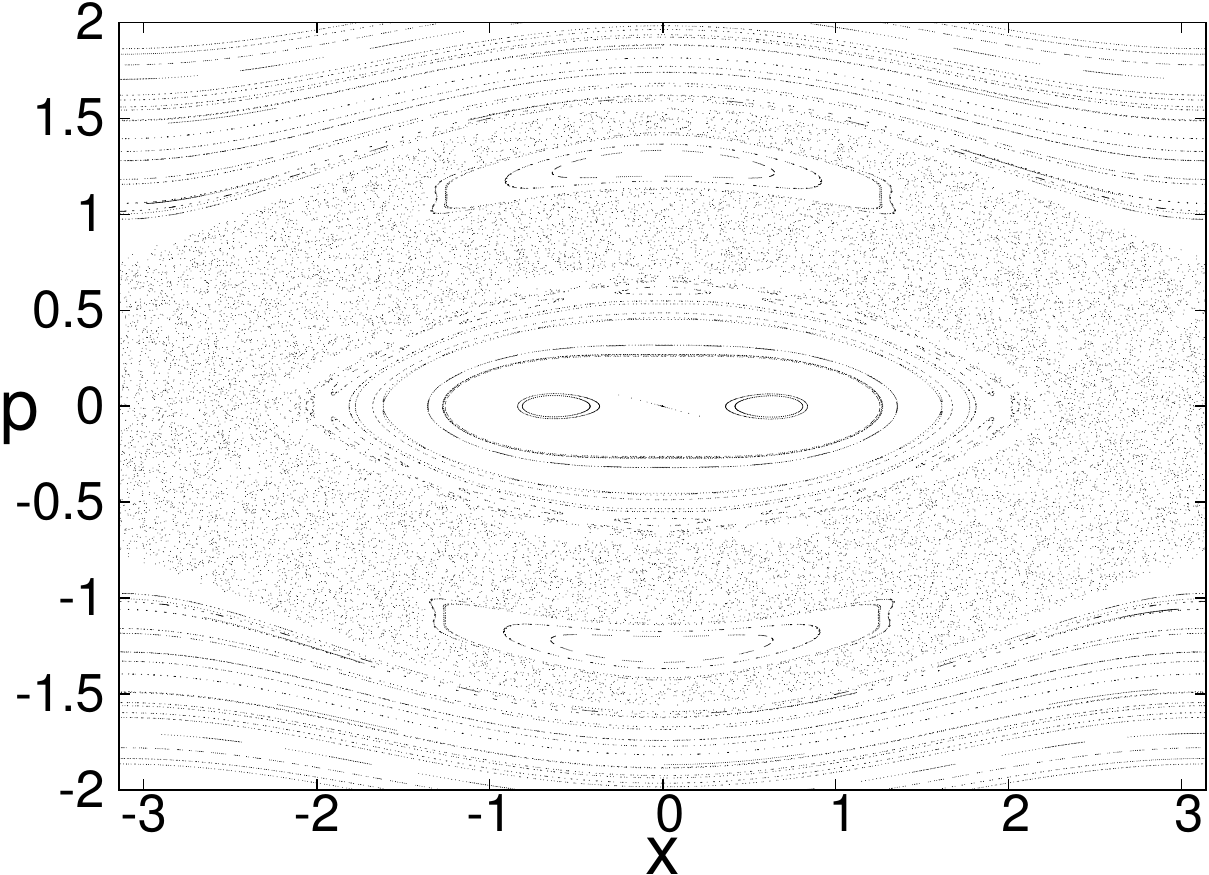}\\
 \includegraphics[width=.4\textwidth]{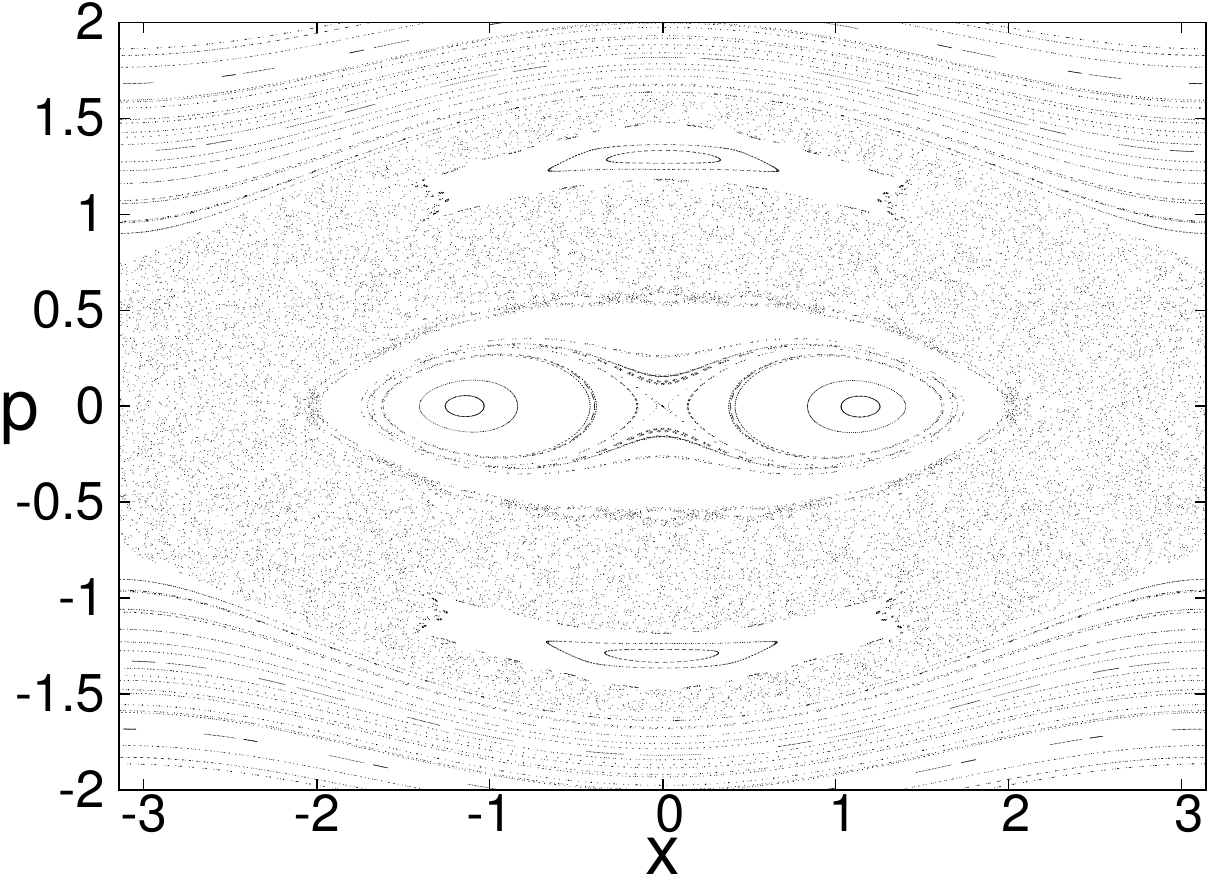}
 \includegraphics[width=.4\textwidth]{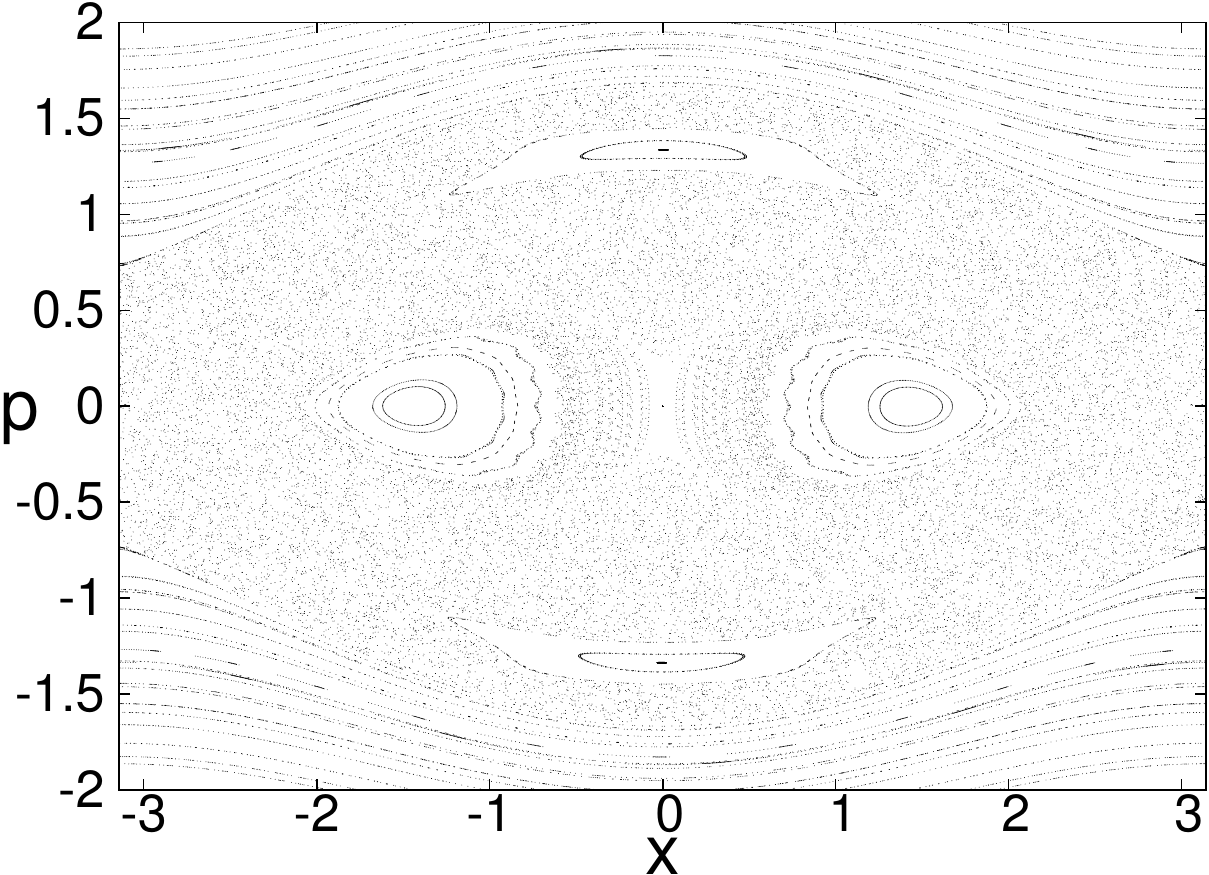}
\caption{Emergence of the bifurcation of the central fixed point. Poincar\'e SOS for $\varepsilon=0.29$, and increasing values of $\gamma$. Top left: $\gamma=0.203$. Top right: $\gamma=0.23$. Bottom left: $\gamma=0.26$. Bottom right: $\gamma=0.29$.}
\label{portraitph}
\end{figure*}

\begin{figure}
\centering
 \includegraphics[width=\linewidth]{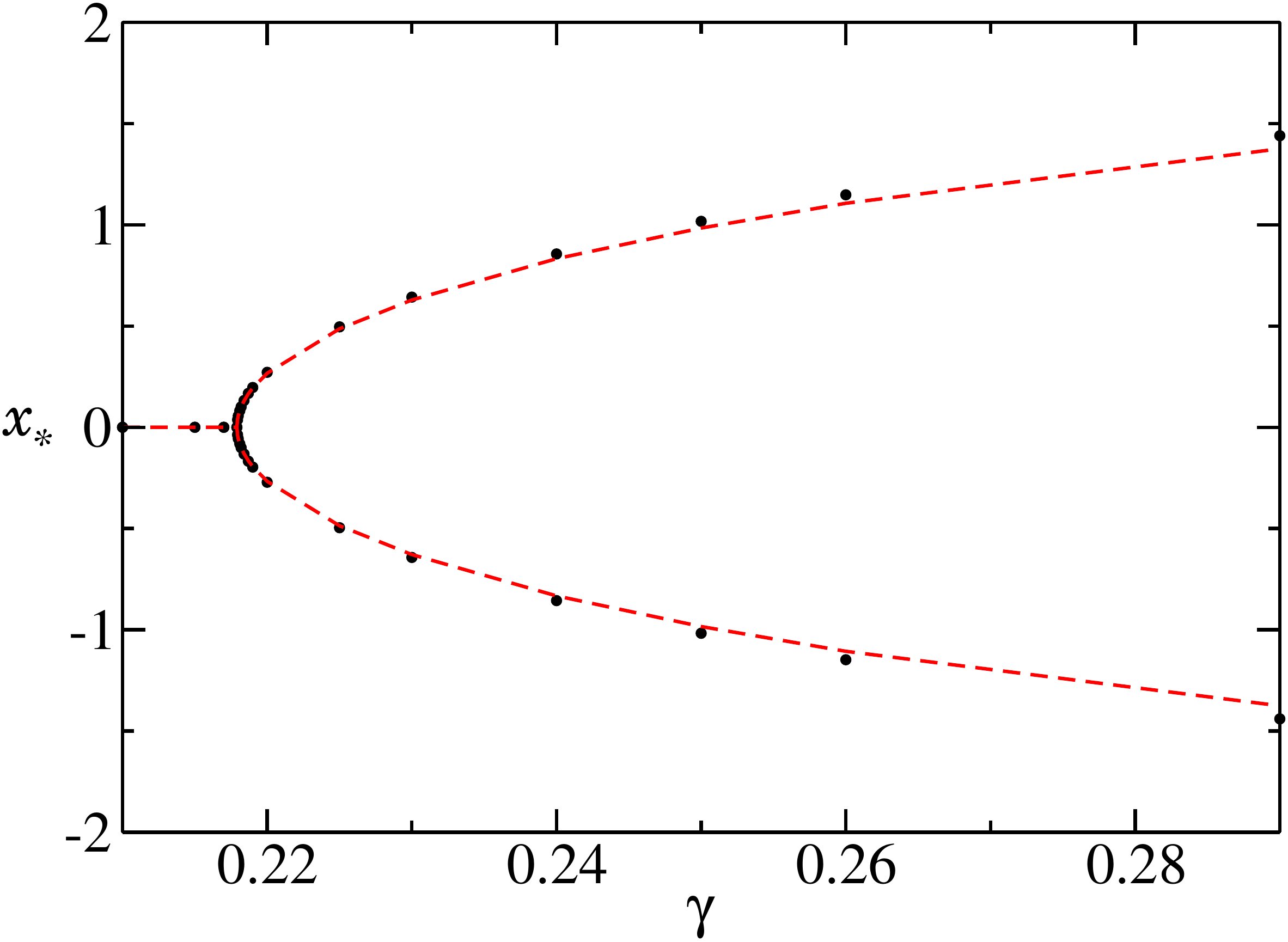}
\caption{(Color online) Bifurcation diagram for the classical modulated pendulum defined by (\ref{Hclas}) for $\varepsilon=0.29$. The vertical axis shows the center{s} of the stable island{s} for $p=0$. The horizontal axis stands for $\gamma$. Black circles: numerical data. Red dashed line: Approximate theory of the bifurcation  $x_* \sim \sqrt{\gamma-\gamma_b}$ with $\gamma_b \approx 0.217 $ (see text).}
\label{classbifu}
\end{figure}

In order to simply illustrate our arguments, we will keep $\varepsilon=0.29$ in this Section. From Fig.~\ref{degree_chaos} it can be seen that, when increasing $\gamma$ from $0.1$ to $0.3$ for this value of $\varepsilon$, the size of the chaotic region increases significantly.

Spatially symmetric islands come from a pitchfork bifurcation \cite{Meyer, delande2,MouchetLeboeuf} of the stable island at the origin $x=0$ for small $\gamma$. 
As shown by numerical simulations in Fig.~\ref{portraitph}, the bifurcation happens, for $\varepsilon=0.29$, around the value $\gamma_b\approx 0.2179$. For $\gamma< \gamma_b$ the origin of the phase space coordinate $(x,p)=(0,0)$ is a stable fixed point, surrounded by a regular island. For $\gamma > \gamma_b$ this point becomes unstable and two new stable points arise at $(x,p)=(\pm x_*,0)$ for the evolution operator over two periods (period doubling bifurcation). The abscissa of the center of the new wells $x_*$ evolve with $\gamma$ following the bifurcation diagram shown in Fig.~\ref{classbifu}.

The detailed analysis of the bifurcation is described by the theory of the normal forms \cite{Meyer, delande2,MouchetLeboeuf}.
A simple sketch of the mechanisms underlying the phenomenon is as follows: the instability of the central point $(x=0,p=0)$ can be understood by approximating the trigonometric potential of (\ref{Hclas}) around its minima using an harmonic potential. Then Eqs~(\ref{clasdyn}) can be mapped onto the Mathieu equation. It is well known that Mathieu equation has bounded or unbounded solutions depending on its parameters. According to this point of view, the central point becomes unstable around $\gamma > 0.217$. The stability of the double wells can be understood by taking the quartic corrections to this harmonic approximation. This leads to classical oscillations, of period $4 \pi$ in our units, between both wells with central points following $x_* \sim \sqrt{\gamma-\gamma_b}$, as shown by the red dashed line in Fig.~\ref{classbifu}.

In the following we are interested in the regime after the bifurcation $\gamma > \gamma_b$. The reason is that the phase space then contains two symmetric stable islands.
An example of such situation is illustrated in the Poincar\'e SOS displayed in Fig.~\ref{portraitph} bottom right. 
Again, it is worth stressing that these islands are not stable under the $2\pi$-period classical map: if a particle is localized on one island (say the left one), one period later the particle is on the other island (the right one). Therefore, there is a classical transport between the islands: at every even multiple of the period the particle stands in the initial well, whereas at every odd multiple of the period it stands in the other well. However, if one looks at the classical dynamics \textit{every two periods}, one is left with two spatial trapping wells. Then it is very similar to the textbook situation of a double well potential, which can lead to tunneling. The main and crucial difference here is that the wells are separated by a region in phase space, where the dynamics can be chaotic.

\subsection{Quantum dynamics}

\begin{figure}
\centering
 \includegraphics[width=\linewidth]{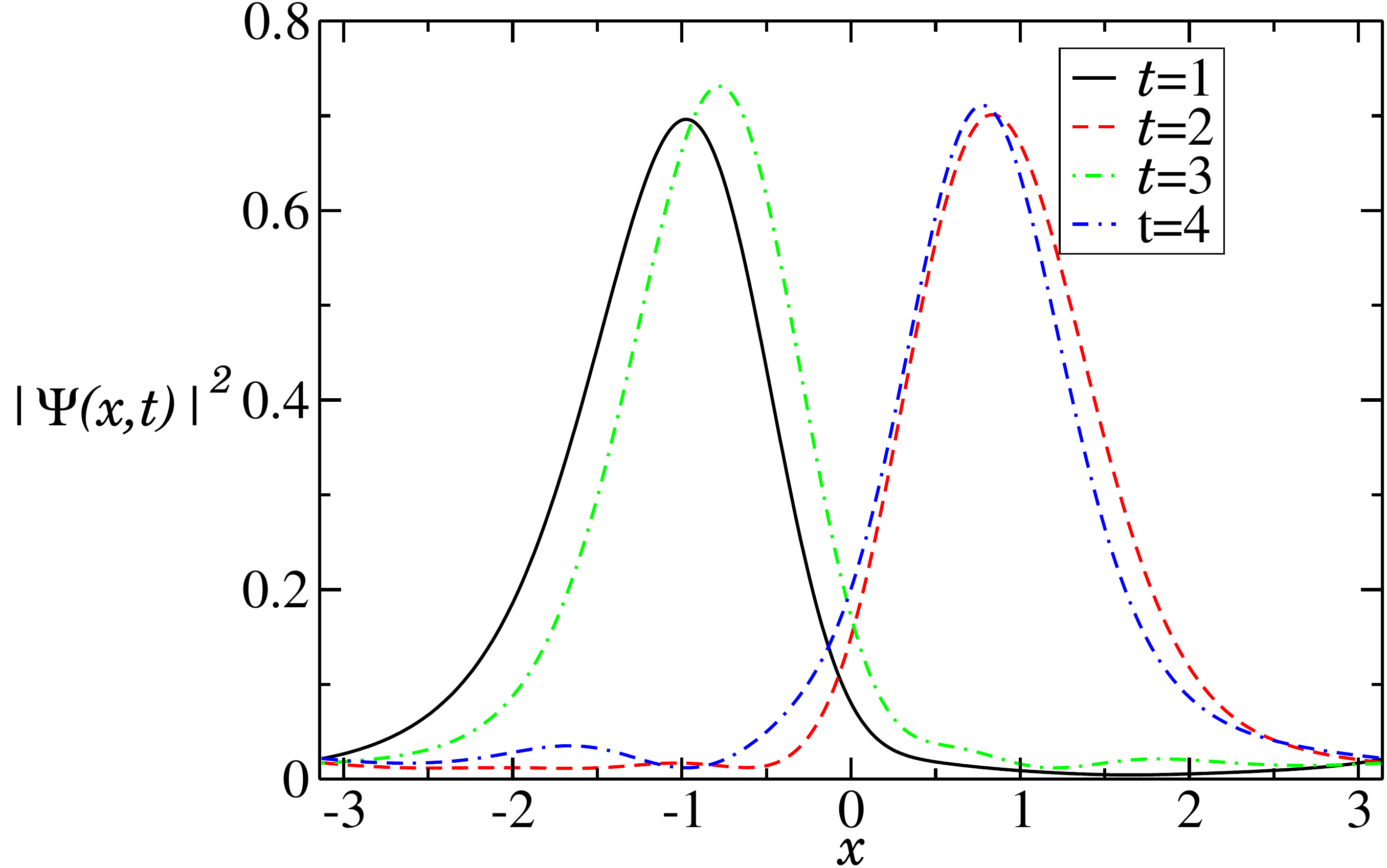}
\caption{ (Color online) Spatial probability density for the quantum atomic modulated pendulum defined by (\ref{Hclas}) for $\varepsilon=0.29$, $\gamma=0.29$ and $\hbar=0.3$. The initial state is a coherent state centered at $(x,p)=(1.2,0)$ with width $\Delta x=2 \pi/10$. 
The
quantum wave packet follows the classical transport between the islands: at every even multiple of the period the wave packet stands in the initial well, whereas at every odd multiple of the period it stands in the other well.
}
\label{QMP}
\end{figure}

\begin{figure}
\centering
 \includegraphics[width=\linewidth]{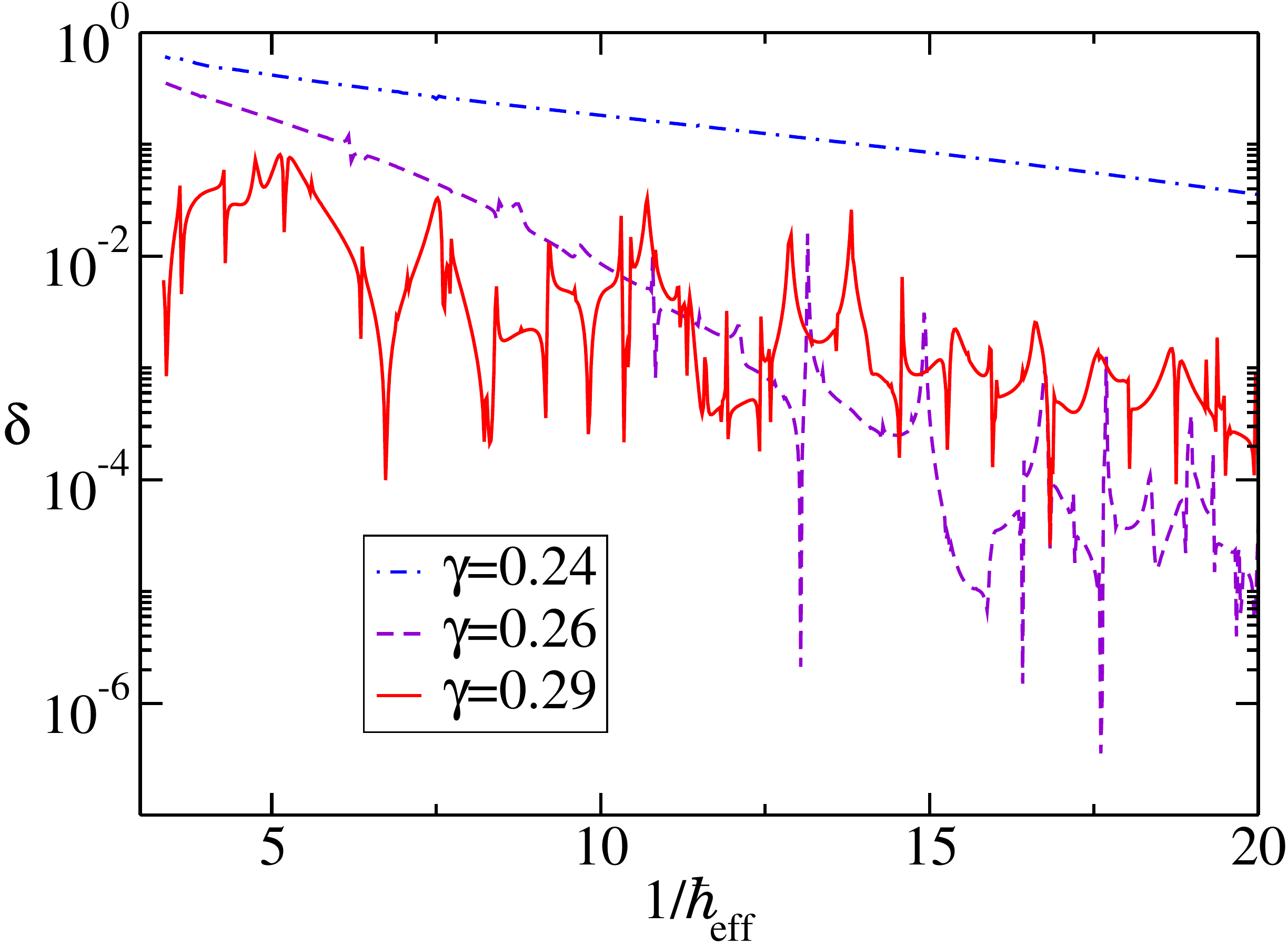}
 \caption{(Color online) Spatial tunneling splitting for the atomic modulated pendulum \eqref{Hclas} as a function of $1/\hbar_{\rm eff}$ for $\varepsilon=0.29$ and different values of $\gamma$. In the regular regime ($\gamma < 0.26$), the energy splitting follows (\ref{splitting_Leg}) 
whereas {it} show{s} strong fluctuations by orders of magnitude in the chaos assisted regime ($\gamma>0.26$).}
\label{deltatunnxvsgam}
\end{figure}

\begin{figure}
\centering
  \includegraphics[width=\linewidth]{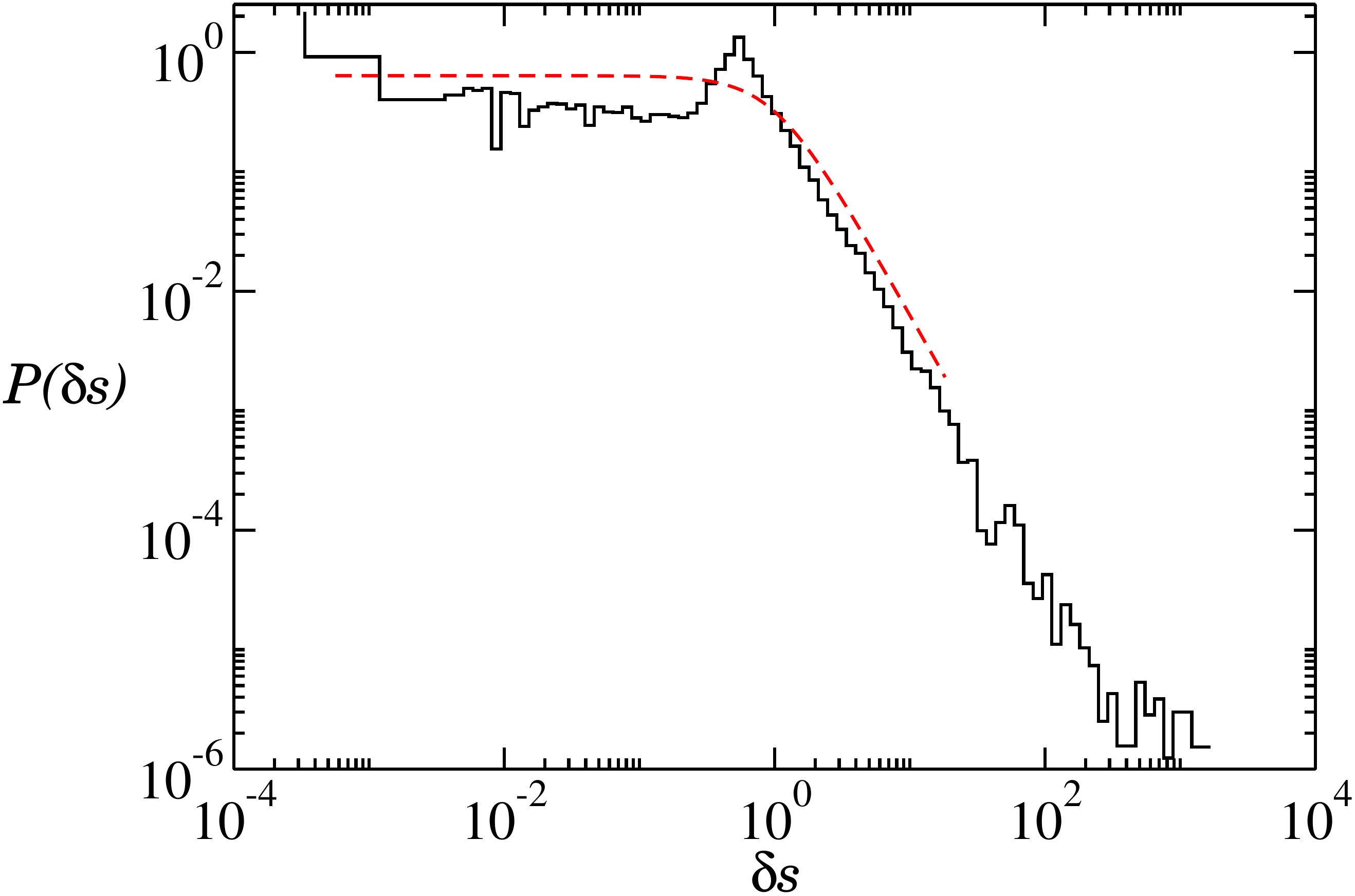}
 \caption{(Color online) Distribution of the spatial splitting fluctuations in the chaotic regime, $\varepsilon=0.29$, $\gamma=0.29$ and $\beta=0$. Black histogram: numerical data. Dashed red line: theoretical prediction (\ref{cauchy}).}
\label{distsplit}
\end{figure}

We ran numerical simulations of the quantum evolution of a coherent state with quasimomentum $\beta=0$, centered on one of the well (see Fig.~\ref{QMP}). We first checked that the
quantum wave packet follows the classical transport after the bifurcation $\gamma>\gamma_b$.

Then we focused on the quantum tunneling between both symmetric stable islands, whose centers are along the position axis. 
In order to get rid of the presence of classical transport between the islands, the quantum dynamics is considered {\it every two periods}. 
The relevant evolution operator is now $\hat{U}^2$, instead of $\hat{U}$ as defined in (\ref{eig_eqU}).
The problem is now very similar to the double well potential problem. Similarly as above, the main focus is on the eigenstates of the quantum propagator $\hat{U}^2$, which are mainly supported by the stable islands. 
Two of them play a central role: 
$\left|\psi_+\right>$, and $\left|\psi_-\right>$. They are associated to the eigenvalues $E_+$ and $E_-$ of the Hamiltonian. We are interested in the energy splitting as defined by (\ref{def_splitting}).

Interestingly, by changing the parameter $\gamma$, one can explore different tunneling regimes from regular to assisted by chaos (see Fig. \ref{deltatunnxvsgam}). 
The variations of the splitting when a parameter like $\hbar_{\rm eff}$ is changed are very different whether a chaotic region of large area as compared to $\hbar_{\rm eff}$ is present in between the islands. In the regular regime, 
the splitting $\delta$ depends on $\hbar_{\rm eff}$ like (\ref{splitting_Leg}).
On the contrary, chaos assisted tunneling shows very strong fluctuations, by orders of magnitude, of the splitting over small parameter intervals \cite{tomul,leyvraz}.

We can further analyze the previous results by looking at the fluctuations of the energy splitting $\delta$ in a more systematic way. 
In the chaotic regime ($\gamma=0.29$), the distribution of the splitting fluctuations with respect to its typical value agrees well with the theoretical prediction (\ref{cauchy}) \cite{tomul,leyvraz}.

\subsection{Tunneling oscillations in space: sensitivity towards the quasimomentum}

\begin{figure}
\centering
 \includegraphics[width=\linewidth]{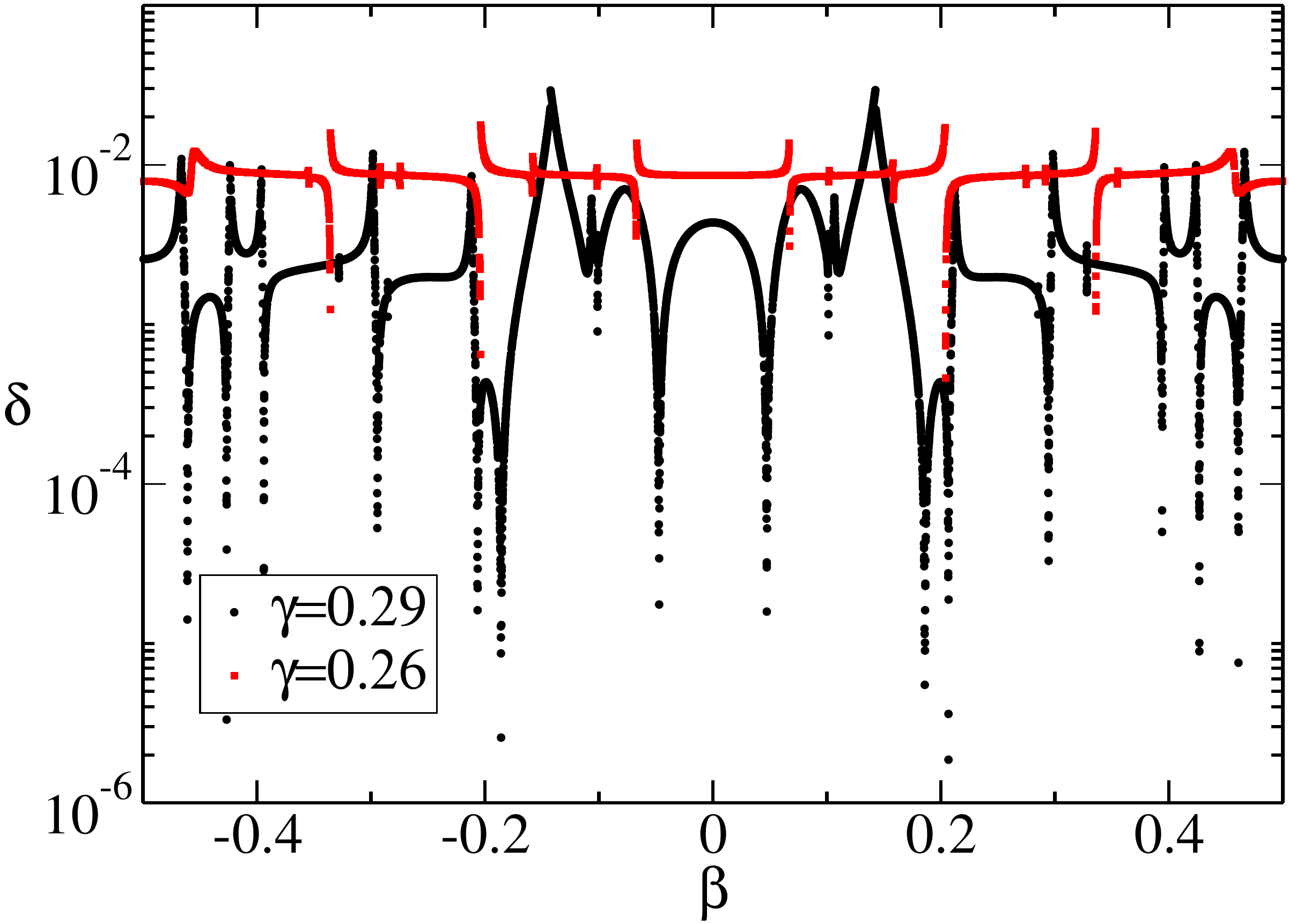}
 \caption{(Color online) Sensitivity of the spatial tunneling splitting as a function of the quasimomentum. In the regular regime $\gamma=0.26$ (red square points), the variations consist in few sharp resonances whereas in the chaotic regime $\gamma=0.29$ (black circle points), the fluctuations are by orders of magnitude, as expected from chaos assisted tunneling theory. Note however that a correlation scale in quasimomentum is clearly visible which scales as the inverse of the density of states in the chaotic sea. {In the present case relevant for experiments,} the correlation scale is larger than the initial distribution of quasimomentum $\Delta \beta\approx 0.02$. The other parameters are $\hbar_{\rm eff}=0.1$ and $\varepsilon=0.29$.}
\label{spasplitvsbeta}
\end{figure}

\begin{figure}
\centering
 \includegraphics[width=\linewidth]{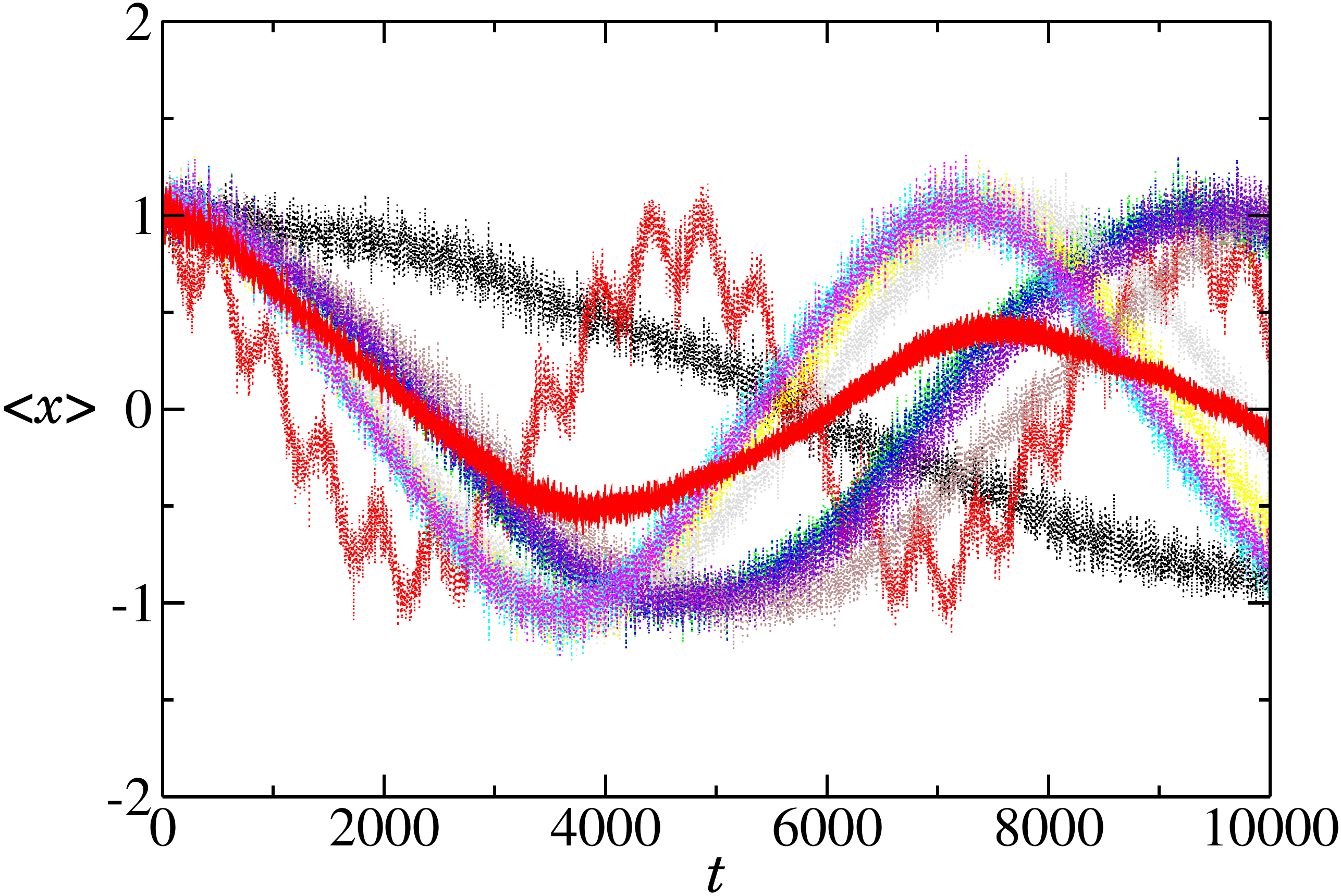}
 \caption{(Color online) Average position $\langle x(t)\rangle$ for the quantum modulated pendulum defined by (\ref{Hclas}) for $\varepsilon=0.29$ and $\gamma=0.29$ as a function of time. The dotted curves stand for different random values of $\beta$ uniformly distributed in $[-0.02, 0.02]$. The red full curve is an average of $180$ of such random values of $\beta$.}
\label{oscil_h0.11_beta_av}
\end{figure}

\begin{figure}
\centering
 \includegraphics[width=\linewidth]{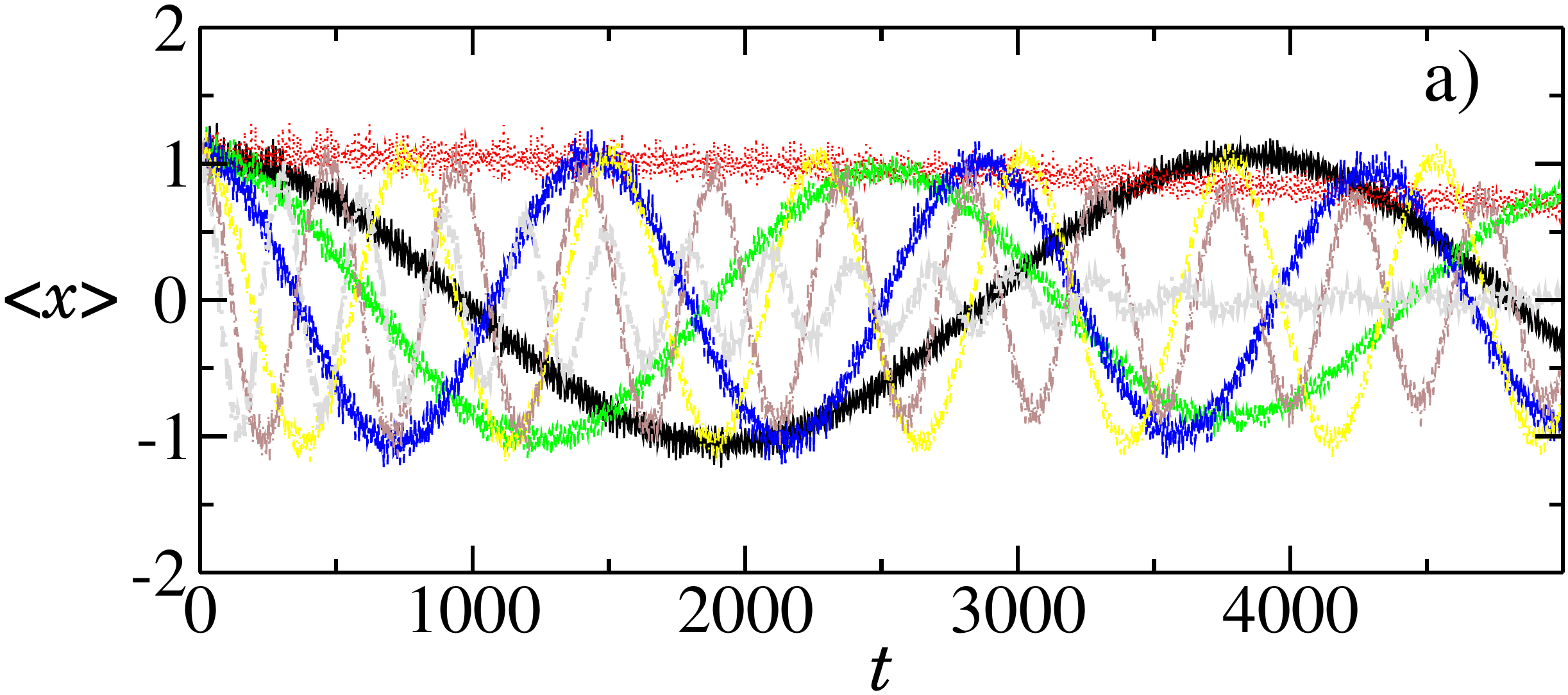}
  \includegraphics[width=\linewidth]{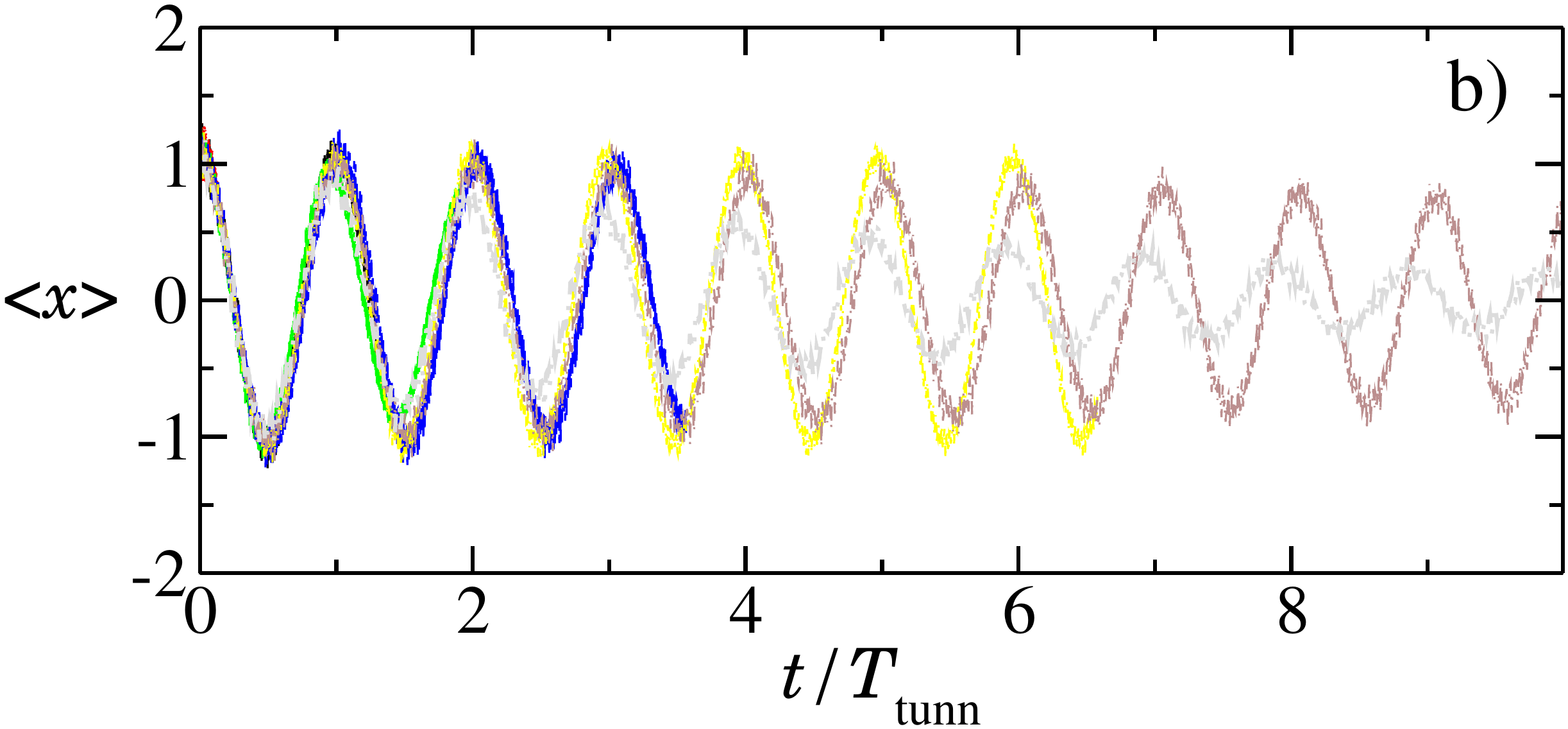}
 \caption{(Color online) a) Average position $\langle x(t)\rangle$ for the quantum modulated pendulum defined by (\ref{Hclas}) for $\varepsilon=0.29$ and $\gamma=0.29$ as a function of time. b) $\langle x(t)\rangle$ with the time rescaled by the tunneling period $T_{\rm tunn}$ defined in (\ref{Ttunnspa}). {The curves are periodic functions of period $1$ (in units of $T_{\rm tunn}$) and of similar amplitude $x_*\approx 1$. }
 Each curve corresponds to an average over $180$ random values of $\beta$ between $-0.02$ and $0.02$. Black line: $\hbar_{\rm eff}=0.144$. Red dotted line: $\hbar_{\rm eff}=0.149$. Green dashed line:  $\hbar_{\rm eff}=0.154$. Blue long dashed line: $\hbar_{\rm eff}=0.16$. Yellow dashed dotted line: $\hbar_{\rm eff}=0.166$. Brown long dashed dotted line: $\hbar_{\rm eff}=0.173$. Gray dashed double dotted line: $\hbar_{\rm eff}=0.18$. }
\label{oscillx_vs_hbar_beta_av}
\end{figure}

If an initial wave packet starts in one well, it will mainly overlap with the two symmetric and antisymmetric eigenstates $|\psi_-\rangle$ and $|\psi_+\rangle$. This leads to the seminal tunneling oscillations, as explained in Sect.~\ref{tunnelp_oscill}. Similarly to \eqref{deltatoT}, we define the period of the oscillations as:
\begin{equation}
  T_{\rm tunn}=\frac{2 \hbar_{\rm eff}}{\delta} \; ,
\label{Ttunnspa}
\end{equation}
where the new factor $2$ comes from the evolution operator being over two periods. 
Note that the oscillations are now displayed by considering the average position $\langle x(t)\rangle$, in the same vein as we considered the average momentum $\langle p(t)\rangle$ in Sect.~\ref{tunnelp_oscill}.

The main question about this new tunneling in space 
is to estimate how much the tunneling oscillations are sensitive towards the quasimomentum $\beta$. Contrary to the case of tunneling in momentum space, a finite quasimomentum does not break the spatial symmetry $x\mapsto -x$. However, due to chaos assisted tunneling, the spatial splitting depends on the quasimomentum because $\beta$ is a parameter that changes the energies of the chaotic states (see Fig. \ref{spasplitvsbeta}). In the regular regime ($\gamma<0.26$), the variations are concentrated around few sharp resonances whereas in the chaotic regime ($\gamma>0.26$), the spatial splitting fluctuates by orders of magnitude, with a correlation scale in $\beta$ which scales like the inverse of the density of states in the chaotic sea, thus as $\sim \hbar_{\rm eff}$. In the case of $\gamma=0.29$ and $\hbar=0.1$, this correlation scale is larger than the width of the experimentally achievable velocity distribution $\Delta \beta \approx 0.02$. Therefore, one expects that averaging over an initial distribution of quasimomentum will not make the spatial tunneling oscillations vanish.

In Fig.~\ref{oscil_h0.11_beta_av} the tunneling oscillations for different quasimomenta are shown for $\varepsilon=0.29$ and $\gamma=0.29$. 
It can be clearly seen that the tunneling oscillations remain when averaging over a small window of quasimomenta uniformly sampled in $[-0.02,0.02]$. In another words having a distribution of quasimomentum to model the atomic cloud does not lead to significant changes in the observation of tunneling oscillations.

The next step is to check whether, within a realistic range of parameters, one can see large fluctuations of the tunneling period (\ref{Ttunnspa}). In complete analogy with Sect.~\ref{tunnelp_oscill}, we chose $\hbar_{\rm eff}$ as the varying parameter. The results are shown in Fig.~\ref{oscillx_vs_hbar_beta_av} a). Large fluctuations by orders of magnitude of the tunneling period can be very well identified.

In order to further validate our approach, the curves shown in Fig.~\ref{oscillx_vs_hbar_beta_av} a) can also be displayed when, for each curve, the time is rescaled by $T_{\rm tunn}$ defined in \eqref{Ttunnspa} and extracted from the spatial splitting at $\beta=0$. Fig.~\ref{oscillx_vs_hbar_beta_av} b) shows such rescaled curves. One can see that for a vast majority of values of $\hbar_{\rm eff}$, all the curves follow the same oscillating trend, namely they are periodic functions of period $1$, in units of $T_{\rm tunn}$ \eqref{Ttunnspa}, and of similar amplitude.

\section{Experimental protocol to observe chaos assisted tunneling in position space using cold atoms}

\label{protocol_exp}

In the previous Section we checked that, if one considers the tunneling between two classical islands, whose center sits along the position axis, then the large variations of the tunneling period due to the presence of classical chaos can be clearly seen. Besides these large fluctuations remain when averaging over a small window of quasimomentum.
Here it is explained how this idea constitutes an original and highly promising suggestion to observe chaos assisted tunneling in a cold atom experiment.
One important requirement is the careful preparation of the initial state. Another great challenge is that we ask to measure the wave function in the {\it position representation}. The way how to achieve both these goals is detailed now.

\subsection{Loading a single classical island along the position axis}

\begin{figure}
\centering
\includegraphics[width=0.7\linewidth,angle=90]{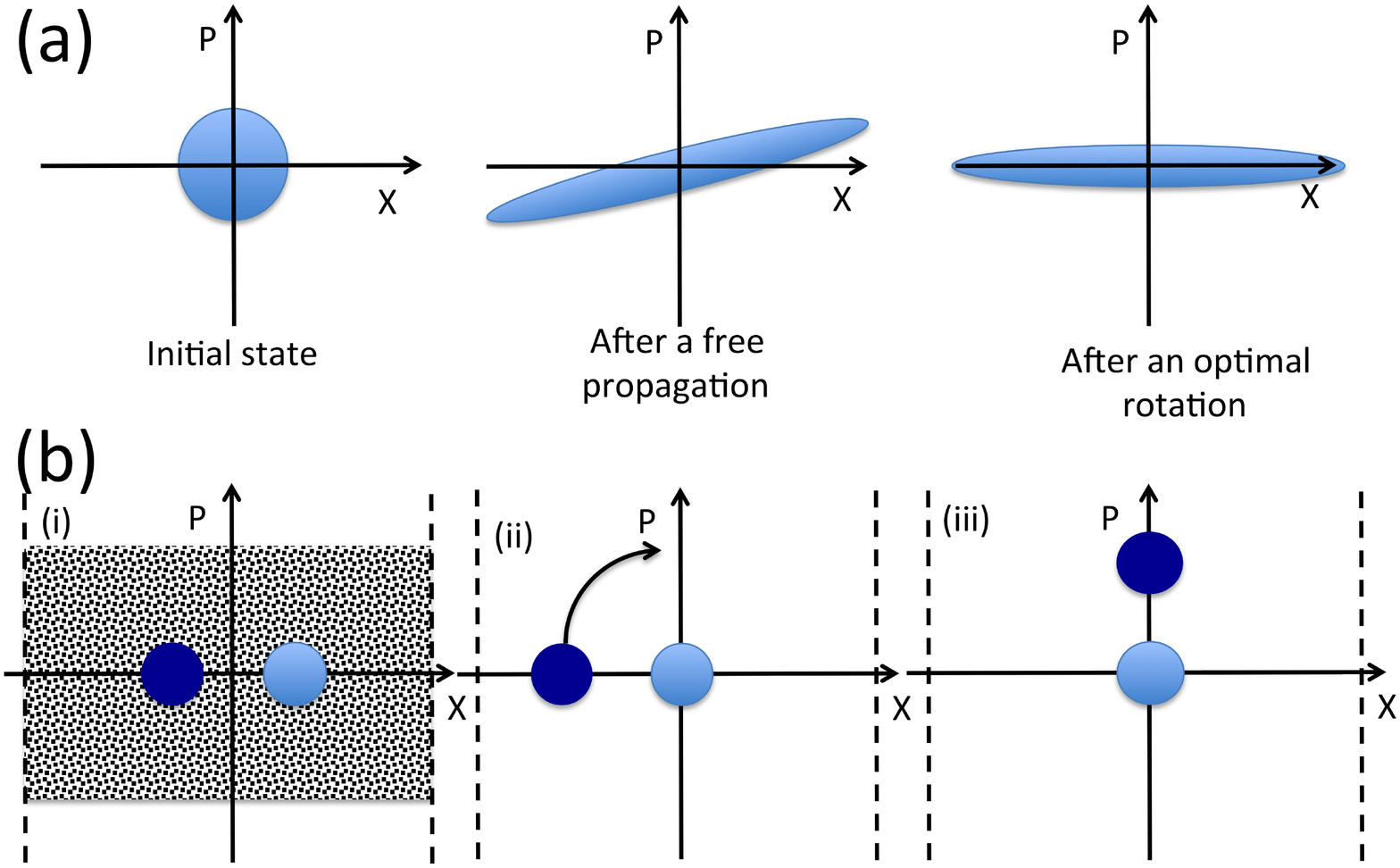}
\caption{(a) Sketch of the 1D delta kick cooling technique. (i) Consider a given initial phase space distribution. (ii) By removing the confinement, the distribution expands along the $X$ direction and position-momentum correlation builds up. (iii) By applying a linear force for a short and appropriate amount of time, the distribution rotates in phase space by the optimal angle so to minimize the momentum and thus the velocity dispersion. (b) Rotation in phase space to measure the information encoded in position space at a submicronic scale. (i) Consider two populated stable islands immersed in a chaotic sea. The different populations are represented by different colors. (ii) Phase space distribution obtained just after the suppression of the modulation and a sudden shift the lattice position (shift by an angle $\varphi_m$). (iii) Free rotation during a quarter of period. The information (population) encoded in the dark disk is then transferred in momentum space.}
\label{protocol}
\end{figure} 

When an interaction free Bose-Einstein condensate is loaded adiabatically into a deep ($s \gg 1$) optical lattice of lattice spacing $d$, the wave function in each well can be approximated by the ground state of an harmonic oscillator 
of angular frequency $\omega_h$: $\psi(x)= e^{-x^2/2a_0^2}/(\pi a_0^2)^{1/4}$ with $a_0=(\hbar/m\omega_h)^{1/2}$.
The expression for $\omega_h$ is obtained by expanding the periodic potential about its minima:
$$ U(x)=-U_0 \cos^2 \left(  \frac{\pi x}{d} \right) \simeq 
-U_0 + \frac{1}{2}m\omega_h^2 x^2\ ,
$$
which leads to:
$$\omega_h = \left( \frac{2\pi^2U_0}{md^2} \right)^{1/2}.$$
We deduce that
$$ \frac{a_0}{d} = \frac{ s^{-1/4}}{\sqrt{2\pi^2}},\quad s=\dfrac{U_0}{E_L} $$ 
Another key parameter is the momentum width, which is related to the width of the velocity distribution of the initial atomic cloud, see e.g.~(\ref{betamin}). In order to achieve the best localization both in position and momentum, it is crucial to use the delta kick cooling technique \cite{deltakick}. It enables one to dilute the sample and thus to decrease the role of interactions and the momentum width. This protocol has been recently implemented in cold atom interferometry experiments, which require a very narrow momentum distribution \cite{deltakick2}. It can be implemented even in the presence of interactions \cite{deltakick3}. The different steps of this protocol are illustrated in Fig.~\ref{protocol} top. One starts with the atom cloud with a vanishing average velocity and a given finite width in velocity. As a result of a free expansion step, the cloud develops a position-velocity correlation. Once the cloud is sufficiently diluted by this expansion, one applies a harmonic force during a short time to induce a rotation in phase space and end up with a much more peaked velocity distribution. This protocol will be applied before loading adiabatically the atoms into the optical lattice.
The next step is to load only a single island to start the tunneling oscillations. This is achieved by applying a sudden phase shift to the optical lattice followed by the setting of the modulation amplitude at the required value.


\subsection {Observation: phase space rotation for analysis in real space}

\begin{figure}
\centering
 \includegraphics[width=\linewidth]{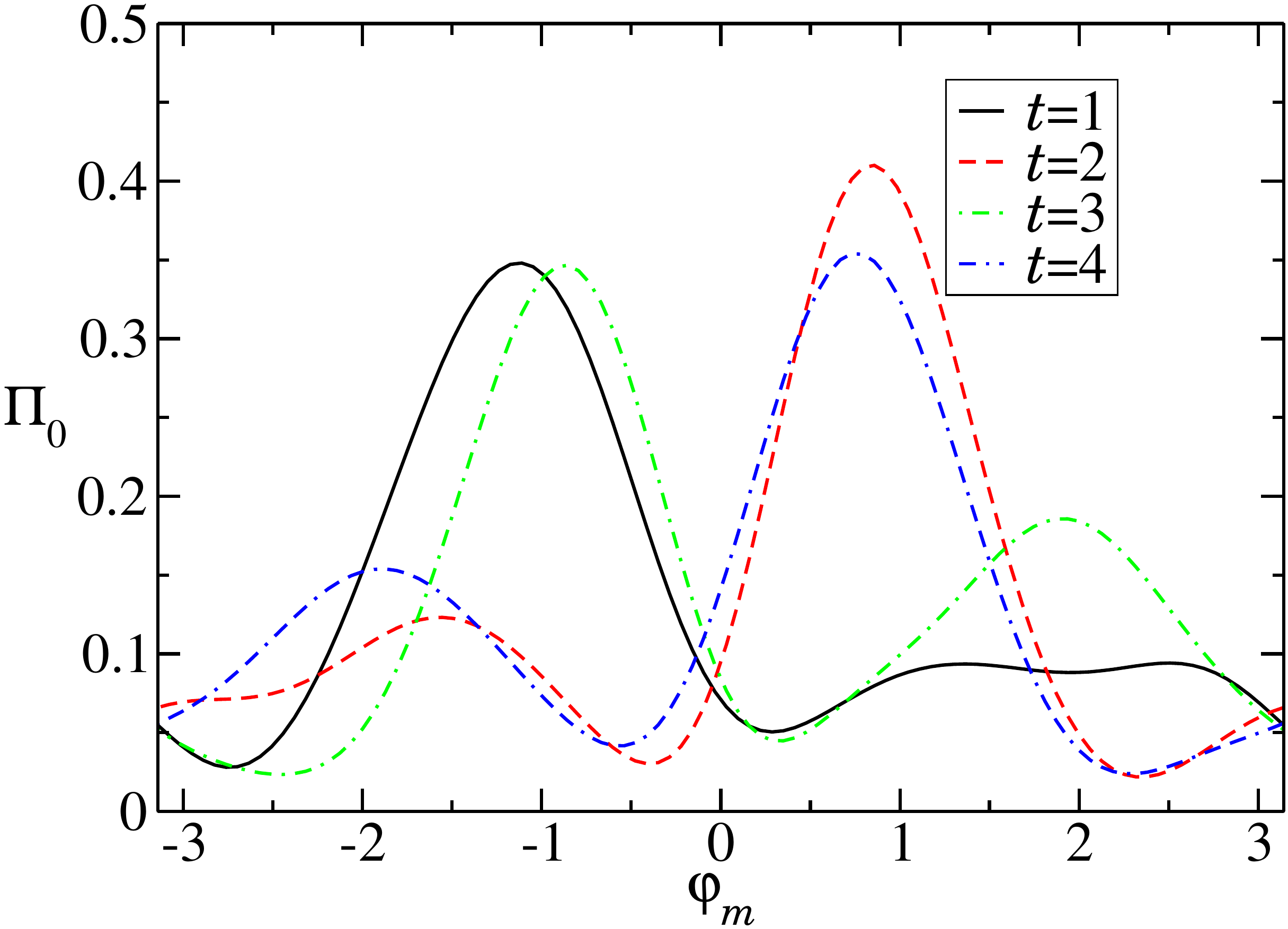}
\caption{(Color online) Phase space rotation of the spatial density probabilities shown in Fig.~\ref{QMP}. The population of the zero velocity class is plotted as a function of the phase shift $\varphi_m$ for different times (see text): $t=0$ corresponds to the initial state, $t=1$ after one period, and so on. The distribution is quite close to the one in position space revealing that the phase space rotation should be an efficient method to observe tunneling in position space at a sub-wavelength scale. The parameters are the following: $\varepsilon=0.29$, $\gamma=0.29$ and $\hbar=0.3$. The initial state is a coherent state centered at $(x,p)=(1.2,0)$ with width $\Delta x=2 \pi/10$. }
\label{Rota}
\end{figure}

Due to the sub-wavelength distance between the wells, the intra-site dynamics in the double wells cannot be observed through in-situ imaging. To observe this dynamics, we propose to use a phase space rotation technique that allows to transfer in momentum space the information encoded in position space at a sub-wavelength size scale. The procedure we propose is summarized in Fig.~\ref{protocol}{ b)}. The initial state corresponds for instance to that of a mixed state with two populated stable islands. Let us assume for instance that the population in each island is different. This is represented in Fig.~\ref{protocol} b) with two different colors (gray scales) for the two islands. When the modulation is abruptly stopped, the phase space distribution is given by the population occupying both stable islands. By shifting suddenly the lattice position by an appropriate angle $\varphi_m$ the population of one island is located about $x=0$ and {$p=0$} while the population of the other island is translated further apart from the center of the lattice sites. The subsequent evolution corresponds to a rotation of the out of center population. Once the rotating distribution  has reached the $p$ axis (after a quarter of period), the lattice is removed and the signal can be inferred from a long time of flight. The displacement resulting from the finite momentum can then be precisely measured. The full protocol requires to repeat the experiment for various phase shifts $\varphi_m$. One can then span a $2\pi$ interval and deduce quantitatively the population occupying each island.

\section{Conclusion}
\label{conclusion}

In this work, we have studied in detail the possibility to observe and characterize both the strong fluctuations and the distribution of tunneling splittings associated with chaos assisted tunneling in a cold atom experiment. We focused on a well-known model, called the atomic modulated pendulum
. This model theoretically displays the phenomenon of chaos assisted tunneling, and has been implemented theoretically \cite{delande1} and experimentally \cite{RaizenScience,PhillipsNature}.

We have studied three different routes in order to observe the strong fluctuations of chaos assisted tunneling using a cold atom implementation of this system. The first one envisions to directly observe the tunneling oscillations between regular islands symmetric in momentum, as in \cite{RaizenScience}. Our results show that, although it was used for the first observations of tunneling phenomena in such systems, the fragility of the $p \mapsto -p$ symmetry with respect to a quasimomentum $\beta$ change makes this method difficult to use to obtain accurate results {in the semiclassical regime relevant for chaos assisted tunneling}. We then assessed a method presented in \cite{delande1}, which proposed to use a Landau-Zener process to force atoms to tunnel by making them cross the $\beta=0$ value. Our study of this scheme shows that the presence of other avoided crossings at different quasimomentum values complicates the picture and requires drastic selection processes that do not leave enough reliable splittings to gather sensible distributions.

At last, we propose{d} a third route to observe chaos assisted tunneling in this system, which should be notably more efficient than the two preceding ones. We use the atomic modulated pendulum in another regime where islands symmetric in space are present. This has the great benefit of being much less sensitive towards the quasimomentum distribution. We confirmed this insight through numerical simulations, and proposed precise answers to experimental challenges with this setting. In particular, the initial state can be precisely prepared using delta kick cooling techniques combined with very careful loading of the wells of the optical lattice. The 
measurement of the wave function of the atomic cloud in position representation
can be achieved by a diabatic translation of the optical lattice, which induces a rotation in the phase space. 

Our analytical estimates and numerical simulations indicate that our protocol is very promising to observe the large fluctuations induced by classical chaos in a quantum tunneling event and to compute their distribution, which have not been observed so far in a quantum system. Our proposal could be implemented with state-of-the-art technology. It will enable to check in a quantum set-up the subtleties of the theoretical predictions of chaos assisted tunneling. Once realized, this proposal will also open the possibilities to use this system for cold atom applications: the optical lattice we studied will present sharp resonances of the tunneling rate $J$, which can be modified by varying the classical and quantum parameters of the system. This opens the way to the use of such an optical lattice dressed by chaos to control cold atoms in a completely new fashion.  For instance, our approach provides for many-body systems in optical lattices a new method to change the ratio $U/J$ (interaction over tunneling energies) in a way that is independent of the specific internal structure of the atomic species under consideration and does not generate atom losses.  Another fascinating perspective includes the control of the range of hoppings in an optical lattice by a proper engineering  of the chaotic sea properties.

\acknowledgements

We thank CalMiP for access to its
supercomputers. This work was supported by Programme Investissements d'Avenir under the program ANR-11-IDEX-
0002-02, reference ANR-10-LABX-0037-NEXT, by the
ANR grant K-BEC No ANR-13-BS04-0001-01, by the ARC grant QUANDROPS 12/17-02 and by the CONICET-
CNRS bilateral project PICS06303R.


\begin{thebibliography}{99}

\bibitem{landau} L. D. Landau, E. M. Lifshitz, {\it Quantum mechanics}, 3rd ed., Pergamon Press (1977)

\bibitem{LZ} O. Morsch and M. Oberthaler, Rev. Mod. Phys., \textbf{78}, 179 (2006).

\bibitem{Santos} L. Santos and L. Roso, Phys. Rev. A, \textbf{60}, 2312 (1999).

\bibitem{Carusotto}  I. Carusotto, D. Embriaco and G. C. La Rocca, Phys. Rev. A, \textbf{65} 053611 (2002).

\bibitem{EPLSG} P. Cheiney, F. Damon, G. Condon, B. Georgeot and D. Gu\'ery-Odelin, EuroPhys. Lett. \textbf{103}, 50006 (2013).

\bibitem{PRAGAPS} F. Damon, G. Condon, P. Cheiney, A. Fortun, B. Georgeot, J. Billy and D. Gu\'ery-Odelin, Phys. Rev. A \textbf{92}, 033614 (2015).

\bibitem{semicl} S. Keshavamurthy and P. Schlagheck, {\it Dynamical Tunneling: Theory and Experiment} (CRC Press, Singapore) 2011.


\bibitem{heller} M. J. Davis, E. J. Heller, J. Chem. Phys. {\bf 75}, 246 (1981)

\bibitem{physrep} O. Bohigas, S. Tomsovic and D. Ullmo, 
Phys. Rep. \textbf{223}, 43 (1993).

\bibitem{tomul} S. Tomsovic and D. Ullmo, 
Phys. Rev. E \textbf{50}, 145 (1994).

\bibitem{leyvraz} F. Leyvraz and D. Ullmo, 
J. Phys. A \textbf{29}, 2529 (1996).

\bibitem{narim} V. A. Podolskiy and E. E. Narimanov, 
Phys. Rev. Lett. \textbf{91}, 263601 (2003).

\bibitem{aberg} S. Aberg, 
Phys. Rev. Lett. \textbf{82}, 299 (1999).

\bibitem{peter} S. Wimberger, P. Schlagheck, C. Eltschka, A. Buchleitner, Phys. Rev. Lett. {\bf 97}, 043001 (2006)

\bibitem{mushroom} A. B\"acker, R. Ketzmerick, S. L\"ock, M. Robnik, G. Vidmar, R. H\"ohmann, U. Kuhl, and H.-J. St\"ockmann,
Phys. Rev. Lett. \textbf{100}, 174103 (2008).

\bibitem{microwave1} C. Dembowski, H.-D. Gr\"af, A. Heine, R. Hofferbert, H. Rehfeld and A. Richter, 
Phys. Rev. Lett. \textbf{84}, 867 (2000)

\bibitem{microwave} R. Hofferbert, H. Alt,, C. Dembowski, H.-D. Gr\"af, H. L. Harney, A. Heine, H. Rehfeld and A. Richter, 
Phys. Rev. E \textbf{71}, 046201 (2005)

\bibitem{barbara} B. Dietz, T. Guhr, B. Gutkin, M. Miski-Oglu, A. Richter, Phys. Rev. E {\bf 90}, 022903 (2014)

\bibitem{delande1} A. Mouchet, C. Miniatura, R. Kaiser, B. Gr\'emaud, and D. Delande,  
Phys. Rev. E \textbf{64}, 016221 (2001)

\bibitem{delande2} A. Mouchet and D. Delande, 
Phys. Rev. E \textbf{67}, 046216 (2003).

\bibitem{artuso} R. Artuso and L. Rebuzzini, 
Phys. Rev. E \textbf{68}, 036221 (2003); S. W\"uster, B. J. Da̧browska-W\"uster and M. J. Davis, Phys. Rev. Lett. \textbf{109}, 080401 (2012).

\bibitem{RaizenScience} D. A. Steck, W. H. Oskay, and M. G. Raizen, 
Science \textbf{293}, 274 (2001).


\bibitem{PhillipsNature} W. K. Hensinger, H. H\"affner, A. Browaeys, N. R. Heckenberg, K. Helmerson, C. McKenzie, G. J. Milburn, W. D. Phillips, S. L. Rolston, H. Rubinsztein-Dunlop and B. Upcroft, 
Nature {\bf 412}, 52 (2001).

\bibitem{PhillipsPRA} W. K. Hensinger, A. Mouchet, P. S. Julienne, D. Delande, N. R. Heckenberg, and H. Rubinsztein-Dunlop, 
Phys. Rev. A \textbf{70}, 013408 (2004)

\bibitem{lenz} M. Lenz, S. W\"uster, C. J. Vale, N. R. Heckenberg, H. Rubinsztein-Dunlop, C. A. Holmes, G. J. Milburn, M. J. Davies, Phys. Rev. A {\bf 88}, 013635 (2013)


\bibitem{splitstep}
R. H. Hardin and F. D. Tappert, 
SIAM Rev. Chronicle {\bf 15}, 423 (1973).


\bibitem{ucf1} S. Washburn and R. Webb, Adv. in Phys. {\bf 35}, 375 (1986)

\bibitem{ucf2} S. Feng and P. A. Lee, Science {\bf 25}, 633 (1991).

\bibitem{deltakick} H. Ammann, N. Christensen, Phys. Rev. Lett. {\bf 78}, 2088 (1997).

\bibitem{JosseCBS} F. Jendrzejewski, K. M\"uller, J. Richard, A. Date, T. Plisson, P. Bouyer, A. Aspect, V. Josse, 
Phys. Rev. Lett. {\bf 109}, 195302 (2012) 

\bibitem{Salomon} M. Ben Dahan, E. Peik, J. Reichel, Y. Castin, C. Salomon, Phys. Rev. Lett. {\bf 76}, 4508 (1996)

\bibitem{Zener} C. Zener, Proc. Roy. Soc A {\bf 137}, 696 (1932)

\bibitem{LZ_3bands} C. E. Carroll, F. T. Hioe, J. Phys. A {\bf 19}, 1151 (1986); J. Phys. A {\bf 19}, 2061 (1986)

\bibitem{Bloch}
S. F\"olling, S. Trotzky, P. Cheinet, M. Feld, R. Saers,  A. Widera, T. M\"uller, I. Bloch, Nature {\bf 448}, 1029 (2007).
\bibitem{GueryOdelin16}
 A. Fortun, C. Cabrera-Gutierrez, G. Condon, E. Michon, J. Billy and D. Gu\'ery-Odelin, ArXiv: 1603.03655v1 [quant-ph] (2016)
 
\bibitem{MouchetLeboeuf} A. Mouchet, P. Leboeuf, Ann. Phys. {\bf 275}, 54 (1999).

\bibitem{Meyer}
 K. Meyer, G. Hall, and D. Offin, {\it Introduction to Hamiltonian dynamical systems and the N-body problem}, vol. 90, Springer Science $\&$ Business Media (2008)

\bibitem{deltakick2} 
H. M\"untinga et al., Phys. Rev. Lett. \textbf{110}, 093602 (2013); T. Kovachy, J. M. Hogan, A. Sugarbaker, S. M. Dickerson, C. A. Donnelly, C. Overstreet, M. A. Kasevich, Phys. Rev. Lett. \textbf{114}, 143004 (2015).

\bibitem{deltakick3} 
G. Condon, A. Fortun, J. Billy, and D. Gu\'ery-Odelin, Phys. Rev. A \textbf{90}, 063616 (2014).
 




\end{thebibliography}
\end{document}